# ReclAIm: A Multi-Agent Framework for Monitoring and Correcting Performance Decline in Medical Imaging AI

Technical Development

**Eleftherios Tzanis, PhD[1,*], Michail E. Klontzas, MD, PhD[1,2,3]**

1. Artificial Intelligence and Translational Imaging (ATI) Lab, Department of Radiology, School of Medicine, University of Crete, Heraklion, Crete, Greece
2. Computational Biomedicine Laboratory, Institute of Computer Science Foundation for Research and Technology Hellas (ICS - FORTH), Heraklion, Crete, Greece
3. Division of Radiology, Department of Clinical Science, Intervention and Technology (CLINTEC), Karolinska Institute, Huddinge, Sweden

***Corresponding author**

**Address for correspondence**

**Eleftherios Tzanis, PhD**
Postdoctoral Researcher
Artificial Intelligence and Translational Imaging (ATI) Lab
Department of Radiology, School of Medicine,
University of Crete, Voutes, 71003, Heraklion, Crete, Greece
E-mail: tzaniseleftherios@gmail.com; etzanis@uoc.gr; eleftherios.tzanis@ki.se
ORCID: 0000-0003-0353-481X

## Abstract

### Purpose

To develop and evaluate a multi-agent framework (ReclAIm) for automated monitoring, detection, and correction of performance decline in medical image classification models.

### Materials and Methods

ReclAIm is a large language model-based multi-agent system that operates through natural language interaction. A master agent coordinating three task-specific agents performed performance evaluation and triggered fine-tuning when substantial performance declines were detected. The fine-tuning workflow incorporated data augmentation, class imbalance handling, and a parameter-anchoring regularization strategy to limit catastrophic forgetting. The system was benchmarked using multiple imaging datasets, including brain MRI, chest CT, and chest radiography, partitioned into model development, inference (performance monitoring), and fine-tuning subsets (60%:20%:20%).

### Results

ReclAIm successfully orchestrated training, evaluation, and performance monitoring across all datasets. Performance discrepancies between test and inference data were detected in 8 of 18 models, prompting fine-tuning workflows that reduced performance gaps. In cases with declines of up to 40.6% (cardiomegaly dataset, InceptionV3), fine-tuning restored performance metrics to within ±2% of baseline values.

### Conclusion

ReclAIm provides a prototype framework for automated monitoring and targeted fine-tuning of medical image classification models, with a natural language interface designed to support accessibility in research and potential clinical applications.

## Introduction

The development and adoption of artificial intelligence (AI) models in clinical practice have accelerated in recent years (1, 2), introducing the challenge of ensuring sustained performance over time. Models that perform well during development may degrade in real-world settings due to data distribution shifts, variations in imaging quality, equipment changes, or modifications in clinical protocols (3, 4). Monitoring model performance, alerting users to performance degradation, and implementing corrective interventions are essential for patient safety and long-term reliability. This need is further reinforced by the EU AI Act (Article 72), which mandates postmarket surveillance for high-risk AI systems (5).

Agentic systems have recently emerged as AI frameworks capable of automating complex tasks in medical AI, ranging from data preprocessing and radiomic feature extraction to model development, evaluation, and deployment (6, 7). In this study, we developed and benchmarked ReclAIm, an accessible, open-source, natural language-interfaced multi-agent framework designed to monitor the performance of medical image classification models, detect performance degradation, and, when necessary, perform fine-tuning to mitigate performance loss.

## Materials and Methods

Data used in this study were retrospectively obtained from publicly available open-source datasets; therefore, no local ethics committee approval was required. The study was conducted between February and October 2025.

### Agentic system

A multi-agent system capable of monitoring the performance of medical image classification models and triggering fine-tuning when necessary was developed using the smolagents library (8). **Figure 1** illustrates the intended functionality of the system, including performance monitoring, identification of performance decline, and triggering of fine-tuning when a substantial drop is detected. The architecture consists of a master agent that orchestrates a team of task-specific agents (**Figure 2**), each equipped with a set of strictly defined Python-based tools. **Figure 3** illustrates an example of user interaction with the system to complete a task. Detailed documentation of the system architecture, agent roles, tools, and prompting strategy is provided in Supplementary Material Part A.

### Datasets

Five publicly available datasets spanning MRI, CT, and radiography were used. For brain tumor classification, a contrast-enhanced T1-weighted MRI dataset (9) comprising 3,064 2D slices from 233 patients (meningioma, glioma, pituitary tumor) was used. For COVID-19 classification, a chest CT dataset (10) of 4,173 2D slices from 210 patients was used for binary COVID-19/non–COVID-19 classification. For 3D imaging, the NoduleMNIST3D dataset (11) containing 1,633 chest CT volumes, was used for binary benign/malignant lung nodule classification. For radiography classification, the PneumoniaMNIST dataset (11), comprising 5,645 chest radiographs, was used for binary pneumonia/normal classification. To assess temporal model degradation, a chest radiography dataset was derived from MIMIC-CXR (12, 13), comprising 35,436 images labeled as no finding or cardiomegaly. Further details on all datasets, including partitioning strategies and temporal splitting, are provided in Supplementary Material Part B.

### Experimental setup

To evaluate system functionality and its ability to assess medical image classification model performance, the following strategy was employed. First, queries were created and provided as input to the master agent to initiate model development. The system was instructed to use the model development subsets of each imaging modality to train ResNet50 (14), InceptionV3 (15), VGG16 (16), EfficientNet (17), and DenseNet121 (18) models. During this step, the master agent invoked the image classification agent, which employed the corresponding tool classes to train and evaluate the models. Training was performed using the training and validation sets, with evaluation conducted on the corresponding test partition. Results obtained on the test data were subsequently compared with inference results on unseen data to assess potential performance discrepancies. Prompts were then issued to the master agent to:

I. Utilize trained models to classify the inference subsets of each modality and compute evaluation metrics (e.g., accuracy, precision, recall, F1score) using the available ground truth labels.

II. Detect potential performance discrepancies by comparing performance on model development test data with that on the inference subset using a heuristic threshold.

III. Analyze the comparison results to determine whether fine-tuning was required and recommend an appropriate fine-tuning strategy.

IV. If necessary, trigger fine-tuning workflows using the fine-tuning dataset to retrain models demonstrating performance discrepancies.

V. Re-evaluate the fine-tuned models on the inference data and compare results with initial test performance to verify improvement.

**System monitoring**

An important component of system evaluation was inspection of execution traces from the master agent and task-specific agents. This ensured that the master agent invoked the appropriate task-specific agents, that each agent selected the correct tool for its assigned task, and that inputs and outputs of each interaction were consistent with the intended workflow. To achieve this, the OpenTelemetry framework (19) was integrated into the system.

**System failure analysis, error handling, and reproducibility**

To generate the reported results, a total of 73 natural language prompts were issued to the agentic system, covering model development, evaluation, deployment, performance comparison, and fine-tuning tasks across the included datasets. These prompts are provided with the open-source code accompanying this manuscript. To further assess system behavior under diverse prompt formulations, an additional set of 59 heterogeneous prompts was designed and executed. These prompts, along with representative OpenTelemetry traces illustrating agent coordination, sequential task execution, and error-handling behavior, are included in Supplementary Material Part C. To evaluate reproducibility under non-deterministic large language model (LLM) conditions, 12 representative prompts spanning different task types were each executed independently five times.

**Computational Performance and LLM API Cost**

Computational performance and resource utilization of the agentic pipeline were assessed across all 18 model–dataset combinations, including model training, inference on unseen data, performance comparison, and fine-tuning when triggered. On a per-model basis, the training phase required an average of 9.5 ± 7.7 minutes of wall-clock time and 8.3 ± 7.1 GPU-minutes. A single monitoring cycle included inference on unseen data, requiring an average of 17.0 ± 20.0 seconds of GPU time per model, and performance comparison, requiring an average of 3.7 ± 1.0 seconds of wall-clock time with no GPU involvement. When fine-tuning was triggered, additional wall-clock time averaged 5.0 ± 4.6 minutes and 4.1 ± 3.7 GPU-minutes per model, followed by an average of 23.0 ± 33.0 GPU-seconds for post–fine-tuning inference and 2.8 ± 1.6 seconds of wall-clock time for post–fine-tuning comparison. In terms of LLM API usage, the end-to-end workflow per model–dataset case consumed an average of 141,622 tokens across all interactions: 41,020 tokens for training (input: 38,080; output: 2,940), 47,870 tokens for inference (input: 43,908; output: 3,962), 15,247 tokens for performance comparison (input: 14,104; output: 1,143), and 37,485 tokens for the fine-tuning phase. This resulted in an average end-to-end LLM API cost of $0.35 per model–dataset case.

**Code and data availability**

The development and implementation code for the ReclAIm project is available at: https://figshare.com/s/e48141f24f297d95f7bb and at: https://github.com/eltzanis/ReclAIm. These repositories include the complete agentic system, the task-specific tools utilized by the agents, and the prompts provided to the master agent to generate the results presented in this study.

All trained and fine-tuned models, along with the corresponding evaluation metrics, comparative analyses, and performance decline detection results, are available at:

https://drive.google.com/drive/folders/12sUAKBMz2asNK5s3GrzatiN3KMFxxfVs?usp=sharing. The repository also includes all plots and JSON files generated from each agent run, documenting the detailed outputs and the specific fine-tuning strategies applied in each case.

## Results

### Classification model development

The agentic system successfully orchestrated the development and evaluation of 18 classification models, including 2D and 3D ResNet50, InceptionV3, VGG16, EfficientNet, and DenseNet121, across five datasets. In all cases, the master agent invoked the appropriate task-specific agent, which analyzed the task and executed the training workflow by inspecting the dataset, verifying class distribution, and selecting suitable configurations, including augmentation strategy, imbalance handling method, evaluation metrics, and hyperparameters. Upon completion, the best-performing and last-epoch models, a confusion matrix, training plots, and a test metrics file reporting accuracy, precision, recall, and F1 scores on the test partition of the model development dataset were saved. These test metrics served as the baseline for comparison with inference on unseen data, enabling the detection of potential performance discrepancies. **Figure 4** illustrates an example of the outputs generated for the EfficientNet model trained on the brain tumor dataset.

### Inference on unseen data

Following training, the master agent initiated inference workflows for all models on the inference datasets, generating confusion matrices, evaluation metrics in JSON format, and per-case CSV files. **Tables 1** and **2** present the test and inference performance for the MRI/CT and radiography datasets, respectively.

### Performance discrepancy detection and fine-tuning

Test and inference metrics were provided to the master agent to identify potential performance discrepancies. The performance comparison agent evaluated macro and per-class accuracy, precision, recall, and F1score across the two stages, generating visualization plots and exporting detailed comparisons and fine-tuning recommendations in JSON and CSV file formats.

When a heuristic threshold of greater than 5% decline was identified in at least one metric, the fine-tuning agent was invoked, initiating a retraining workflow based on the comparison agent's recommendations. After retraining, the updated model was redeployed on inference data, and performance was compared with the original baseline to confirm improvement.

Fine-tuning was required for 8 of 18 models: EfficientNet, InceptionV3, and ResNet50 for the brain tumor MRI dataset; InceptionV3 for the COVID-19 CT dataset; DenseNet121 for the NoduleMNIST3D CT dataset; VGG16 for the pneumonia radiography dataset; and InceptionV3 and VGG16 for the cardiomegaly radiography dataset.

**Effect of fine-tuning**

Per-class percentage differences between test and inference performance before and after fine-tuning for the eight degraded models are shown in **Figures 5 and 6**. For InceptionV3 on the cardiomegaly dataset, substantial pre-fine-tuning discrepancies were observed, particularly for recall (class_0 = -40.6%, class_1 = +20.3%) and F1score (class_0 = -21.2%). Precision also showed changes in opposite directions (class_0 = +13.3%, class_1 = -21.5%). After fine-tuning, these differences were mitigated, with all metric differences reduced to small magnitudes (precision: -2.3% to +1.7%, recall: +1.4% to -2.1%, F1: -0.5% to -0.3%). For VGG16, pre-fine-tuning degradation was more moderate but consistent across metrics (precision: -4.1% to -3.9%, recall: -5.1% to -2.9%, F1: -4.6% to -3.4%). Fine-tuning resulted in partial improvement, reducing the magnitude of these differences (precision: -3.9% to -2.6%, recall: -3.3% to -3.0%, F1: -3.6% to -2.8%). For InceptionV3 on the COVID-19 CT dataset, substantial performance discrepancies were observed prior to fine-tuning. Recall for class_0 decreased markedly, representing an 81.0% relative decline (from 0.922 to 0.175), while class_1 showed a smaller positive change (+9.4%, from 0.908 to 0.993). Precision exhibited changes in opposite directions (class_0 = +6.3%, class_1 = -38.9%), and F1score declined for both classes (class_0 = -67.5%, class_1 = -21.4%). Following fine-tuning, these discrepancies were resolved, with all metrics approaching parity between test and inference performance (precision: +2.1% and -1.4%, recall: -1.9% and +2.3%, F1: +0.1% and +0.4% for class_0 and class_1, respectively).

**Discussion**

This manuscript presents ReclAIm, a prototype multi-agent framework for detecting and mitigating performance drop in medical image classification models. Across 18 model–dataset combinations spanning MRI, CT, and radiography, the system successfully executed model development, performance discrepancy detection, and fine-tuning workflows without human intervention. Fine-tuning was required in 8 of 18 models following detection of performance decline exceeding a 5% threshold in at least one metric, and subsequent retraining reduced discrepancies between test and inference performance to near parity in representative cases. Users interact with the system via natural language prompts, while the master agent interprets tasks and coordinates the appropriate agents. ReclAIm supports both human-in-the-loop operation, in which the system returns a performance assessment and awaits user input, and fully automated end-to-end execution. In both modes, memory capabilities enable retention of prior context and linkage of sequential tasks. In addition to performance-driven adaptation, the framework also supports proactive performance enhancement, allowing users to directly fine-tune models using new data, with parameters specified via natural language or automatically selected based on predefined defaults. While this study focuses on performance-aware fine-tuning to demonstrate monitoring and correction, the same agentic infrastructure can be used to improve initial model performance in a user-guided manner.

Continual learning is increasingly recognized as necessary in medical image classification (20, 21), as model performance may degrade over time due to data distribution shifts, changes in clinical protocols, and modifications in imaging equipment (3, 4). Despite this need, practical implementation remains limited. Key barriers include challenges in reliably detecting and quantifying clinically meaningful changes in model performance and the risk of catastrophic forgetting during incremental updates. In addition, the technical complexity of designing, validating, and maintaining robust monitoring, retraining, and deployment pipelines presents a considerable challenge, particularly in clinical settings with limited AI infrastructure or expertise.

Recent literature demonstrates expanding application of AI agents in the medical domain (7, 22-24), although most existing systems focus on text-based tasks such as diagnostic report generation or clinical summarization (22-24). The effectiveness of agentic systems in more autonomous, interactive settings, where agents manipulate their environment, process data, and orchestrate end-to-end AI workflows, remains largely unexplored. mAIstro demonstrated end-to-end multi-agent functionality for medical imaging, including exploratory data analysis, radiomics feature extraction, model development, evaluation, and deployment (6), but did not address continual learning. The present study extends this work by demonstrating that agentic systems can also monitor and maintain deployed AI models. A multi-agent architecture was selected over a single-agent design to reduce context overload, improve reasoning reliability, and minimize incorrect tool invocation by constraining each agent to a focused set of tasks. In addition, agent coordination enables implicit error checking and adaptive recovery through re-invocation with updated constraints.

ReclAIm interacts with users through natural language. In a clinical scenario, personnel responsible for monitoring AI system performance could periodically prompt ReclAIm to evaluate deployed models. If a substantial performance decline is detected, the system can initiate a multi-agent workflow that performs targeted performance analysis and controlled fine-tuning. Human oversight remains important to approve model updates and critically evaluate post–fine-tuning performance.

The U.S. Food and Drug Administration (FDA) has recognized the need for continuous monitoring of machine learning model performance and adaptive retraining, issuing guidance for Predetermined Change Control Plans (PCCP) based on good machine learning practice principles (25). This guidance supports iterative improvement of medical AI algorithms, provided predefined constraints are met. Although the current ReclAIm prototype does not yet incorporate all elements required for regulatory compliance, such as version control linked to deployed models or formal change-control specifications, it serves as an exploratory framework illustrating how agentic, tool-based workflows could support future PCCP-style processes.

ReclAIm has several strengths and limitations. The ability to perform automated performance monitoring through natural language interaction, combined with the use of state-of-the-art fine-tuning strategies, represents a key strength of this work. Additionally, although ReclAIm is built on LLMs, the use of strictly defined Python tools reduces the risk of hallucinations. Nonetheless, the current implementation requires ground truth labels for inference data to assess performance. In real-world clinical settings, such labels are often unavailable or delayed. A complementary system for automatic extraction of diagnostic labels from free-text radiology reports could help address this limitation and enable more autonomous monitoring.

Another limitation concerns the handling of clustered data (i.e., multiple images or scans originating from the same patient) and the use of fixed thresholds for performance change detection. ReclAIm computes metrics at the image or scan level and does not account for intra-patient correlation when multiple samples belong to the same patient. Therefore, careful dataset curation is required to avoid inclusion of duplicate studies. In addition, the comparison agent applies a single predefined global threshold across all evaluated metrics to identify meaningful changes, without supporting metric-specific thresholds. Furthermore, ReclAIm was evaluated on only one dataset with temporal splitting. For the remaining datasets, the inference subsets were derived from random partitions of the same distribution, as temporal annotations and scanner or protocol metadata were unavailable. Consequently, these experiments primarily demonstrate workflow functionality, confirm that the system detects and responds to metric changes, and assess performance discrepancies attributable to sampling variability. Evaluation on additional temporally annotated datasets would strengthen the evidence base. The study also does not include direct evaluation of catastrophic forgetting or comparison with scripted or single-agent pipelines. Finally, as ChatGPT-4.1 was accessed via API, full local execution was not performed. Assessment with locally deployable LLMs remains important for addressing data privacy concerns and enabling fully local deployment.

To the best of our knowledge, this is the first multi-agent system specifically developed for monitoring and adaptive maintenance of medical image classification models. Through its integrated tools, the system can orchestrate preprocessing, training, deployment, and performance monitoring, and can trigger corrective fine-tuning workflows when performance degradation is detected. Its design, relying on natural language interaction, enables accessible implementation and represents a step toward broader adoption in research settings and, potentially, clinical environments. Future directions may include extending the framework to other types of AI models and evaluating fully local implementations using smaller LLMs to enhance data privacy.

**Acknowledgements**

None.

**Table 1.** Test and inference evaluation metrics for the MRI and CT datasets.

| **Brain Tumor MRI dataset** | | | | | | | | | | | | | | |
|---|---|---|---|---|---|---|---|---|---|---|---|---|---|---|
| **Model** | **Evaluation Dataset** | **Accuracy** | **Precision** | | | | **Recall** | | | | **F1** | | | |
| | | | Macro | Class 0 | Class 1 | Class 2 | Macro | Class 0 | Class 1 | Class 2 | Macro | Class 0 | Class 1 | Class 2 |
| **EfficientNet** | Test set | 96.5% | 0.947 | 0.995 | 0.862 | 0.984 | 0.968 | 0.944 | 0.976 | 0.984 | 0.956 | 0.969 | 0.915 | 0.984 |
| | Inference set | 96.1% | 0.962 | 0.941 | 0.961 | 0.983 | 0.959 | 0.977 | 0.925 | 0.975 | 0.960 | 0.958 | 0.943 | 0.979 |
| **Fine-tuned EfficientNet** | Inference set | 96.4% | 0.964 | 0.954 | 0.962 | 0.975 | 0.962 | 0.969 | 0.935 | 0.983 | 0.963 | 0.962 | 0.948 | 0.979 |
| **InceptionV3** | Test set | 96.7% | 0.951 | 0.990 | 0.872 | 0.989 | 0.971 | 0.963 | 0.988 | 0.963 | 0.960 | 0.977 | 0.927 | 0.976 |
| | Inference set | 92.5% | 0.926 | 1.000 | 0.874 | 0.905 | 0.928 | 0.869 | 0.972 | 0.942 | 0.924 | 0.930 | 0.920 | 0.923 |
| **Fine-tuned InceptionV3** | Inference set | 95.8% | 0.958 | 0.962 | 0.911 | 1.000 | 0.957 | 0.977 | 0.953 | 0.942 | 0.957 | 0.969 | 0.932 | 0.970 |
| **ResNet50** | Test set | 95.3% | 0.932 | 0.991 | 0.837 | 0.967 | 0.947 | 0.981 | 0.928 | 0.931 | 0.938 | 0.986 | 0.880 | 0.949 |
| | Inference set | 93.6% | 0.935 | 0.939 | 0.932 | 0.935 | 0.934 | 0.954 | 0.897 | 0.950 | 0.934 | 0.947 | 0.914 | 0.943 |
| **Fine-tuned ResNet50** | Inference set | 96.4% | 0.965 | 0.977 | 0.981 | 0.937 | 0.963 | 0.962 | 0.944 | 0.983 | 0.964 | 0.969 | 0.962 | 0.960 |
| **VGG16** | Test set | 94.3% | 0.916 | 1.000 | 0.748 | 1.000 | 0.955 | 0.912 | 1.000 | 0.952 | 0.928 | 0.954 | 0.856 | 0.976 |
| | Inference set | 95.8% | 0.958 | 0.969 | 0.904 | 1.000 | 0.958 | 0.969 | 0.963 | 0.942 | 0.957 | 0.969 | 0.932 | 0.970 |

| **Covid-19 CT dataset** | | | | | | | | | | | |
|---|---|---|---|---|---|---|---|---|---|---|---|
| **Model** | **Evaluation Dataset** | **Accuracy** | **Precision** | | | **Recall** | | | **F1** | | |
| | | | Macro | Class 0 | Class 1 | Macro | Class 0 | Class 1 | Macro | Class 0 | Class 1 |
| **EfficientNet** | Test set | 97.1% | 0.971 | 0.977 | 0.965 | 0.970 | 0.961 | 0.979 | 0.971 | 0.969 | 0.972 |
| | Inference set | 97.4% | 0.974 | 0.977 | 0.970 | 0.973 | 0.968 | 0.979 | 0.974 | 0.972 | 0.975 |
| **InceptionV3** | Test set | 91.5% | 0.914 | 0.902 | 0.927 | 0.915 | 0.922 | 0.908 | 0.915 | 0.912 | 0.917 |
| | Inference set | 60.0% | 0.762 | 0.959 | 0.566 | 0.584 | 0.175 | 0.993 | 0.508 | 0.296 | 0.721 |
| **Fine-tuned InceptionV3** | Inference set | 91.7% | 0.917 | 0.921 | 0.914 | 0.917 | 0.905 | 0.928 | 0.917 | 0.913 | 0.921 |
| **ResNet50** | Test set | 95.2% | 0.952 | 0.926 | 0.978 | 0.953 | 0.978 | 0.928 | 0.952 | 0.951 | 0.953 |
| | Inference set | 93.2% | 0.932 | 0.911 | 0.952 | 0.932 | 0.950 | 0.915 | 0.932 | 0.930 | 0.933 |
| **VGG16** | Test set | 97.1% | 0.971 | 0.967 | 0.974 | 0.971 | 0.972 | 0.969 | 0.971 | 0.970 | 0.972 |
| | Inference set | 96.5% | 0.965 | 0.965 | 0.965 | 0.965 | 0.963 | 0.968 | 0.965 | 0.964 | 0.967 |

| **NoduleMNIST3D CT dataset** | | | | | | | | | | | |
|---|---|---|---|---|---|---|---|---|---|---|---|
| **Model** | **Evaluation Dataset** | **Accuracy** | **Precision** | | | **Recall** | | | **F1** | | |
| | | | Macro | Class 0 | Class 1 | Macro | Class 0 | Class 1 | Macro | Class 0 | Class 1 |
| **3D ResNet50** | Test set | 84.5% | 0.790 | 0.894 | 0.686 | 0.784 | 0.902 | 0.667 | 0.787 | 0.898 | 0.676 |
| | Inference set | 84.7% | 0.791 | 0.912 | 0.670 | 0.810 | 0.882 | 0.738 | 0.800 | 0.897 | 0.702 |
| **DenseNet121** | Test set | 85.8% | 0.815 | 0.889 | 0.742 | 0.784 | 0.929 | 0.639 | 0.797 | 0.908 | 0.687 |
| | Inference set | 83.4% | 0.776 | 0.897 | 0.655 | 0.785 | 0.882 | 0.688 | 0.780 | 0.889 | 0.671 |
| **Fine-tuned DenseNet121** | Inference set | 85.6% | 0.811 | 0.890 | 0.732 | 0.786 | 0.923 | 0.650 | 0.797 | 0.906 | 0.689 |

95% confidence intervals (CIs) for all reported metrics are provided in Supplementary Material Part D (Supplementary Table S1)

**Table 2**. Test and inference evaluation for radiography datasets

| **Cardiomegaly** | | | | | | | | | | | |
|---|---|---|---|---|---|---|---|---|---|---|---|
| **Model** | **Evaluation Dataset** | **Accuracy** | **Precision** | | | **Recall** | | | **F1** | | |
| | | | Macro | Class 0 | Class 1 | Macro | Class 0 | Class 1 | Macro | Class 0 | Class 1 |
| **EfficientNet** | Test set | 86.5% | 0.865 | 0.863 | 0.866 | 0.865 | 0.866 | 0.863 | 0.865 | 0.865 | 0.864 |
| | Inference set | 84.1% | 0.841 | 0.826 | 0.856 | 0.841 | 0.854 | 0.829 | 0.841 | 0.840 | 0.842 |
| **InceptionV3** | Test set | 83.2% | 0.834 | 0.809 | 0.860 | 0.832 | 0.870 | 0.794 | 0.832 | 0.838 | 0.825 |
| | Inference set | 74.1% | 0.795 | 0.916 | 0.675 | 0.736 | 0.517 | 0.955 | 0.726 | 0.661 | 0.791 |
| **Fine-tuned InceptionV3** | Inference set | 82.8% | 0.832 | 0.790 | 0.874 | 0.830 | 0.882 | 0.777 | 0.828 | 0.834 | 0.823 |
| **ResNet50** | Test set | 84.2% | 0.842 | 0.851 | 0.833 | 0.842 | 0.828 | 0.855 | 0.842 | 0.839 | 0.844 |
| | Inference set | 83.0% | 0.830 | 0.829 | 0.831 | 0.830 | 0.821 | 0.838 | 0.830 | 0.825 | 0.835 |
| **VGG16** | Test set | 85.7% | 0.857 | 0.853 | 0.861 | 0.857 | 0.863 | 0.851 | 0.857 | 0.858 | 0.856 |
| | Inference set | 82.3% | 0.823 | 0.818 | 0.827 | 0.823 | 0.819 | 0.827 | 0.823 | 0.819 | 0.827 |
| **Fine-tuned VGG16** | Inference set | 82.9% | 0.829 | 0.820 | 0.839 | 0.830 | 0.834 | 0.825 | 0.829 | 0.827 | 0.832 |
| **PneumoniaMNIST** | | | | | | | | | | | |
| **Model** | **Evaluation Dataset** | **Accuracy** | **Precision** | | | **Recall** | | | **F1** | | |
| | | | Macro | Class 0 | Class 1 | Macro | Class 0 | Class 1 | Macro | Class 0 | Class 1 |
| **EfficientNet** | Test set | 94.2% | 0.937 | 0.927 | 0.947 | 0.915 | 0.856 | 0.975 | 0.925 | 0.890 | 0.961 |
| | Inference set | 93.0% | 0.924 | 0.912 | 0.936 | 0.895 | 0.820 | 0.971 | 0.908 | 0.863 | 0.953 |
| **InceptionV3** | Test set | 94.2% | 0.921 | 0.872 | 0.971 | 0.936 | 0.924 | 0.949 | 0.928 | 0.897 | 0.960 |
| | Inference set | 94.5% | 0.934 | 0.909 | 0.958 | 0.927 | 0.886 | 0.967 | 0.930 | 0.897 | 0.963 |
| **ResNet50** | Test set | 93.3% | 0.913 | 0.868 | 0.958 | 0.919 | 0.890 | 0.949 | 0.916 | 0.879 | 0.954 |
| | Inference set | 92.5% | 0.907 | 0.868 | 0.945 | 0.902 | 0.851 | 0.952 | 0.904 | 0.859 | 0.949 |
| **VGG16** | Test set | 94.9% | 0.936 | 0.907 | 0.965 | 0.936 | 0.907 | 0.965 | 0.936 | 0.907 | 0.965 |
| | Inference set | 93.7% | 0.931 | 0.920 | 0.942 | 0.906 | 0.839 | 0.973 | 0.917 | 0.877 | 0.957 |
| **Fine-tuned VGG16** | Inference set | 94.1% | 0.933 | 0.916 | 0.950 | 0.916 | 0.861 | 0.971 | 0.924 | 0.887 | 0.960 |

95% confidence intervals (CIs) for all reported metrics are provided in Supplementary Material Part D (Supplementary Table S2)

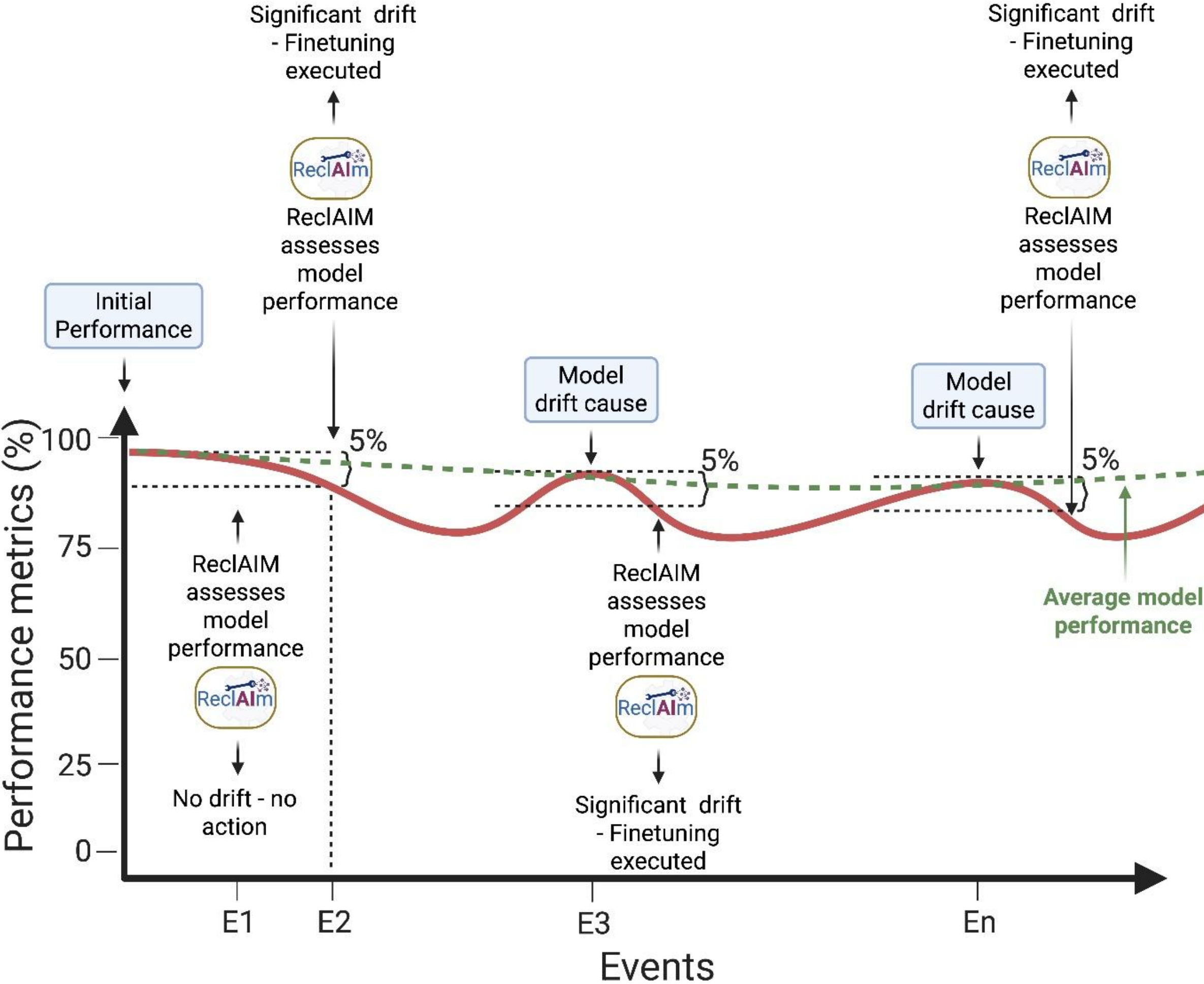

**Figure 1.** Schematic representation of potential uses of ReclAIm. The model is initially trained, achieving certain performance. Various events can cause model performance drop (e.g. deployment at another site, seasonal changes of disease prevalence). ReclAIm can be asked in natural language to assess model performance at regular time points (Events - E1-n). When it detects a considerable discrepancy in model performance, it executes a finetuning process that aims to keep a relatively stable average model performance.

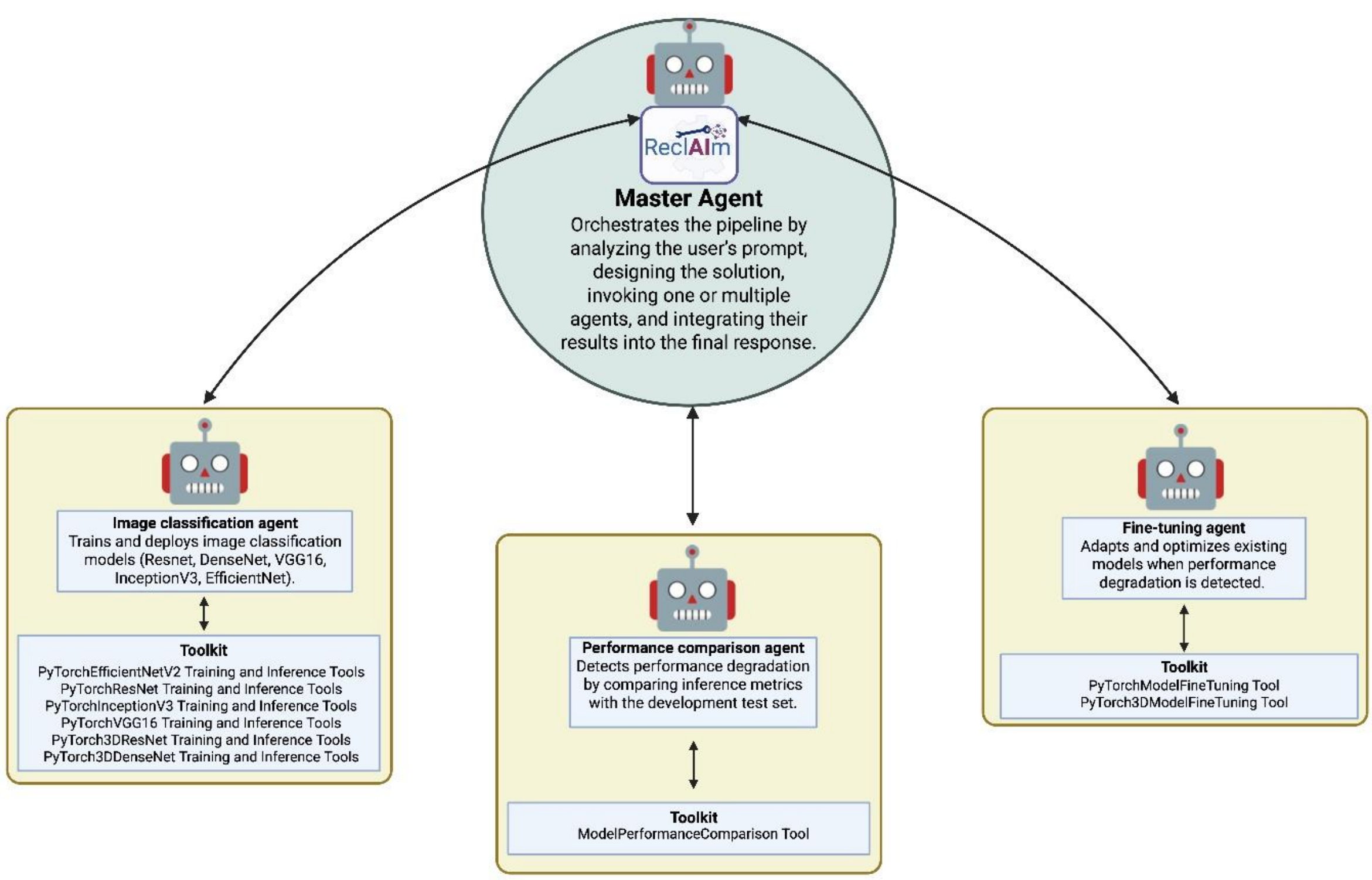

**Figure 2.** Master agent and task specific agents with their respective toolkits.

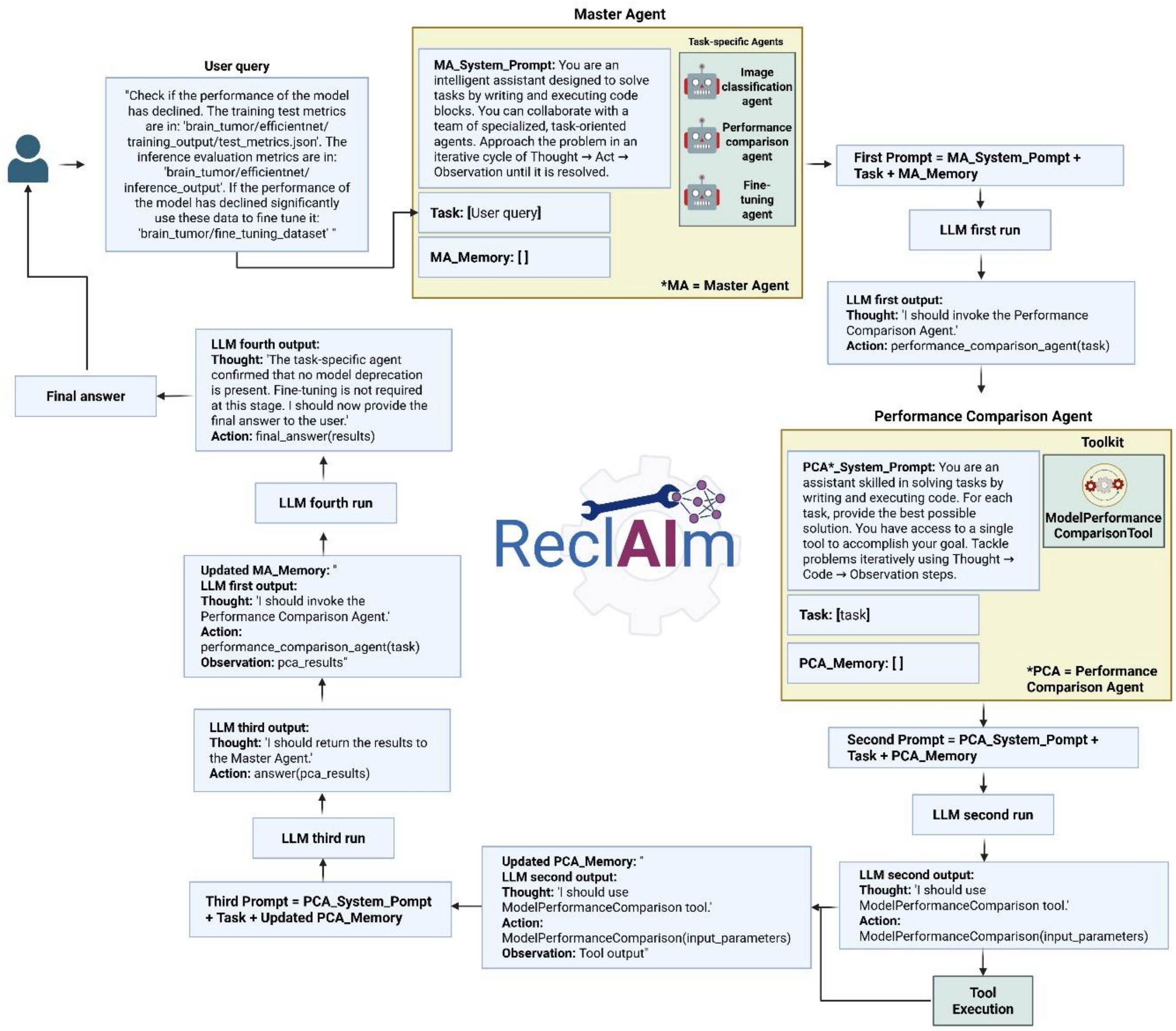

**Figure 3.** Flowchart demonstrating an example of a complete agentic workflow indicating the interaction between ReclAIm and the user. MA: Master Agent, LLM: Large Language Model, PCA: Performance Comparison Agent.

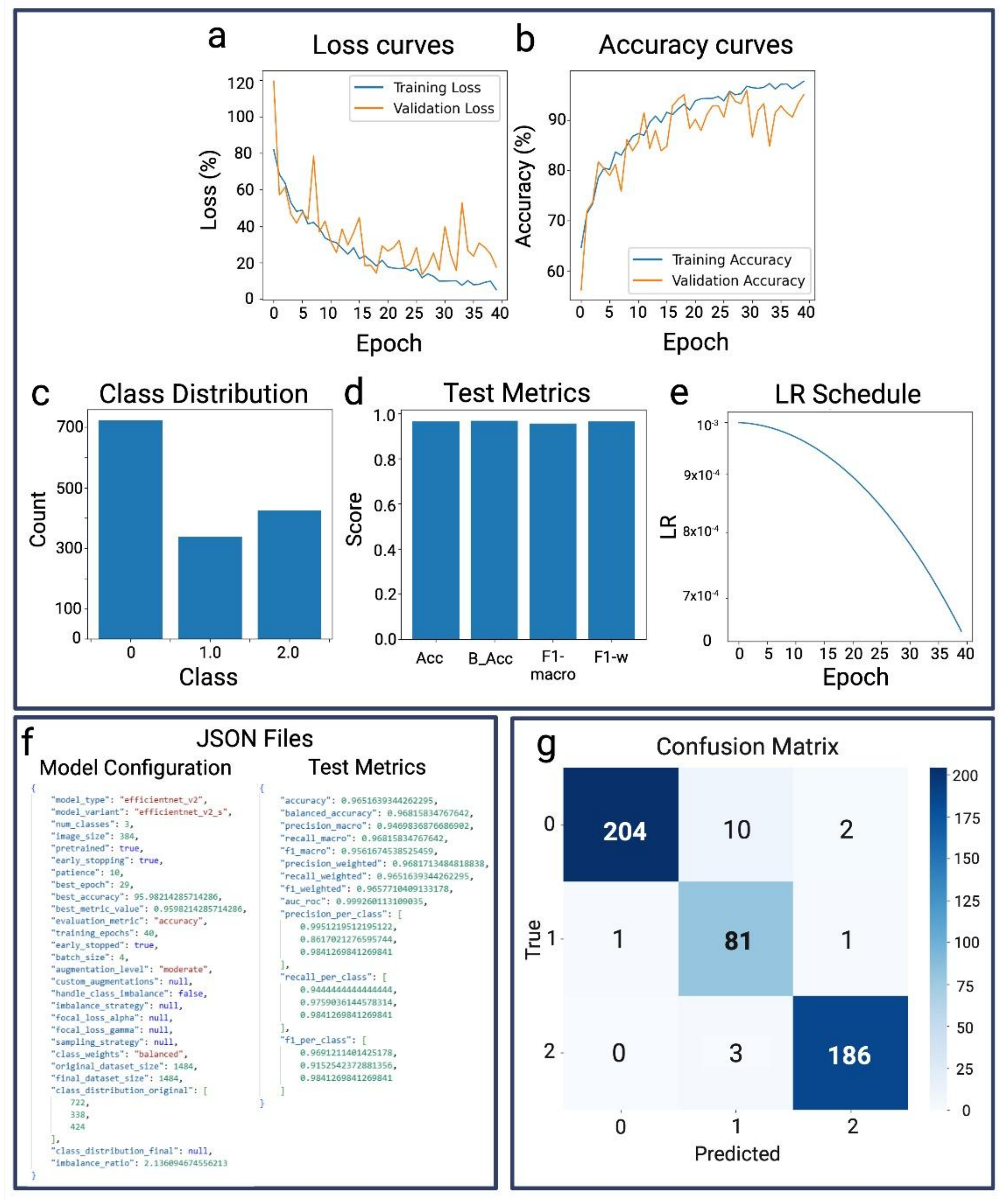

**Figure 4.** Indicative outputs of ReclAIm for an EfficientNet model trained on the brain tumor MRI dataset. LR: Learning Rate.

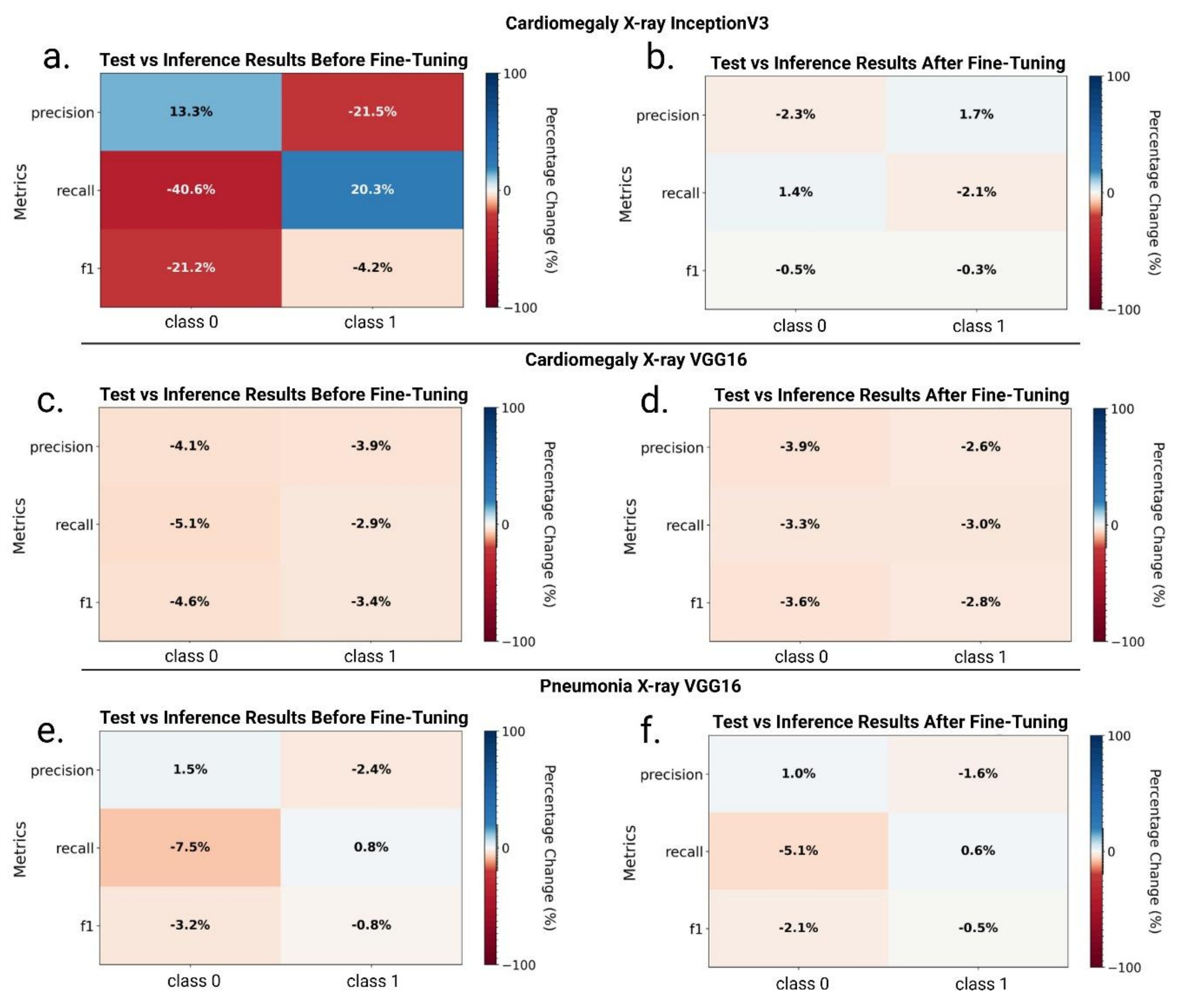

**Figure 5.** Test and inference results before and after fine-tuning with ReclAIm for Cardiomegaly X-ray InceptionV3 (a, b), Cardiomegaly X-ray VGG16 (c, d) and Pneumonia X-ray VGG16 (e, f) models.

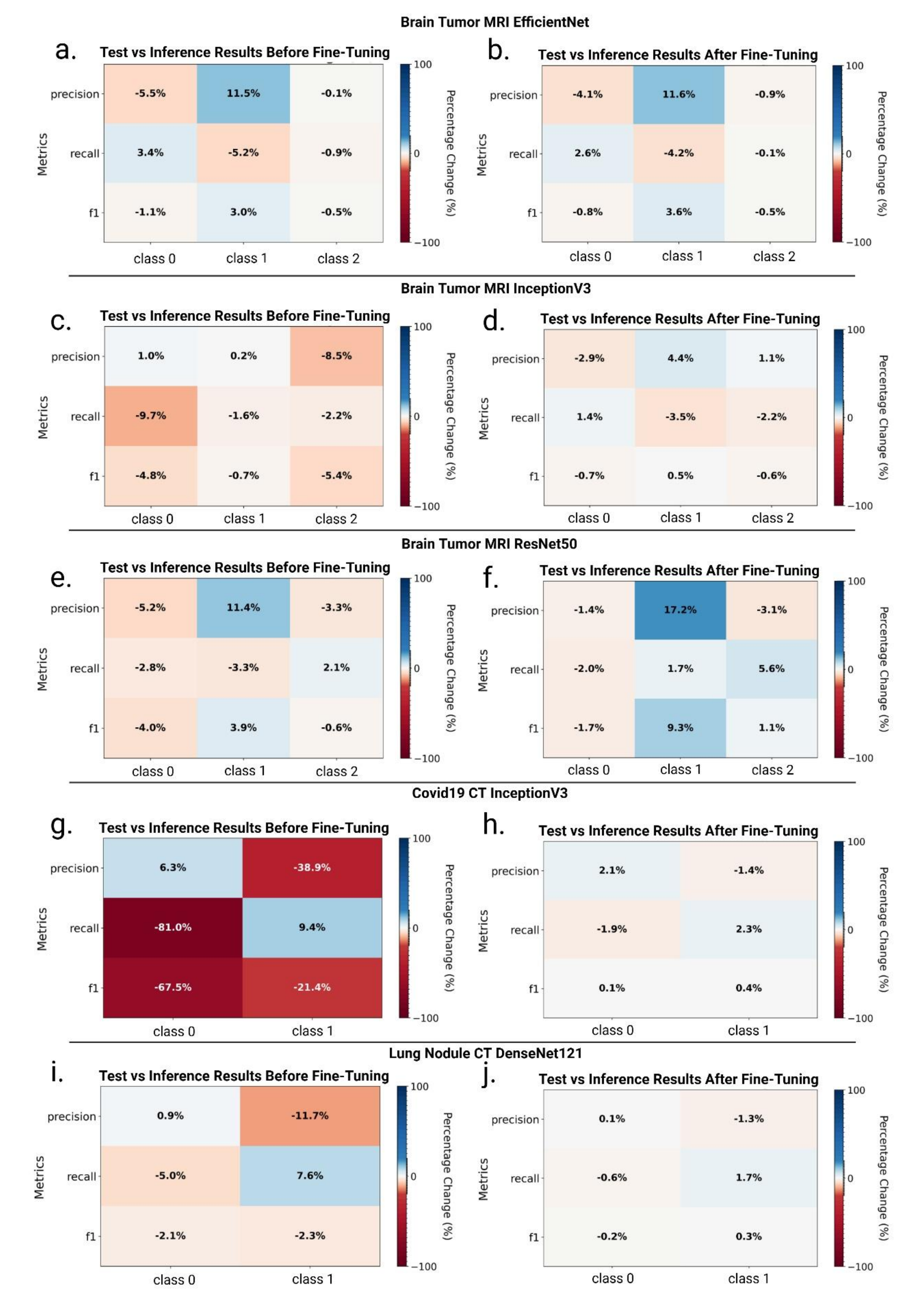

**Figure 6.** Test and inference results before and after fine-tuning with ReclAIm for Brain Tumor MRI EfficientNet (a, b), Brain Tumor MRI InceptionV3 (c, d), Brain Tumor MRI ResNet50 (e, f), Covid-19 CT InceptionV3 (g, h) and Lung Nodule CT DenseNet121 (i, j) models.

**Supplementary Material Part A - ReclAIm: A Multi-Agent Framework for Monitoring and Correcting Performance Decline in Medical Imaging AI**

This section provides a detailed description of the multi-agent system introduced in Section 2.1, including the role and tools of each task-specific agent, the catastrophic forgetting prevention strategy, and the system's prompting architecture and operational workflow.

I. The *image classification agent* is responsible for training and deploying a variety of deep learning models, including 2D and 3D ResNet (1), InceptionV3 (2), VGG16 (3), EfficientNet (4) and DenseNet (5). This agent can be used by users who wish to train custom deep learning models with in-house data. It utilizes dedicated python tool classes that support model training and deployment, provide configurable augmentation pipelines ranging from basic to advanced or custom transformations, and implement multiple strategies for handling class imbalance, including weighted or focal loss functions as well as over- or under-sampling.

II. The *performance comparison agent* evaluates inference outputs against the corresponding test results to assess model accuracy and generalizability. It compares a series of global metrics such as accuracy, balanced accuracy, area under the ROC curve, precision, recall, and F1-score, as well as per-class performance, to identify potential degradation when a model is applied to new data. By quantifying considerable declines and generating recommendations, including fine-tuning or focused retraining on underperforming classes, the agent provides a mechanism for monitoring and maintenance of deployed models.

III. The *fine-tuning agent* adapts and optimizes existing models when performance degradation is detected. It applies strategies such as full or partial layer retraining, head-only adaptation, and gradual unfreezing, while supporting differential learning rates and optimizer reinitialization (6, 7). Moreover, it integrates data augmentation, class imbalance handling, and a parameter-anchoring regularization strategy (8-10) designed to limit catastrophic forgetting during adaptation.

To limit catastrophic forgetting during fine-tuning, a parameter-anchoring regularization strategy was implemented that constrains the updated model weights to remain close to those of the original network, following principles similar to L2-SP regularization and Elastic Weight Consolidation (9, 11). Technically, this was implemented by adding a regularization term to the loss function that penalizes deviations between the current parameters $\theta$ and the original parameters $\theta_0$. The total optimization objective is expressed as:

$$L_{total} = L_{task} + \lambda \sum_i \|\theta_i - \theta_{0,i}\|_2$$

where $L_{task}$ is the standard training loss and λ controls the strength of the penalty. This mechanism limits destructive updates, ensuring that the model retains previously acquired knowledge while adapting to new data. An important advantage of this approach is that it does not require access to the original training data, enabling fine-tuning even for models whose training datasets are unavailable, restricted, or proprietary, such as in commercial or privacy-sensitive medical imaging applications. This lightweight regularization technique offers a practical and effective way to balance adaptation and stability during continual learning.

The system is implemented in Python and relies on the following software components: the smolagents library (12) provides the agentic orchestration framework and the CodeAgent class (https://huggingface.co/docs/smolagents/v1.24.0/en/reference/agents#smolagents.CodeAgent), LiteLLM (13) serves as the intermediate model abstraction layer, routing API calls to GPT-4.1 via the OpenAI API (accessed as openai/gpt-4.1 through smolagents' LiteLLMModel interface - https://huggingface.co/docs/smolagents/v1.24.0/en/reference/models#smolagents.LiteLLMModel), PyTorch (14) underpins all classification model training, inference, and fine-tuning tools and OpenTelemetry (15) provides observability and execution tracing. GPT-4.1 serves as the core reasoning engine for both the master agent and all task-specific agents, using the default temperature setting (temperature = 1).

The operation of the agentic system relies on the system prompt, which serves as the permanent set of instructions guiding agent behavior, defining agent roles, available tools, and the prompting strategy. The ReAct (16) prompting approach was adopted, where the system operates in a continuous cycle of reasoning, acting, and observing. All agents use the smolagents CodeAgent implementation, in which the LLM generates executable Python code snippets that invoke the appropriate tools with the required parameters. This is the primary mechanism by which LLM reasoning is translated into tool execution.

The control loop operates as follows: the user prompt, the agent's system prompt, and any information stored in memory are passed to GPT-4.1 via LiteLLM (13). The LLM generates a reasoning trace followed by a Python code block that calls the relevant tool. The smolagents framework executes this code, captures the tool's structured return object (a dictionary containing outputs, file paths, and status information), and appends the result to the conversation context. The LLM then evaluates the observation and either produces the final response or initiates a new reasoning-acting cycle.

Runtime errors raised during tool execution are captured by the CodeAgent framework and appended to the conversation context as structured error messages. The LLM analyzes the error, infers the cause, and generates a revised code block with adjusted parameters.

The tools available to each specialized agent are documented below. This manuscript was written according to the CLAIM guidelines (17).

# Tools documentation

# PyTorch Training Tools Documentation

## Overview

The **PyTorch Training Tools** are a comprehensive suite of deep learning training utilities designed for medical image classification tasks. The suite includes six specialized tools, each optimized for different CNN architectures while maintaining consistent interfaces and functionality. These tools provide state-of-the-art training capabilities with advanced data augmentation, early stopping, comprehensive evaluation, class imbalance and error handling.

## Supported Architectures

### 1. PyTorchResNetTrainingTool

**Supported Models**: ResNet18, ResNet34, ResNet50, ResNet101, ResNet152 - **Image Size**: 224×224 pixels.

### 2. PyTorchInceptionV3TrainingTool

**Supported Models**: Inception V3 (with/without auxiliary logits) - **Image Size**: 299×299 pixels.

### 3. PyTorchVGG16TrainingTool

**Supported Models**: VGG16, VGG16 with Batch Normalization - **Image Size**: 224×224 pixels.

### 4. PyTorchEfficientNetV2TrainingTool

**Supported Models**: EfficientNetV2-S, EfficientNetV2-M, EfficientNetV2-L - **Image Sizes**: 384×384 (S) and 480×480 (M/L) pixels.

### 5. PyTorch3DResNetTrainingTool

**Supported Models:** 3D-ResNet10, 3D-ResNet18, 3D-ResNet34, 3D-ResNet50 - Volume Size: Configurable

### 6. PyTorch3DDenseNetTrainingTool

**Supported Models:** DenseNet121, DenseNet169, DenseNet201, DenseNet264 - Volume Size: Configurable

## Universal Input Parameters

All training tools share the same interface with architecture-specific optimizations:

### Required Parameters

| Parameter | Type | Description |
|---|---|---|
| data_dir | string | Directory containing dataset with train/val/test folders |
| output_dir | string | Directory where trained model and results will be saved |
| num_classes | integer | Number of classes for classification |

### Optional Parameters

| Parameter | Type | Default | Description |
|---|---|---|---|
| num_epochs | integer | 10 | Number of training epochs |
| batch_size | integer | 16 | Batch size for training |
| pretrained | boolean | True | Whether to use pretrained ImageNet weights |
| early_stopping | boolean | True | Whether to use early stopping |
| patience | integer | 5 | Number of epochs to wait before early stopping |
| augmentation_level | string | "basic" | Data augmentation: 'none', 'basic', 'moderate', 'heavy', 'custom' |
| custom_augmentations | object | None | Custom augmentation settings when level is 'custom' |

### Architecture-Specific Parameters

#### ResNet-Specific

| Parameter | Type | Default | Description |
|---|---|---|---|
| model_type | string | "resnet50" | ResNet variant: resnet18, resnet34, resnet50, resnet101, resnet152 |

#### Inception V3-Specific

| Parameter | Type | Default | Description |
|---|---|---|---|
| aux_logits | boolean | True | Whether to use auxiliary logits during training |

#### VGG16-Specific

| Parameter | Type | Default | Description |
|---|---|---|---|
| use_batch_norm | boolean | False | Whether to use VGG16 with batch normalization |

#### EfficientNetV2-Specific

| Parameter | Type | Default | Description |
|---|---|---|---|
| model_variant | string | "efficientnet_v2_s" | Variant: efficientnet_v2_s, efficientnet_v2_m, efficientnet_v2_l |
| use_custom_image_size | boolean | False | Whether to use custom image size |
| custom_image_size | integer | 224 | Custom image size (width=height) |

## **Data Structure Requirements** (2D medical imaging data)

All training tools expect data organized in the standard PyTorch ImageFolder structure:

### Standard Directory Structure

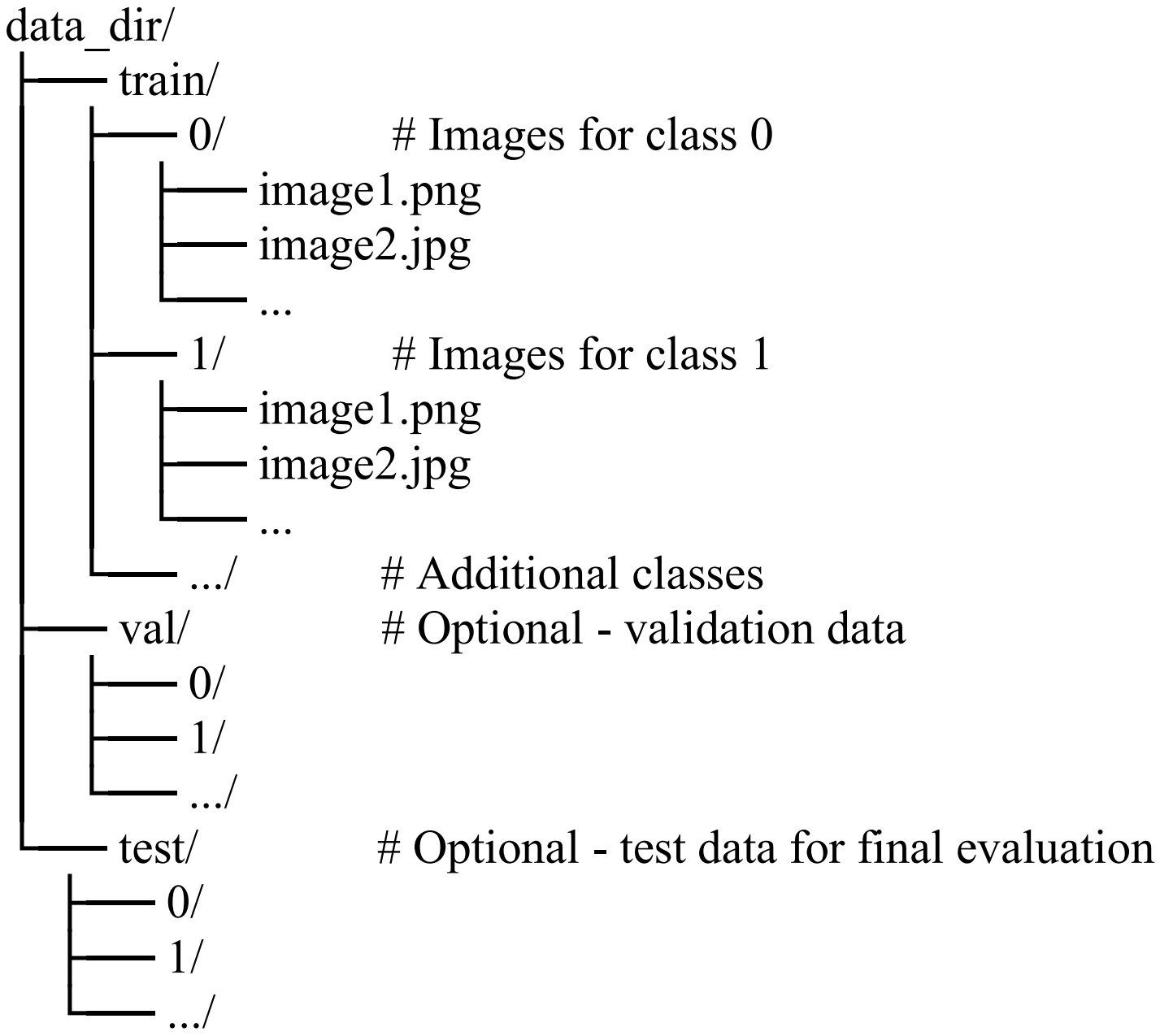

### Alternative: CSV-Based Labels

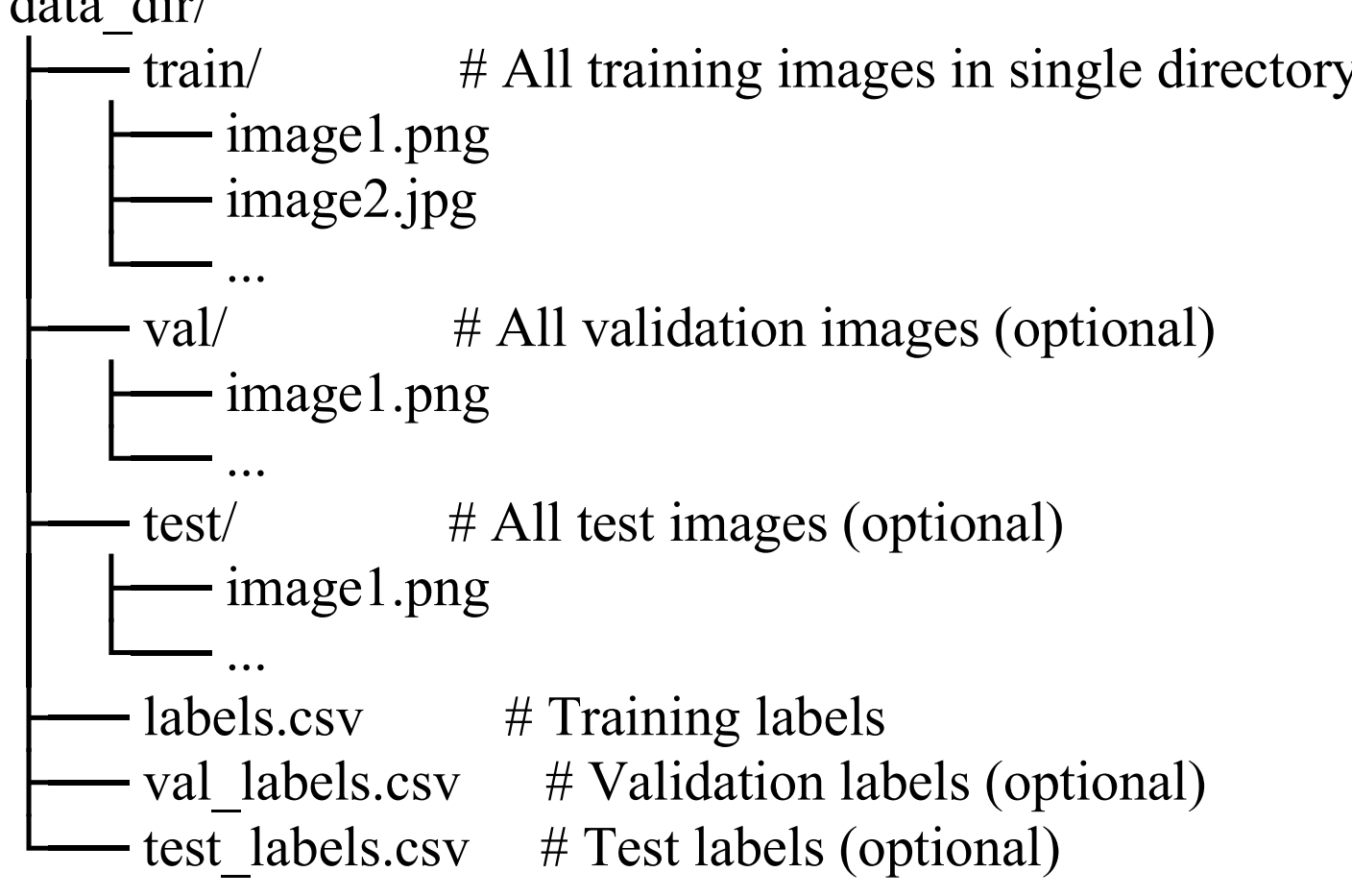

#### CSV Format

filename,label
image1.png,0
image2.jpg,1
image3.png,0

### Automatic Data Handling

- **Validation Split**: If no val/ directory exists, automatically splits training data 80/20
- **Format Detection**: Automatically detects CSV vs directory structure
- **Class Detection**: Automatically counts classes from directory structure or CSV labels
- **Flexible Naming**: Supports both 'filename' and 'image_name' columns in CSV files

## **Data Structure Requirements** (3D medical imaging data)

The tools PyTorch3DResNetTrainingTool and PyTorch3DDenseNetTrainingTool work with numpy .npz files containing 3D medical imaging data:

Single File Format:

data.npz containing:

- train_images: shape (N, 28, 28, 28) or (N, 1, 28, 28, 28)

- train_labels: shape (N,) or (N, 1)

- val_images: shape (M, 28, 28, 28) - optional

- val_labels: shape (M,) - optional

- test_images: shape (K, 28, 28, 28) - optional

- test_labels: shape (K,) - optional

Directory Format:

```
data_dir/
├── train.npz
│   ├── images: (N, 28, 28, 28)
│   ├── labels: (N,)
├── val.npz              # Optional
│   ├── images: (M, 28, 28, 28)
│   └── labels: (M,)
├── test.npz             # Optional
    ├── images: (K, 28, 28, 28)
    └── labels: (K,)
```

### Automatic Data Handling

- **Validation Split**: If no val_images exist, automatically splits training data using validation_split ratio
- **Format Detection**: Automatically detects single file vs directory structure
- **Channel Addition:** Automatically adds channel dimension if not present
- **Key Detection:** Automatically detects data keys in .npz files

## Data Augmentation System

### Augmentation Levels

#### None (augmentation_level="none")

```
transforms.Compose([
    transforms.Resize(image_size),
    transforms.ToTensor(),
    transforms.Normalize([0.485, 0.456, 0.406], [0.229, 0.224, 0.225])
])
```

#### Basic (augmentation_level="basic") - Default

```
transforms.Compose([
    transforms.Resize(image_size),
    transforms.RandomHorizontalFlip(p=0.5),
    transforms.RandomRotation(degrees=10),
    transforms.ToTensor(),
    transforms.Normalize([0.485, 0.456, 0.406], [0.229, 0.224, 0.225])
])
```

#### Moderate (augmentation_level="moderate")

```
transforms.Compose([
    transforms.Resize(image_size),
    transforms.RandomHorizontalFlip(p=0.5),
    transforms.RandomRotation(degrees=15),
    transforms.ColorJitter(brightness=0.2, contrast=0.2, saturation=0.2, hue=0.1),
    transforms.RandomAffine(degrees=0, translate=(0.1, 0.1), scale=(0.9, 1.1)),
    transforms.ToTensor(),
    transforms.Normalize([0.485, 0.456, 0.406], [0.229, 0.224, 0.225])
])
```

#### Heavy (augmentation_level="heavy")

```
transforms.Compose([
    transforms.Resize(int(image_size[0] * 1.1)),
    transforms.RandomCrop(image_size),
    transforms.RandomHorizontalFlip(p=0.5),
    transforms.RandomVerticalFlip(p=0.3),
    transforms.RandomRotation(degrees=20),
    transforms.ColorJitter(brightness=0.3, contrast=0.3, saturation=0.3, hue=0.15),
    transforms.RandomAffine(degrees=0, translate=(0.15, 0.15), scale=(0.85, 1.15), shear=10),
    transforms.RandomPerspective(distortion_scale=0.2, p=0.5),
    transforms.RandomGrayscale(p=0.1),
    transforms.GaussianBlur(kernel_size=3, sigma=(0.1, 2.0)),
    transforms.RandomErasing(p=0.1, scale=(0.02, 0.1)),
    transforms.ToTensor(),
    transforms.Normalize([0.485, 0.456, 0.406], [0.229, 0.224, 0.225])
])
```

### Custom (augmentation_level="custom")

```python
# Example custom augmentation settings
custom_augmentations = {
    "horizontal_flip": True,
    "horizontal_flip_p": 0.5,
    "vertical_flip": False,
    "rotation": True,
    "rotation_degrees": 15,
    "color_jitter": True,
    "brightness": 0.2,
    "contrast": 0.2,
    "saturation": 0.1,
    "hue": 0.05,
    "random_crop": False,
    "affine": True,
    "affine_translate": (0.1, 0.1),
    "affine_scale": (0.9, 1.1),
    "perspective": False,
    "grayscale": False,
    "gaussian_blur": False,
    "random_erasing": True,
    "erasing_p": 0.1
}
```

## Class Imbalance Handling

## Configuration Parameters

### Core Parameters

```python
# Enable/disable imbalance handling
handle_class_imbalance: bool = False

# Strategy selection
imbalance_strategy: str = "weighted_loss"
# Options: "weighted_loss", "focal_loss", "oversampling", "undersampling""

# Evaluation metric for model selection
evaluation_metric: str = "accuracy"
# Options: "accuracy", "f1_macro", "f1_weighted", "balanced_accuracy", "auc_roc"
```

### Focal Loss Parameters

```python
# Focal loss weighting factor
focal_loss_alpha: float = 1.0

# Focal loss focusing parameter
focal_loss_gamma: float = 2.0
```

### Sampling Parameters

```
# Sampling strategy for resampling techniques
sampling_strategy: str = "auto"
# Options: "auto", "minority", "majority", "not_minority", "not_majority", or custom dict

# Custom class weights
class_weights: Union[dict, str] = None
# Options: None, "balanced", or custom dictionary {class_id: weight}
```

## Imbalance Detection

### Automatic Detection

The tools automatically analyze class distribution and calculate imbalance metrics:

```
# Class distribution analysis
train_class_dist = get_class_distribution(train_dataset)
max_class_count = np.max(train_class_dist)
min_class_count = np.min(train_class_dist)
imbalance_ratio = max_class_count / min_class_count
```

### Warning System

```
# Automatic warning when imbalance is detected
if not handle_class_imbalance and imbalance_ratio > 2.0:
    logging.warning(f"High class imbalance detected (ratio: {imbalance_ratio:.2f}). "
                   "Consider enabling handle_class_imbalance=True")
```

### Reporting

- **Class distribution**: Sample counts per class
- **Imbalance ratio**: Ratio between majority and minority classes
- **Recommendations**: Automatic suggestions for handling strategies

## Imbalance Strategies

### 1. Weighted Loss (weighted_loss)

**Description**: Applies class weights to the loss function to penalize misclassification of minority classes more heavily.

**How it works**: - Calculates balanced class weights: weight_i = n_samples / (n_classes * n_samples_i) - Applies weights to CrossEntropyLoss - No change to dataset size.

**Configuration**:

```
handle_class_imbalance=True,
imbalance_strategy="weighted_loss",
class_weights="balanced"  # or custom dict
```

### 2. Focal Loss (focal_loss)

**Description**: A loss function that focuses learning on hard-to-classify examples by down-weighting easy examples.

**Mathematical Formula**:

$$FL(pt) = -\alpha(1-pt)^\gamma * \log(pt)$$

Where: - pt is the predicted probability for the true class - α (alpha) is the weighting factor for rare class - γ (gamma) is the focusing parameter.

**How it works**: - Easy examples (high pt) get down-weighted by $(1-pt)^\gamma$ - Hard examples (low pt) receive more focus - Alpha parameter balances positive/negative examples.

**Configuration**:

```
handle_class_imbalance=True,
imbalance_strategy="focal_loss",
focal_loss_alpha=1.0,    # Class weighting
focal_loss_gamma=2.0     # Focusing parameter
```

### 3. Oversampling

**Description**: Increases minority class samples by random duplication.

**How it works**: - Randomly duplicates minority class samples - Balances dataset to specified ratio - Increases total dataset size

**Configuration**:

```
handle_class_imbalance=True,
imbalance_strategy="oversampling",
sampling_strategy="auto"  # or custom ratio
```

### 4. Undersampling

**Description**: Reduces majority class samples to balance the dataset.

**How it works**: - Randomly removes majority class samples - Reduces total dataset size - Balances class distribution

**Advantages**: - Faster training - Reduces storage requirements - Simple implementation

**Configuration**:

```
handle_class_imbalance=True,
imbalance_strategy="undersampling",
sampling_strategy="auto"
```

### Early Stopping Mechanism

All tools implement intelligent early stopping: - **Monitoring Metric**: Validation accuracy (higher is better) - **Patience**: Configurable epochs to wait for improvement - **Best Model Saving**: Automatically saves best performing model - **Training Continuation**: Option to disable for full epoch training

### Memory Optimization

- **Pin Memory**: Automatically enabled when CUDA is available
- **Gradient Accumulation**: Handled automatically for large batch equivalents
- **Mixed Precision**: Can be enabled for compatible hardware
- **DataLoader Workers**: Optimized for parallel data loading

## Output Files and Results

### Model Files

#### Best Model

**File**: best_model.pt

```
{
    'epoch': 15,
    'model_state_dict': model.state_dict(),
    'optimizer_state_dict': optimizer.state_dict(),
    'loss': 0.234,
    'accuracy': 94.5
}
```

#### Final Model

**File**: final_model.pt - Contains final epoch state.

### Configuration Files

#### Model Configuration

**File**: model_config.json

```
{
    "model_type": "resnet50",
    "num_classes": 2,
    "image_size": 224,
    "pretrained": true,
    "early_stopping": true,
    "patience": 5,
    "best_epoch": 15,
    "best_accuracy": 94.5,
    "training_epochs": 18,
```

```
  "early_stopped": true,
  "batch_size": 16,
  "augmentation_level": "basic",
  "custom_augmentations": null
}
```

### Training History

**File**: training_history.json

```
{
  "train_loss": [0.693, 0.456, 0.342, ...],
  "val_loss": [0.651, 0.423, 0.356, ...],
  "train_acc": [65.2, 78.4, 85.1, ...],
  "val_acc": [68.1, 79.2, 84.8, ...],
  "learning_rates": [0.001, 0.001, 0.0005, ...]
}
```

## Evaluation Results

### Test Metrics (if test directory exists)

**File**: test_metrics.json

```
{
  "accuracy": 0.9245,
  "precision_macro": 0.9187,
  "recall_macro": 0.9156,
  "f1_macro": 0.9171,
  "precision_per_class": [0.9456, 0.8918],
  "recall_per_class": [0.8934, 0.9378],
  "f1_per_class": [0.9189, 0.9144]
}
```

## Visualizations

### Training Plots

**File**: training_plots.png

Three-panel visualization: 1. **Loss Curves**: Training and validation loss over epochs 2. **Accuracy Curves**: Training and validation accuracy over epochs
3. **Learning Rate Schedule**: Learning rate changes over training

### Confusion Matrix (if test data available)

**File**: confusion_matrix.png - Heatmap visualization of model performance per class - Annotations showing actual counts - Color-coded for easy interpretation.

## Logs

**File**: training.log

```
2024-01-15 10:30:15,123 - INFO - Starting ResNet training with configuration:
2024-01-15 10:30:15,124 - INFO - Model type: resnet50
2024-01-15 10:30:15,125 - INFO - Number of classes: 2
2024-01-15 10:30:15,126 - INFO - Using device: cuda:0
...
2024-01-15 10:35:42,567 - INFO - Epoch 1/20 - Train Loss: 0.4562, Train Acc: 78.43%, Val Loss: 0.4123, Val Acc: 81.56%, LR: 0.001000
```

### Return Object

All training tools return a comprehensive results dictionary:

```python
{
  "status": "success",
  "best_model_path": "output/best_model.pt",
  "final_model_path": "output/final_model.pt",
  "config_path": "output/model_config.json",
  "plots_path": "output/training_plots.png",
  "history_path": "output/training_history.json",
  "test_metrics_path": "output/test_metrics.json",  # If test data available
  "confusion_matrix_path": "output/confusion_matrix.png",  # If test data available
  "best_accuracy": 94.5,
  "test_accuracy": 92.1,  # If test data available
  "training_time_seconds": 3847.2,
  "epochs_completed": 18,
  "early_stopped": True,
  "early_stopping_used": True,
  "augmentation_level": "basic"
}
```

# PyTorch Inference Tools Documentation

## Overview

The **PyTorch Inference Tools** are a comprehensive suite of deep learning inference utilities designed for medical image classification tasks. These tools provide inference capabilities for models trained with the corresponding training tools, featuring batch processing and comprehensive performance evaluation.

## Universal Input Parameters

All inference tools share the same interface with architecture-specific optimizations:

### Required Parameters

| Parameter | Type | Description |
|---|---|---|
| image_dir | string | Directory containing images for inference |
| model_path | string | Path to the trained model file (.pt format) |
| output_dir | string | Directory where prediction results will be saved |

### Optional Parameters

| Parameter | Type | Default | Description |
|---|---|---|---|
| config_path | string | None | Path to model configuration JSON file |
| ground_truth_file | string | None | CSV file with ground truth labels for evaluation |
| class_names | array | None | List of class names for human-readable output |
| num_classes | integer | None | Number of classes (auto-detected from config/class_names) |
| batch_size | integer | 32 | Batch size for inference processing |
| case_column | string | "case" | Column name for case identifiers in ground truth |
| label_column | string | "label" | Column name for labels in ground truth |
| file_extension | string | ".png" | Extension to add to case IDs if needed |

### Architecture-Specific Parameters

#### ResNet-Specific

| Parameter | Type | Default | Description |
|---|---|---|---|
| model_type | string | Auto-detect | ResNet variant (auto-detected from config) |

#### Inception V3-Specific

| Parameter | Type | Default | Description |
|---|---|---|---|
| aux_logits | boolean | True | Whether model was trained with auxiliary logits |

#### VGG16-Specific

| Parameter | Type | Default | Description |
|---|---|---|---|
| use_batch_norm | boolean | False | Whether model uses batch normalization |

#### EfficientNetV2-Specific

| Parameter | Type | Default | Description |
|---|---|---|---|
| model_variant | string | Auto-detect | EfficientNetV2 variant (auto-detected from config) |
| use_custom_image_size | boolean | False | Whether to use custom image size |
| custom_image_size | integer | Auto-detect | Custom image size override |

## Data Structure Requirements

### Inference-Only Mode (No Ground Truth)

```
image_dir/
├── image1.png
├── image2.jpg
├── image3.bmp
```

```
├── patient_001.tiff
└── scan_042.jpeg
```

### Evaluation Mode (With Ground Truth)

#### Option 1: Ground Truth CSV File

```
image_dir/
├── image1.png
├── image2.jpg
└── ...

ground_truth.csv:
case,label
image1,0
image2,1
patient_001,0
scan_042,1
```

#### Option 2: Automatic Case ID Matching

```
ground_truth.csv:
case,label
001,0      # Matches patient_001.png (with file_extension=".png")
042,1      # Matches scan_042.jpg
```

### Supported Image Formats

- **Primary**: PNG, JPEG, JPG
- **Additional**: BMP, TIFF, TIF
- **Color Modes**: RGB (automatically converted from grayscale)
- **Bit Depths**: 8-bit, 16-bit (normalized to 8-bit)
- **3D Imaging data**: numpy .npz file containing test images and ground truth labels

## Model Configuration Integration

### Automatic Configuration Loading

All inference tools automatically load configuration from the training output:

```json
{
  "model_type": "resnet50",
  "num_classes": 3,
  "image_size": 224,
  "pretrained": true,
  "use_batch_norm": false,
  "aux_logits": true
}
```

### Configuration Priority

1. **Explicit Parameters**: Directly provided parameters take highest priority
2. **Config File**: Values from model_config.json
3. **Defaults**: Tool-specific defaults as fallback

### Config File Benefits

- **Consistency**: Ensures inference uses same settings as training
- **Automation**: Reduces manual parameter specification
- **Reproducibility**: Maintains exact model specifications

## Output Files and Results

### Prediction Results

#### Main Predictions File

**File**: predictions.csv

**Columns Explanation**: - filename: Original image filename - case_id: Extracted case identifier - predicted_class: Numeric class prediction (0-indexed) - confidence: Confidence score for predicted class - true_label: Ground truth label (if available) - correct: Whether prediction matches ground truth - predicted_class_name: Human-readable class name - true_class_name: Human-readable true class name - prob_[class]: Probability for each class

### Performance Metrics (When Ground Truth Available)

#### Comprehensive Metrics File

**File**: metrics.json

```
{
  "accuracy": 0.8509,
  "total_images": 3421,
  "correct_predictions": 2911,
  "precision_macro": 0.8804,
  "recall_macro": 0.8128,
  "f1_macro": 0.8113,
  "precision_class_0": 0.8592,
  "recall_class_0": 1.0000,
  "f1_class_0": 0.9242,
  "support_class_0": 244,
  "sensitivity_class_0": 1.0000,
  "specificity_class_0": 0.9874,
  "auc_class_0": 0.9998,
  "precision_class_1": 0.9905,
  "recall_class_1": 0.9984,
  "f1_class_1": 0.9944,
  "support_class_1": 624,
  "sensitivity_class_1": 0.9984,
```

```
  "specificity_class_1": 0.9979,
  "auc_class_1": 0.9997
}
```

#### Metrics Summary

**File**: metrics_summary.csv

```
metric,value
accuracy,0.8509
precision_macro,0.8804
recall_macro,0.8128
f1_macro,0.8113
sensitivity_Normal,1.0000
specificity_Normal,0.9874
auc_Normal,0.9998
```

### Visualizations

#### Confusion Matrix

**File**: confusion_matrix.png - Heatmap showing actual vs predicted classifications - Annotated with actual counts - Color-coded for easy interpretation - Class names on axes (if provided)

#### ROC Curves (Binary Classification)

**File**: roc_curve.png - ROC curve with AUC score - Diagonal reference line.

### Logs

**File**: inference.log

```
2024-01-15 14:30:15,123 - INFO - Starting ResNet inference with settings:
2024-01-15 14:30:15,124 - INFO - Model path: models/pneumonia/best_model.pt
2024-01-15 14:30:15,125 - INFO - Model type: resnet50
2024-01-15 14:30:15,126 - INFO - Using device: cuda:0
2024-01-15 14:30:15,127 - INFO - Found 1250 images for inference
2024-01-15 14:32:45,890 - INFO - Processed 320/1250 images
2024-01-15 14:35:12,456 - INFO - Inference completed in 177.33 seconds
2024-01-15 14:35:15,123 - INFO - Saved predictions to predictions.csv
```

## Return Object

All inference tools return a comprehensive results dictionary:

```
{
  "status": "success",
  "predictions_path": "output/predictions.csv",
  "log_file": "output/inference.log",
  "metrics": {
    "accuracy": 0.8509,
    "precision_macro": 0.8804,
```

```
        # ... complete metrics dict
    },
    "metrics_path": "output/metrics.json",
    "confusion_matrix_path": "output/confusion_matrix.png",
    "roc_curve_path": "output/roc_curve.png",  # Binary classification only
    "image_dir": "data/test_images",
    "model_path": "models/best_model.pt",
    "output_dir": "inference_results",
    "num_images_processed": 1250,
    "has_ground_truth": True,
    "processing_time_seconds": 177.33,
    "model_type": "resnet50"
}
```

## Performance Evaluation Capabilities

### Binary Classification Metrics

- **Accuracy**: Overall correctness percentage
- **Precision**: True positives / (True positives + False positives)
- **Recall (Sensitivity)**: True positives / (True positives + False negatives)
- **Specificity**: True negatives / (True negatives + False positives)
- **F1-Score**: Harmonic mean of precision and recall
- **AUC-ROC**: Area under the receiver operating characteristic curve

### Multi-Class Classification Metrics

- **Macro-averaged**: Unweighted mean across all classes
- **Per-class metrics**: Individual performance for each class
- **Support**: Number of samples per class
- **One-vs-Rest AUC**: AUC for each class vs all others
- **Confusion Matrix**: Detailed breakdown of classifications

### Clinical Relevance Metrics

- **Sensitivity (Recall)**: Important for medical screening applications
- **Specificity**: Important for avoiding false alarms
- **Positive Predictive Value**: Precision in medical context
- **Negative Predictive Value**: Confidence in negative results

## Model Performance Comparison Tool

## Overview

The **ModelPerformanceComparisonTool** is a monitoring system that compares model performance between training evaluation results and real-world inference results to detect performance degradation over time. It automatically identifies when models need fine-tuning or retraining by analyzing metric differences and providing actionable recommendations.

### Metric Mapping

The tool intelligently maps corresponding metrics between training and inference evaluation files, handling different naming conventions and file structures.

### Decision Thresholds

Configurable percentage thresholds that determine when performance decline is considered significant enough to warrant intervention.

## Input Parameters

### Required Parameters

| Parameter | Type | Description |
| --- | --- | --- |
| training_metrics_path | string | Path to test_metrics.json file from training evaluation |
| inference_metrics_path | string | Path to metrics.json file from inference evaluation |
| output_dir | string | Directory where comparison results will be saved |

### Optional Parameters

| Parameter | Type | Default | Description |
| --- | --- | --- | --- |
| decline_threshold | float | 5.0 | Percentage threshold for significant performance decline |
| class_names | array | None | List of class names for better reporting |

## Data Structure Requirements

### Training Metrics File (test_metrics.json)

Generated by training tools after test set evaluation:

```json
{
  "accuracy": 0.9763,
  "precision_macro": 0.9746,
  "recall_macro": 0.9740,
  "f1_macro": 0.9740,
  "precision_per_class": [1.0, 0.9984, 0.9446],
  "recall_per_class": [0.9549, 1.0, 0.9871],
  "f1_per_class": [0.9769, 0.9992, 0.9654]
}
```

### Inference Metrics File (metrics.json)

Generated by inference tools when ground truth is available:

```json
{
  "accuracy": 0.8509,
```

```
  "total_images": 3421,
  "correct_predictions": 2911,
  "precision_macro": 0.8804,
  "recall_macro": 0.8128,
  "f1_macro": 0.8114,
  "precision_class_0": 0.8592,
  "recall_class_0": 1.0,
  "f1_class_0": 0.9242,
  "support_class_0": 244.0,
  "sensitivity_class_0": 1.0,
  "specificity_class_0": 0.9874,
  "auc_class_0": 0.9998
}
```

## Metric Mapping System

### Global Metrics Mapping

The tool automatically maps corresponding global metrics:

| Training File Key | Inference File Key | Description |
|---|---|---|
| accuracy | accuracy | Overall classification accuracy |
| precision_macro | precision_macro | Macro-averaged precision |
| recall_macro | recall_macro | Macro-averaged recall |
| f1_macro | f1_macro | Macro-averaged F1-score |

### Per-Class Metrics Mapping

Maps between array-based (training) and individual key-based (inference) formats:

| Training Format | Inference Format | Example |
|---|---|---|
| precision_per_class[0] | precision_class_0 | Class 0 precision |
| recall_per_class[1] | recall_class_1 | Class 1 recall |
| f1_per_class[2] | f1_class_2 | Class 2 F1-score |

### Class Detection Logic

1. **From Training File**: Counts length of precision_per_class array
2. **From Inference File**: Identifies class indices from key patterns
3. **Validation**: Ensures consistent class counts between files
4. **Flexible Naming**: Handles different class naming conventions

### Performance Analysis Engine

#### Percentage Change Calculation

For each metric pair:

```
percentage_change = ((inference_value - training_value) / training_value) * 100
absolute_decline = abs(percentage_change) if percentage_change < 0 else 0
is_significant = absolute_decline > decline_threshold
```

#### Status Classification

Each metric receives a status: - **IMPROVED**: percentage_change > 0 - **UNCHANGED**: percentage_change == 0 - **DECLINED**: percentage_change < 0

#### Significance Assessment

Determines if decline warrants intervention: - **Minor Decline**: 0 < decline < threshold - **Significant Decline**: decline ≥ threshold

## Decision Logic System

#### Overall Performance Status

| Condition | Status | Action Required |
|---|---|---|
| No significant declines | GOOD | Continue monitoring |
| Some metrics declined < threshold | MINOR_DECLINE | Increased monitoring |
| Any metric declined ≥ threshold | SIGNIFICANT_DECLINE | Fine-tuning recommended |

### Recommendation Engine

#### 1. Fine-Tuning Required (High Priority)

**Triggered when**: significant_declines > 0

**Recommendation**:

```json
{
  "type": "FINE_TUNING_REQUIRED",
  "priority": "HIGH",
  "description": "Model performance has significantly declined on X metrics. Fine-tuning with new data is recommended.",
  "details": {
    "declined_metrics": 3,
    "max_decline": 8.5,
    "threshold": 5.0
  }
}
```

#### 2. Focus on Specific Classes (Medium Priority)

**Triggered when**: Specific classes show significant decline

**Recommendation**:

```
{
  "type": "FOCUS_ON_CLASSES",
  "priority": "MEDIUM",
  "description": "Focus fine-tuning efforts on classes with the most significant performance declines.",
  "details": {
    "most_problematic_classes": [
      {"class": "class_2", "metric": "precision", "decline": 12.3},
      {"class": "class_0", "metric": "recall", "decline": 8.7}
    ]
  }
}
```

#### 3. Monitoring Recommended (Low Priority)

**Triggered when**: Minor declines detected

**Recommendation**:

```
{
  "type": "MONITORING_RECOMMENDED",
  "priority": "LOW",
  "description": "Some metrics have declined but not significantly. Continue monitoring performance.",
  "details": {
    "declined_metrics": 2,
    "max_decline": 3.2
  }
}
```

#### 4. Performance Maintained (Info)

**Triggered when**: No decline detected

**Recommendation**:

```
{
  "type": "PERFORMANCE_MAINTAINED",
  "priority": "INFO",
  "description": "Model performance has been maintained or improved. No immediate action required."
}
```

## Output Files

### Detailed Report

**File**: performance_comparison_report.json

Complete analysis including: - All metric comparisons - Statistical summaries - Detailed recommendations - Class-specific analysis

```json
{
  "global_metrics": {
    "accuracy": {
      "training_value": 0.9763,
      "inference_value": 0.8509,
      "percentage_change": -12.84,
      "absolute_decline": 12.84,
      "is_significant_decline": true,
      "status": "DECLINED"
    }
  },
  "per_class_metrics": {
    "precision": {
      "class_0": {
        "training_value": 1.0,
        "inference_value": 0.8592,
        "percentage_change": -14.08,
        "absolute_decline": 14.08,
        "is_significant_decline": true,
        "status": "DECLINED"
      }
    }
  },
  "summary": {
    "total_metrics_compared": 12,
    "metrics_declined": 8,
    "significant_declines": 5,
    "max_decline_percentage": 15.2,
    "overall_performance_status": "SIGNIFICANT_DECLINE"
  },
  "recommendations": [...]
}
```

## Summary CSV

**File**: performance_comparison_summary.csv

Tabular format for easy analysis:

| metric_type | metric_name | class | training_value | inference_value | percentage_change | absolute_decline | is_significant_decline | status |
|---|---|---|---|---|---|---|---|---|
| global | accuracy | N/A | 0.9763 | 0.8509 | -12.84 | 12.84 | true | DECLINED |
| per_class | precision | class_0 | 1.0 | 0.8592 | -14.08 | 14.08 | true | DECLINED |

## Visualizations

**File**: global_metrics_comparison.png

Two-panel visualization: 1. **Bar Chart**: Side-by-side comparison of training vs inference metrics 2. **Change Chart**: Percentage change with decline threshold line

**File**: per_class_metrics_heatmap.png

Heatmap showing percentage changes across all classes and metrics: - **Green**: Improved performance - **Yellow**: Minor changes - **Red**: Significant declines

#### Logs

**File**: performance_comparison.log

Detailed execution log with timestamps and analysis steps.

### Advanced Features

#### Flexible Class Naming

Supports various class naming conventions: - **Numeric**: class_0, class_1, class_2 - **Named**: Uses provided class_names parameter - **Mixed**: Handles inconsistent naming patterns

#### Missing Data Handling

- **Graceful degradation**: Continues analysis with available metrics
- **Warning logs**: Reports missing metrics without failing
- **Partial comparison**: Analyzes available subset of metrics

#### Multi-Class Scalability

- **Binary classification**: Handles 2-class problems
- **Multi-class**: Scales to any number of classes
- **Class expansion**: Detects when new classes are added

#### Statistical Robustness

- **Zero-division protection**: Handles edge cases safely
- **Numerical stability**: Prevents overflow/underflow errors
- **Outlier detection**: Identifies extreme changes for investigation

## Return Object

```
{
  "status": "success",
  "overall_performance_status": "SIGNIFICANT_DECLINE",
  "total_metrics_compared": 15,
  "metrics_declined": 8,
  "significant_declines": 4,
  "max_decline_percentage": 15.2,
  "fine_tuning_recommended": True,
  "recommendations": [
```

```
    {
      "type": "FINE_TUNING_REQUIRED",
      "priority": "HIGH",
      "description": "Model performance has significantly declined...",
      "details": {...}
    }
  ],
  "detailed_report_path": "output/performance_comparison_report.json",
  "summary_csv_path": "output/performance_comparison_summary.csv",
  "log_file": "output/performance_comparison.log",
  "output_dir": "output/"
}
```

## PyTorch Model FineTuning Tool

### Overview

The **PyTorchModelFineTuningTool** is a specialized tool designed to adapt pre-trained CNN models to new data distributions while preserving learned knowledge. It implements multiple fine-tuning strategies to prevent catastrophic forgetting and optimize performance on evolving datasets.

### Purpose

This tool addresses scenarios where:

- Model performance has degraded due to distribution shift

- New data from the same domain becomes available

- You need to adapt existing models without retraining from scratch

- Computational resources are limited for full retraining

### Supported Model Architectures

- **ResNet**: ResNet18, ResNet34, ResNet50, ResNet101, ResNet152
- **VGG**: VGG16, VGG16 with Batch Normalization
- **Inception V3**: With or without auxiliary logits
- **EfficientNetV2**: EfficientNetV2-S, EfficientNetV2-M, EfficientNetV2-L

### Input Parameters

#### Required Parameters

| Parameter | Type | Description |
|---|---|---|
| pretrained_model_path | string | Path to the trained model file (.pt format) |
| config_path | string | Path to original model configuration JSON file |
| new_data_dir | string | Directory containing new training and validation data |

| Parameter | Type | Description |
|---|---|---|
| output_dir | string | Directory where fine-tuned model and results will be saved |

### Optional Parameters

| Parameter | Type | Default | Description |
|---|---|---|---|
| fine_tuning_strategy | string | “partial” | Strategy: ‘full’, ‘partial’, ‘head_only’, ‘gradual_unfreezing’ |
| freeze_layers | integer | auto-detected | Number of layers to freeze from the beginning |
| fine_tune_learning_rate | float | 1e-5 | Learning rate for fine-tuning |
| backbone_learning_rate | float | 1e-6 | Lower learning rate for frozen/backbone layers |
| num_epochs | integer | 10 | Number of fine-tuning epochs |
| batch_size | integer | 16 | Batch size for training |
| reset_optimizer | boolean | True | Whether to reset optimizer state |
| warmup_epochs | integer | 3 | Number of warmup epochs for gradual unfreezing |
| early_stopping | boolean | True | Whether to use early stopping |
| patience | integer | 5 | Early stopping patience |
| augmentation_level | string | “basic” | Data augmentation: ‘none’, ‘basic’, ‘moderate’, ‘heavy’ |
| weight_decay | float | 1e-5 | Weight decay for regularization |
| use_cosine_annealing | boolean | True | Whether to use cosine annealing LR scheduler |
| add_new_classes | boolean | False | Whether new data contains additional classes |
| new_num_classes | integer | None | Total number of classes after adding new ones |

## Data Structure Requirements

The tool expects data organized in the standard ImageFolder structure:

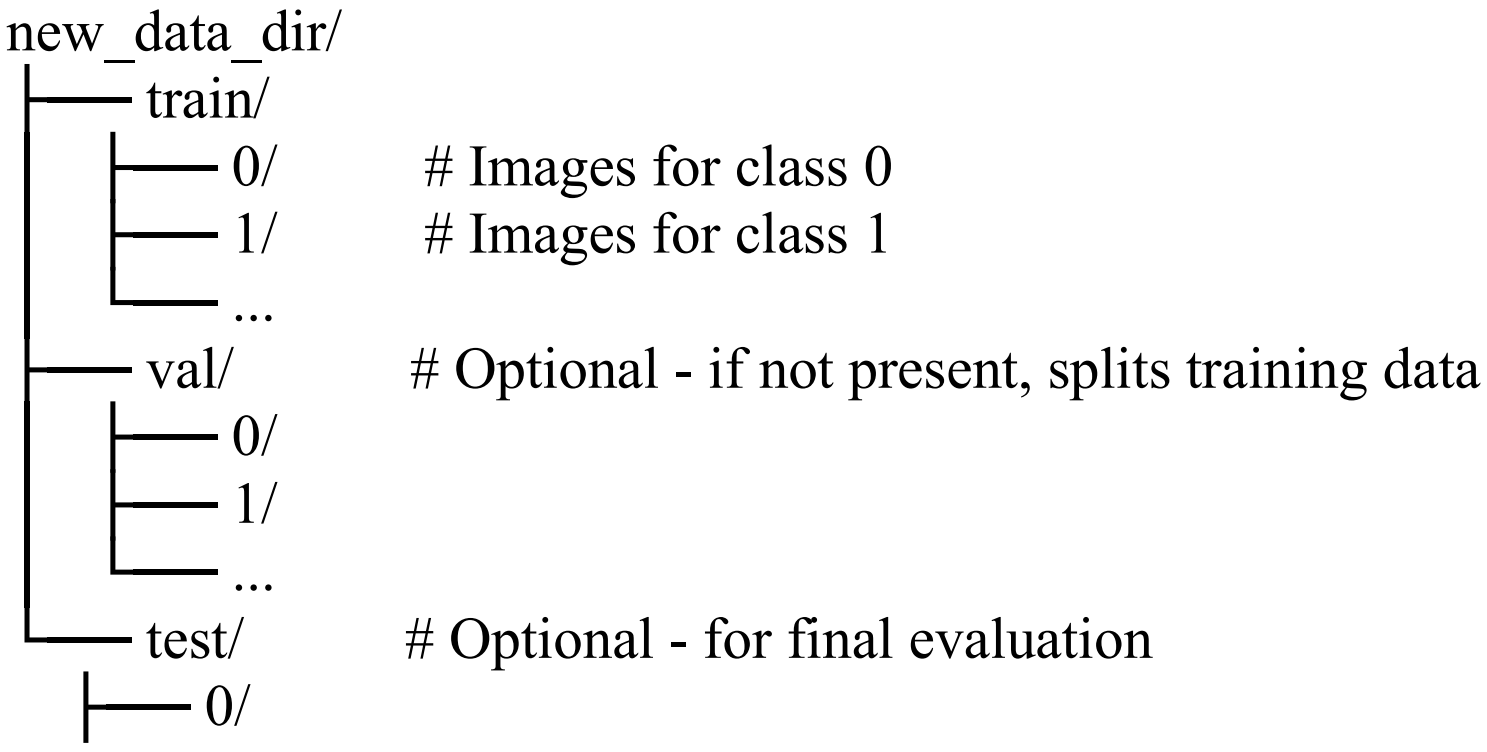

1/
...

**Alternative**: CSV-based labels are also supported with labels.csv, val_labels.csv, and test_labels.csv files.

## Fine-Tuning Strategies

### 1. Full Fine-Tuning (strategy="full")

- **Description**: Updates all model parameters
- **Use Case**: Large amount of new data, significant distribution shift
- **Pros**: Maximum adaptation capability
- **Cons**: Higher risk of catastrophic forgetting

### 2. Head-Only Fine-Tuning (strategy="head_only")

- **Description**: Only trains the final classifier layer
- **Use Case**: Limited new data, minor distribution shift
- **Pros**: Minimal forgetting risk, fast training
- **Cons**: Limited adaptation capability

### 3. Partial Fine-Tuning (strategy="partial")

- **Description**: Freezes early layers, trains later layers and classifier
- **Use Case**: Moderate amount of new data, balanced adaptation needs
- **Pros**: Good balance between adaptation and preservation
- **Cons**: Requires careful layer selection

### 4. Gradual Unfreezing (strategy="gradual_unfreezing")

- **Description**: Progressively unfreezes layers during training
- **Use Case**: Large datasets, when you want controlled adaptation
- **Pros**: Smooth transition, reduced forgetting
- **Cons**: Longer training time

## Differential Learning Rates

When using partial fine-tuning with backbone_learning_rate specified: - **Backbone layers**: Use lower learning rate (e.g., 1e-6) - **Classifier layers**: Use higher learning rate (e.g., 1e-5)

This approach allows: - Gentle adaptation of feature extractors - More aggressive training of task-specific layers

## Data Augmentation Levels

### None (augmentation_level="none")

- Only resize and normalization

### Basic (augmentation_level="basic")

- Horizontal flip (50% probability)
- Random rotation (±10 degrees)

### Moderate (augmentation_level="moderate")

- Basic augmentations +
- Color jitter (brightness, contrast, saturation, hue)
- Random affine transformations

### Heavy (augmentation_level="heavy")

- All moderate augmentations +
- Vertical flip
- Random perspective
- Gaussian blur
- Random erasing

## Learning Rate Scheduling

### Cosine Annealing (Default)

- Smoothly decreases learning rate following cosine curve
- Helps with convergence and prevents overfitting
- Minimum learning rate set to 1e-7

### Reduce on Plateau (Alternative)

- Monitors validation loss
- Reduces LR by factor of 0.5 when loss plateaus
- Patience of 2 epochs before reduction

## Early Stopping Mechanism

- Monitors validation accuracy for improvement
- Stops training if no improvement for patience epochs
- Saves best model based on validation accuracy
- Prevents overfitting and saves computational time

### Output Files

#### Model Files

- best_finetuned_model.pt: Best performing model during training
- final_finetuned_model.pt: Final model state after training

#### Configuration and History

- fine_tuning_config.json: Complete configuration used for fine-tuning
- fine_tuning_history.json: Training metrics for each epoch

#### Evaluation Results

- test_metrics_finetuned.json: Detailed performance metrics on test set
- confusion_matrix_finetuned.png: Confusion matrix visualization

#### Visualizations

- fine_tuning_plots.png: Combined plots showing:
  - Loss curves (training and validation)
  - Accuracy curves (training and validation)
  - Learning rate schedule
  - Number of frozen parameters over time

#### Logs

- fine_tuning.log: Detailed training logs with timestamps

## Return Object

```
{
  "status": "success",
  "best_model_path": "path/to/best_finetuned_model.pt",
  "final_model_path": "path/to/final_finetuned_model.pt",
  "config_path": "path/to/fine_tuning_config.json",
  "plots_path": "path/to/fine_tuning_plots.png",
  "history_path": "path/to/fine_tuning_history.json",
  "test_metrics_path": "path/to/test_metrics_finetuned.json",
  "confusion_matrix_path": "path/to/confusion_matrix_finetuned.png",
  "best_accuracy": 95.2,
  "test_accuracy": 94.8,
  "training_time_seconds": 1245.6,
  "epochs_completed": 8,
  "early_stopped": True,
  "fine_tuning_strategy": "partial",
  "original_num_classes": 2,
  "final_num_classes": 2,
```

```
    "classes_added": False
}
```

**Supplementary Material Part B - ReclAIm: A Multi-Agent Framework for Monitoring and Correcting Performance Decline in Medical Imaging AI**

This section provides detailed information on the five datasets introduced in Section 2.2, including acquisition characteristics, class distributions, patient-level partitioning strategies, and the chronological splitting approach applied to the MIMIC-CXR dataset for temporal degradation assessment.

To cover a broad spectrum of imaging modalities, we selected five publicly available datasets encompassing MRI, CT, and X-ray imaging.

For brain tumor classification, we used a contrast-enhanced T1-weighted MRI dataset (1) comprising 3,064 2D slices from 233 patients, annotated into three classes: meningioma, glioma, and pituitary tumor. The dataset provides explicit patient identifiers, enabling patient-level data partitioning. For covid-19 classification, we used a chest CT dataset (2) consisting of 4,173 2D CT slices derived from 210 patients. A binary covid/non-covid classification task was constructed. Slices are grouped per patient by the dataset authors, allowing patient-based splitting and preventing cross-subset leakage. To evaluate workflows on native 3D imaging data, we used the NoduleMNIST3D dataset from the MedMNIST collection (3), which contains 1,633 chest CT volumes for binary benign/malignant lung nodule classification. Each instance corresponds to a complete CT scan rather than individual slices. For X-ray classification, we used the PneumoniaMNIST dataset from the MedMNIST collection (3), which includes 5,645 chest X-ray images annotated for binary classification into pneumonia and normal cases.

To assess temporal model degradation, we derived a chest X-ray dataset from MIMIC-CXR (4-6). From 96,161 posterior-anterior chest X-ray images, we constructed a binary classification dataset comprising 35,436 images labeled as no finding or cardiomegaly. Patient-level deduplication was applied by retaining a single image per patient, selecting the earliest available study, thereby preventing patient overlap and repeated-measure bias. MIMIC-CXR images include anonymized acquisition dates that preserve relative temporal order. To evaluate temporal performance degradation, we adopted percentage-based chronological splitting rather than fixed calendar intervals, as anonymization stretches the original 2011–2016 acquisition period across a wide date range.

Chronologically ordered data were divided into three segments: (i) a model development period comprising the earliest 40% of the data, (ii) a temporal gap period comprising the subsequent 20%, excluded entirely from this study to amplify drift effects, and (iii) an inference and fine-tuning period comprising the most recent 40% of the data.

For all datasets, patient-based splitting was applied to prevent cross-subset leakage. Data were partitioned into model development (60%), inference (20%), and fine-tuning (20%) cohorts, with the model development cohort further divided into training (70%), validation (15%), and test (15%) sets. Inference datasets were stored in flat directory structures with labels provided via CSV files, while fine-tuning datasets were organized into class-specific folders.

**Supplementary Material Part C - ReclAIm: A Multi-Agent Framework for Monitoring and Correcting Performance Decline in Medical Imaging AI**

This supplementary section documents the prompt-based evaluation of the proposed agentic framework. It includes the full set of 59 additional natural language prompts used to assess system behavior across diverse tasks, as well as representative OpenTelemetry traces illustrating agent coordination, sequential task execution, parameter propagation, and error-handling mechanisms. To assess robustness across hardware configurations, all experiments were repeated on two different systems: a workstation equipped with an Intel Core i9-14900 CPU, 128 GB RAM, and an NVIDIA RTX 5090 GPU (32 GB memory) and a second workstation equipped with an AMD Ryzen 7 5800 CPU, 64 GB RAM, and an NVIDIA RTX 2080 Ti GPU (11 GB memory). All data were stored and processed locally on these systems. Code execution, including tool execution, model training, evaluation, deployment, fine-tuning, and performance monitoring, was performed entirely locally. Only natural language reasoning and agent communication were handled remotely via ChatGPT-4.1 API calls, using the default temperature setting (temperature = 1).

**Reproducibility analysis**

To assess the system's reproducibility under non-deterministic LLM conditions, we introduced a reproducibility evaluation stage. Specifically, we reran 12 representative prompts, covering training, inference, performance evaluation, and fine-tuning workflows across three cases (lung nodule classification, cardiomegaly classification, and brain tumor classification), each repeated 5 times independently. The prompts are available in Supplementary Material Part C under the following identifiers:

nodule_training_prompt_densenet121, nodule_inference_prompt_densenet121, nodule_performance_prompt_densenet121, nodule_fine_tune_prompt_densenet121, nodule_fine_tuned_inference_prompt_densenet121, nodule_fine_tuned_performance_prompt_densenet121, cardiomegaly_dataset_training_prompt_4, cardiomegaly_dataset_inference_prompt_4, cardiomegaly_dataset_performance_prompt_2, brain_tumor_new_dataset_training_prompt_1, brain_tumor_new_dataset_inference_prompt_1, brain_tumor_new_dataset_performance_prompt_1.

Reproducibility was assessed by examining four aspects of each run: the system's reasoning, the agent orchestration decisions (i.e., which task-specific agent was invoked), the tool selection made by the specialized agents, and the arguments passed to each tool call. Across all 60 runs (12 prompts × 5 repetitions), the system consistently interpreted the user's intent correctly, dispatched the appropriate task-specific agent, identified the correct tool, and populated tool arguments accurately as specified in the user request.

**Failure analysis and error handling**

Stress-testing ReclAIm with a diverse set of prompts showed no cases of failure. In a case where model training failed due to GPU exhaustion, the agent analyzed the error, inferred that the failure was due to hardware memory constraints, and reissued the task with a reduced batch size while preserving all other user-specified requirements. When the adjusted configuration still exceeded available memory, the process was repeated until a feasible configuration was identified and training completed successfully.

**Prompts**

**NoduleMNIST3D dataset (CT) 3D-Resnet50**

```
nodule_training_prompt_resnet50 = """
Train a 3D ResNet-50 model using the dataset located at
"splitted_data/nodule_3d_ct/model_development/data.npz".
Configure the task for binary classification (2 classes).
Run training for 50 epochs with a batch size of 64 and early stopping patience set to 5.
Do not use pretrained weights and apply basic data augmentation only.
Save all outputs to: "new_tests/model_development/nodule_3d_ct/3d_resnet50/training_output".
"""
nodule_inference_prompt_resnet50 = """
Run inference using the trained 3D ResNet-50 model located at
"new_tests/model_development/nodule_3d_ct/3d_resnet50/training_output/best_model.pt"
on the dataset: "splitted_data/nodule_3d_ct/inference_dataset/data.npz".
Assume a 2-class classification task and use the available ground truth labels to compute evaluation
metrics. Store all inference outputs in:
"new_tests/model_development/nodule_3d_ct/3d_resnet50/inference_output".
"""
nodule_performance_prompt_resnet50 = """
Evaluate whether the model's performance has degraded by comparing metrics from training-time
testing and inference-time evaluation. The training test metrics are stored in:
"new_tests/model_development/nodule_3d_ct/3d_resnet50/training_output/test_metrics.json"
and the inference metrics are available at:
"new_tests/model_development/nodule_3d_ct/3d_resnet50/inference_output/metrics.json".
Save the comparison results to: "new_tests/compare_performance/nodule_3d_ct/3d_resnet50/".
```

If a significant performance drop is detected, fine-tune the existing model using the dataset:
"splitted_data/nodule_3d_ct/fine_tuning_dataset/data.npz".
Use the model checkpoint:
"new_tests/model_development/nodule_3d_ct/3d_resnet50/training_output/best_model.pt"
and the configuration file:
"new_tests/model_development/nodule_3d_ct/3d_resnet50/training_output/model_config.json".
Use partilal fine-tuning with a learning rate of 0.0001, batch size of 32, for 20 epochs, and early stopping patience of 4.
Store the fine-tuned model in: "new_tests/fine_tuned_models/nodule_3d_ct/3d_resnet50".
"""
nodule_fine_tuned_inference_prompt_resnet50 = """
Run inference using the fine-tuned 3D ResNet-50 model located at
"new_tests/fine_tuned_models/nodule_3d_ct/3d_resnet50/best_finetuned_model.pt"
on the dataset:
"splitted_data/nodule_3d_ct/inference_dataset/data.npz".
Assume a binary (2-class) task and compute evaluation metrics using the provided
ground truth labels.
Save all evaluation outputs to:
"new_tests/fine_tuned_models/nodule_3d_ct/3d_resnet50/inference_output".
"""

nodule_fine_tuned_performance_prompt_resnet50 = """
Determine whether model performance has decreased by comparing the training test metrics at
"new_tests/model_development/nodule_3d_ct/3d_resnet50/training_output/test_metrics.json"
with the fine-tuned model's inference evaluation metrics at
"new_tests/fine_tuned_models/nodule_3d_ct/3d_resnet50/inference_output/metrics.json".
Save the comparison output to:
"new_tests/fine_tuned_models/nodule_3d_ct/3d_resnet50/compare_fine_tuned_with_training_results
".
"""

**NoduleMNIST3D dataset (CT) 3D-Densenet121**

nodule_training_prompt_densenet121 = """
Train a 3D DenseNet-121 model using the dataset located at
"splitted_data/nodule_3d_ct/model_development/data.npz".
Configure the task for binary classification (2 classes).
Train for 60 epochs with a batch size of 16 and early stopping patience set to 8.
Apply moderate data augmentation and explicitly handle class imbalance.
Use accuracy as the evaluation metric.
Save all training outputs to:
"new_tests/model_development/nodule_3d_ct/3d_densenet121/training_output".
"""

nodule_inference_prompt_densenet121 = """
Load the trained 3D DenseNet-121 model from
"new_tests/model_development/nodule_3d_ct/3d_densenet121/training_output/best_model.pt"
and perform inference on the dataset located at
"splitted_data/nodule_3d_ct/inference_dataset/data.npz".
Treat the task as a binary classification problem and compute performance metrics
using the provided ground truth labels.
Store the evaluation results in
"new_tests/model_development/nodule_3d_ct/3d_densenet121/inference_output".
"""

nodule_performance_prompt_densenet121 = """
Assess whether performance has decreased by comparing the training test metrics at
"new_tests/model_development/nodule_3d_ct/3d_densenet121/training_output/test_metrics.json"
with the inference evaluation metrics at
"new_tests/model_development/nodule_3d_ct/3d_densenet121/inference_output/metrics.json".
Write the comparison outputs to:
"new_tests/compare_performance/nodule_3d_ct/3d_densenet121/".
"""

nodule_fine_tune_prompt_densenet121 = """
Use the fine-tuning dataset at
"splitted_data/nodule_3d_ct/fine_tuning_dataset/data.npz"
to further train the model for up to 100 epochs, with an early stopping patience of 10
and a batch size of 64.
Apply heavy augmentation during training and address class imbalance using weighted cross entropy.
Evaluate the fine-tuning process using accuracy as the primary metric.

Start from the checkpoint located at
"new_tests/model_development/nodule_3d_ct/3d_densenet121/training_output/best_model.pt"
and follow the configuration in
"new_tests/model_development/nodule_3d_ct/3d_densenet121/training_output/model_config.json".
Write the fine-tuned model outputs to
"new_tests/fine_tuned_models/nodule_3d_ct/3d_densenet121".
"""

nodule_fine_tuned_inference_prompt_densenet121 = """
Run inference using the fine-tuned 3D DenseNet-121 model located at
"new_tests/fine_tuned_models/nodule_3d_ct/3d_densenet121/best_finetuned_model.pt"
on the dataset:
"splitted_data/nodule_3d_ct/inference_dataset/data.npz".
Assume a binary (2-class) classification task and compute evaluation metrics
using the provided ground truth labels.
Save all evaluation outputs to:
"new_tests/fine_tuned_models/nodule_3d_ct/3d_densenet121/inference_output".
"""

nodule_fine_tuned_performance_prompt_densenet121 = """
Determine whether the model's performance has decreased by comparing the training test metrics at
"new_tests/model_development/nodule_3d_ct/3d_densenet121/training_output/test_metrics.json"
with the fine-tuned model's inference evaluation metrics at
"new_tests/fine_tuned_models/nodule_3d_ct/3d_densenet121/inference_output/metrics.json".
Save the comparison outputs to:
"new_tests/fine_tuned_models/nodule_3d_ct/3d_densenet121/compare_fine_tuned_with_training_res
ults"
"""

**Cardiomegaly classification training prompts (X-ray)**

cardiomegaly_dataset_training_prompt_1 = """
Train an EfficientNet-based classification model using the datasets located at
"splitted_data/cardiomegaly_dataset/model_development"
(train, validation, and test splits).
Configure the task for binary classification (2 classes).
Train for 40 epochs with early stopping patience set to 3 and a batch size of 64.
Apply heavy data augmentation.

Save all training outputs to:
"new_tests/model_development/cardiomegaly/efficientnet/training_output".
"""
cardiomegaly_dataset_training_prompt_2 = """
Use a ResNet-50 architecture to train a binary classification model
on the cardiomegaly dataset provided in
"splitted_data/cardiomegaly_dataset/model_development",
which contains the training, validation, and test splits.
Run training for up to 200 epochs with an early stopping patience of 15
and a batch size of 32.
Apply moderate data augmentation during training.
Store all training results in
"new_tests/model_development/cardiomegaly/resnet50/training_output".
"""
cardiomegaly_dataset_training_prompt_3 = """
Train a VGG-16 classification model using the train, validation, and test datasets located at
"splitted_data/cardiomegaly_dataset/model_development".
Configure the task for binary classification (2 classes).
Train for 150 epochs with early stopping patience set to 10 and a batch size of 16.
Apply basic data augmentation. Enable batch normalization layers in the model.
Save all training outputs to:
"new_tests/model_development/cardiomegaly/vgg16/training_output".
"""
cardiomegaly_dataset_training_prompt_4 = """
Use an InceptionV3 architecture to train a binary classification model
on the cardiomegaly dataset provided in
"splitted_data/cardiomegaly_dataset/model_development",
which contains the training, validation, and test splits.
Ensure that pretrained weights are not used and that auxiliary logits are enabled.
Run training for up to 20 epochs with an early stopping patience of 4
and a batch size of 8.
Apply heavy data augmentation during training.
Store all training results in
"new_tests/model_development/cardiomegaly/inceptionv3/training_output".
"""

**Brain tumor classification training prompts (MRI)**

brain_tumor_new_dataset_training_prompt_1 = """
Train an EfficientNet-based classification model using the train, validation, and test datasets located at
"splitted_data/brain_tumor_new_dataset/model_development".
Configure the task for 3-class classification.
Train for 50 epochs with early stopping patience set to 5 and a batch size of 8.
Handle class imbalance through oversampling and apply basic data augmentation.
Save all training outputs to:
"new_tests/model_development/brain_tumor_new_dataset/efficientnet/training_output".

"""
brain_tumor_new_dataset_training_prompt_2 = """
Develop a resnet50 model for classification tasks.
Dataset paths for training, validation, and testing phases:
"splitted_data/brain_tumor_new_dataset/model_development".
Configure for 3 classification categories. Initialize model without pretrained weights.
Configure early stopping patience at 15 epochs and set maximum training duration to 150 epochs.
Apply moderate-level data augmentation.

Save training outputs to:
new_tests/model_development/brain_tumor_new_dataset/resnet50/training_output
"""
brain_tumor_new_dataset_training_prompt_3 = """
Develop a vgg16 model for classification purposes.
Dataset locations for train, validation, and test splits:
"splitted_data/brain_tumor_new_dataset/model_development".
Configure the model for 3 classification classes. Activate batch normalization layers in the architecture.
Configure early stopping with patience of 3 epochs and run training for a maximum of 20 epochs.
Apply heavy-level data augmentation.
Save training outputs to directory:
new_tests/model_development/brain_tumor_new_dataset/vgg16/training_output
"""
brain_tumor_new_dataset_training_prompt_4 = """
Train an InceptionV3-based classification model using the train, validation, and test datasets located at
"splitted_data/brain_tumor_new_dataset/model_development".
Configure the task for 3-class classification.
Handle class imbalance using a weighted cross entropy loss function.
Enable auxiliary logits and do not use pretrained weights.
Train for 80 epochs with early stopping patience set to 5.
Apply moderate data augmentation.
Save all training outputs to:
"new_tests/model_development/brain_tumor_new_dataset/inceptionv3/training_output".
"""

**Covid classification training prompts (CT)**

covid_new_dataset_training_prompt_1 = """
Train an EfficientNet-based classification model using the dataset located at
"splitted_data/covid_new_dataset/model_development".
Configure the task for binary classification (2 classes).
Do not use pretrained weights and disable early stopping.
Train for 50 epochs.
Save all training outputs to:
"new_tests/model_development/covid_new_dataset/efficientnet/training_output".
"""
covid_new_dataset_training_prompt_2 = """
Train a ResNet-50 classification model using the dataset located at
"splitted_data/covid_new_dataset/model_development".
Configure the task for binary classification (2 classes).
Enable pretrained weights and apply early stopping.
Train for up to 200 epochs.
Save all training outputs to:
"new_tests/model_development/covid_new_dataset/resnet50/training_output".
"""
covid_new_dataset_training_prompt_3 = """
Train a VGG-16 classification model using the datasets located at
"splitted_data/covid_new_dataset/model_development".
Configure the task for binary classification (2 classes).
Apply a moderate level of data augmentation.
Disable early stopping and train for 50 epochs.
Save all training outputs to:

"new_tests/model_development/covid_new_dataset/vgg16/training_output".
"""
covid_new_dataset_training_prompt_4 = """
Train an InceptionV3-based classification model using the images located at
"splitted_data/covid_new_dataset/model_development".
Configure the task for binary classification (2 classes).
Enable early stopping with a patience of 5 and train for up to 30 epochs.
Use auxiliary logits during training.
Save all training outputs to:
"new_tests/model_development/covid_new_dataset/inceptionv3/training_output".
"""

**PneumoniaMNIST classification training prompts (X-ray)**

pneumoniamnist_training_prompt_1 = """
Train an EfficientNet-based classification model using the train, validation, and test datasets located at
"splitted_data/pneumoniamnist_28/model_development".
Configure the task for binary classification (2 classes).
Use pretrained weights and apply basic data augmentation.
Train for 120 epochs with early stopping patience set to 10 and a batch size of 4.
Save all training outputs to:
"new_tests/model_development/pneumoniamnist_28/efficientnet/training_output".
"""
pneumoniamnist_training_prompt_2 = """
Train a ResNet-50 classification model using the dataset located at
"splitted_data/pneumoniamnist_28/model_development".
Configure the task for binary classification (2 classes).
Train for 50 epochs with early stopping patience set to 4 and a batch size of 32.
Disable pretrained weights, apply moderate data augmentation,
and handle class imbalance during training.
Save all training outputs to:
"new_tests/model_development/pneumoniamnist_28/resnet50/training_output".
"""
pneumoniamnist_training_prompt_3 = """
Train a VGG-16 classification model using the train, validation, and test datasets located at
"splitted_data/pneumoniamnist_28/model_development".
Configure the task for binary classification (2 classes).
Train for 100 epochs with early stopping disabled and a batch size of 8.
Do not use pretrained weights and apply basic data augmentation.
Save all training outputs to:
"new_tests/model_development/pneumoniamnist_28/vgg16/training_output".
"""
pneumoniamnist_training_prompt_4 = """
Train an InceptionV3-based classification model using the datasets located at
"splitted_data/pneumoniamnist_28/model_development".
Configure the task for binary classification (2 classes).
Do not apply any class imbalance handling.
Use pretrained weights and apply heavy data augmentation.
Train for 70 epochs with early stopping patience set to 5.
Save all training outputs to:
"new_tests/model_development/pneumoniamnist_28/inceptionv3/training_output".
"""

**Cardiomegaly classification inference prompts (X-ray)**

cardiomegaly_dataset_inference_prompt_1 = """
Use the trained EfficientNet model located at
"new_tests/model_development/cardiomegaly/efficientnet/training_output/best_model.pt"
to classify images in the directory:
"splitted_data/cardiomegaly_dataset/inference_dataset/inference_test".
Assume a binary classification task (2 classes).
Evaluate predictions using the ground truth labels from:
"splitted_data/cardiomegaly_dataset/inference_dataset/inference_labels.csv".
Save all evaluation outputs to:
"new_tests/model_development/cardiomegaly/efficientnet/inference_output".
"""

cardiomegaly_dataset_inference_prompt_2 = """
Load the resnet50 model from this location:
"new_tests/model_development/cardiomegaly/resnet50/training_output/best_model.pt",
and perform classification on images located in:
"splitted_data/cardiomegaly_dataset/inference_dataset/inference_test".
Configure for 2 classification categories.
Assess prediction performance against ground truth labels found at:
"splitted_data/cardiomegaly_dataset/inference_dataset/inference_labels.csv".
Store evaluation results in directory:
"new_tests/model_development/cardiomegaly/resnet50/inference_output".
"""

cardiomegaly_dataset_inference_prompt_3 = """
Load the vgg16 model from this location:
"new_tests/model_development/cardiomegaly/vgg16/training_output/best_model.pt",
and perform classification on images located in:
"splitted_data/cardiomegaly_dataset/inference_dataset/inference_test". Enable batch normalization.
Configure for 2 classification categories. Model configuration filepath:
new_tests/model_development/cardiomegaly/vgg16/training_output/model_config.json. Disable
batch normalization.
Assess prediction performance against ground truth labels found at:
"splitted_data/cardiomegaly_dataset/inference_dataset/inference_labels.csv".
Store evaluation results in directory:
"new_tests/model_development/cardiomegaly/vgg16/inference_output".
"""

cardiomegaly_dataset_inference_prompt_4 = """
Use the trained InceptionV3 model located at
"new_tests/model_development/cardiomegaly/inceptionv3/training_output/best_model.pt"
to classify images in the directory:
"splitted_data/cardiomegaly_dataset/inference_dataset/inference_test".
Configure the task for binary classification (2 classes) and enable auxiliary logits.
Evaluate predictions using the ground truth labels from:
"splitted_data/cardiomegaly_dataset/inference_dataset/inference_labels.csv".
Save all evaluation outputs to:
"new_tests/model_development/cardiomegaly/inceptionv3/inference_output".
"""

**Brain tumor classification inference prompts (MRI)**

brain_tumor_new_dataset_inference_prompt_1 = """
Use the trained EfficientNet model located at

"new_tests/model_development/brain_tumor_new_dataset/efficientnet/training_output/best_model.pt
"
to classify images in the directory:
"splitted_data/brain_tumor_new_dataset/inference_dataset/inference_test".
Configure the task for 3-class classification.
Evaluate predictions using the ground truth labels from:
"splitted_data/brain_tumor_new_dataset/inference_dataset/inference_labels.csv".
Save all evaluation outputs to:
"new_tests/model_development/brain_tumor_new_dataset/efficientnet/inference_output".
"""

brain_tumor_new_dataset_inference_prompt_2 = """
Load the resnet50 model from this location:
"new_tests/model_development/brain_tumor_new_dataset/resnet50/training_output/best_model.pt",
and perform classification on images located in:
"splitted_data/brain_tumor_new_dataset/inference_dataset/inference_test".
Configure for 3 classification categories.
Assess prediction performance against ground truth labels found at:
"splitted_data/brain_tumor_new_dataset/inference_dataset/inference_labels.csv".
Store evaluation results in directory:
"new_tests/model_development/brain_tumor_new_dataset/resnet50/inference_output".
"""

brain_tumor_new_dataset_inference_prompt_3 = """
Use the trained VGG 16 model located at
"new_tests/model_development/brain_tumor_new_dataset/vgg16/training_output/best_model.pt"
to classify images in the directory:
"splitted_data/brain_tumor_new_dataset/inference_dataset/inference_test".
Configure the task for 3-class classification and load the model configuration from:
"new_tests/model_development/brain_tumor_new_dataset/vgg16/training_output/model_config.json"
.
Evaluate predictions using the ground truth labels from:
"splitted_data/brain_tumor_new_dataset/inference_dataset/inference_labels.csv".
Save all evaluation outputs to:
"new_tests/model_development/brain_tumor_new_dataset/vgg16/inference_output".
"""

brain_tumor_new_dataset_inference_prompt_4 = """
Load the inceptionv3 model from this location:
"new_tests/model_development/brain_tumor_new_dataset/inceptionv3/training_output/best_model.pt
",
and perform classification on images located in:
"splitted_data/brain_tumor_new_dataset/inference_dataset/inference_test".
Configure for 3 classification categories. Enable auxiliary logits.
Assess prediction performance against ground truth labels found at:
"splitted_data/brain_tumor_new_dataset/inference_dataset/inference_labels.csv".
Store evaluation results in directory:
"new_tests/model_development/brain_tumor_new_dataset/inceptionv3/inference_output".
"""

**Covid classification inference prompts (CT)**

covid_new_dataset_inference_prompt_1 = """
Load the efficientnet model from this location:
"new_tests/model_development/covid_new_dataset/efficientnet/training_output/best_model.pt",

and perform classification on images located in:
"splitted_data/covid_new_dataset/inference_dataset/inference_test".
Configure for 2 classification categories.
Assess prediction performance against ground truth labels found at:
"splitted_data/covid_new_dataset/inference_dataset/inference_labels.csv".
Store evaluation results in directory:
"new_tests/model_development/covid_new_dataset/efficientnet/inference_output".
"""

covid_new_dataset_inference_prompt_2 = """
Classify images in
"splitted_data/covid_new_dataset/inference_dataset/inference_test"
using the ResNet 50 model at
"new_tests/model_development/covid_new_dataset/resnet50/training_output/best_model.pt".
Use 2 classes and evaluate against
"splitted_data/covid_new_dataset/inference_dataset/inference_labels.csv".
Write outputs to
"new_tests/model_development/covid_new_dataset/resnet50/inference_output".
"""

covid_new_dataset_inference_prompt_3 = """
Load the vgg16 model from this location:
"new_tests/model_development/covid_new_dataset/vgg16/training_output/best_model.pt",
and perform classification on images located in:
"splitted_data/covid_new_dataset/inference_dataset/inference_test". Enable batch normalization.
Configure for 2 classification categories. Model configuration filepath:
new_tests/model_development/covid_new_dataset/vgg16/training_output/model_config.json
Assess prediction performance against ground truth labels found at:
"splitted_data/covid_new_dataset/inference_dataset/inference_labels.csv".
Store evaluation results in directory:
"new_tests/model_development/covid_new_dataset/vgg16/inference_output".
"""

covid_new_dataset_inference_prompt_4 = """
Load the InceptionV3 model from
"new_tests/model_development/covid_new_dataset/inceptionv3/training_output/best_model.pt"
and perform image classification on the dataset stored in
"splitted_data/covid_new_dataset/inference_dataset/inference_test".
Treat the task as a binary classification problem, enable auxiliary logits,
and assess performance using the ground truth labels provided in
"splitted_data/covid_new_dataset/inference_dataset/inference_labels.csv".
Store the evaluation results in
"new_tests/model_development/covid_new_dataset/inceptionv3/inference_output".
"""

**PneumoniaMNIST classification inference prompts (X-ray)**

pneumoniamnist_inference_prompt_1 = """
Use the trained EfficientNet model located at
"new_tests/model_development/pneumoniamnist_28/efficientnet/training_output/best_model.pt"
to classify images in the directory:
"splitted_data/pneumoniamnist_28/inference_dataset/inference_test".
Configure the task for binary classification (2 classes).
Evaluate predictions using the ground truth labels from:
"splitted_data/pneumoniamnist_28/inference_dataset/inference_labels.csv".

Save all evaluation outputs to:
"new_tests/model_development/pneumoniamnist_28/efficientnet/inference_output".
"""

pneumoniamnist_inference_prompt_2 = """
Load the resnet50 model from this location:
"new_tests/model_development/pneumoniamnist_28/resnet50/training_output/best_model.pt",
and perform classification on images located in:
"splitted_data/pneumoniamnist_28/inference_dataset/inference_test".
Configure for 2 classification categories.
Assess prediction performance against ground truth labels found at:
"splitted_data/pneumoniamnist_28/inference_dataset/inference_labels.csv".
Store evaluation results in directory:
"new_tests/model_development/pneumoniamnist_28/resnet50/inference_output".
"""

pneumoniamnist_inference_prompt_3 = """
Use the trained VGG-16 model located at
"new_tests/model_development/pneumoniamnist_28/vgg16/training_output/best_model.pt"
to classify images in the directory:
"splitted_data/pneumoniamnist_28/inference_dataset/inference_test".
Configure the task for binary classification (2 classes).
Evaluate predictions using the ground truth labels from:
"splitted_data/pneumoniamnist_28/inference_dataset/inference_labels.csv".
Save all evaluation outputs to:
"new_tests/model_development/pneumoniamnist_28/vgg16/inference_output".
"""

pneumoniamnist_inference_prompt_4 = """
Classify images in
"splitted_data/pneumoniamnist_28/inference_dataset/inference_test"
using the InceptionV3 model at
"new_tests/model_development/pneumoniamnist_28/inceptionv3/training_output/best_model.pt".
Use 2 classes and evaluate against
"splitted_data/pneumoniamnist_28/inference_dataset/inference_labels.csv".
Write outputs to
"new_tests/model_development/pneumoniamnist_28/inceptionv3/inference_output".
"""

**Cardiomegaly classification performance check and fine-tuning prompts (X-ray)**

cardiomegaly_dataset_performance_prompt_1 = """
Assess whether the model’s performance has degraded by comparing the training test metrics at
"new_tests/model_development/cardiomegaly/efficientnet/training_output/test_metrics.json"
with the inference evaluation metrics at
"new_tests/model_development/cardiomegaly/efficientnet/inference_output/metrics.json".

Save the comparison results to:
"new_tests/compare_performance/cardiomegaly/efficientnet/".

If a significant performance drop is detected, fine-tune the model using the dataset:
"splitted_data/cardiomegaly_dataset/fine_tuning_dataset/".

Apply full fine-tuning with the following settings:
- Batch size: 4

- Epochs: 30
- Early stopping patience: 5

Initialize fine-tuning from the checkpoint:
"new_tests/model_development/cardiomegaly/efficientnet/training_output/best_model.pt"
using the configuration file:
"new_tests/model_development/cardiomegaly/efficientnet/training_output/model_config.json".

Save the fine-tuned model to:
"new_tests/fine_tuned_models/cardiomegaly/efficientnet".
"""

cardiomegaly_dataset_performance_prompt_2 = """
Determine whether model performance has degraded.
Training test metrics location:
new_tests/model_development/cardiomegaly/inceptionv3/training_output/test_metrics.json
Inference evaluation metrics location:
new_tests/model_development/cardiomegaly/inceptionv3/inference_output/metrics.json
Results destination: new_tests/compare_performance/cardiomegaly/inceptionv3/
Should significant performance degradation be detected, apply fine-tuning using data from:
splitted_data/cardiomegaly_dataset/fine_tuning_dataset/
Fine-tuning configuration: train head only, 20 epochs, early stopping patience of 4
Model file path:
"new_tests/model_development/cardiomegaly/inceptionv3/training_output/best_model.pt".
Configuration file path:
new_tests/model_development/cardiomegaly/inceptionv3/training_output/model_config.json'
Store fine-tuned model at: new_tests/fine_tuned_models/cardiomegaly/inceptionv3
"""

cardiomegaly_dataset_performance_prompt_3 = """
Evaluate whether the model's performance has degraded by comparing the training test metrics at
"new_tests/model_development/cardiomegaly/resnet50/training_output/test_metrics.json"
with the inference evaluation metrics at
"new_tests/model_development/cardiomegaly/resnet50/inference_output/metrics.json".

Save the comparison results to:
"new_tests/compare_performance/cardiomegaly/resnet50/".

If a significant performance drop is detected, fine-tune the model using the dataset:
"splitted_data/cardiomegaly_dataset/fine_tuning_dataset/".

Apply partial fine-tuning with a moderate augmentation level.
Initialize from the checkpoint:
"new_tests/model_development/cardiomegaly/resnet50/training_output/best_model.pt"
using the configuration file:
"new_tests/model_development/cardiomegaly/resnet50/training_output/model_config.json".

Save the fine-tuned model to:
"new_tests/fine_tuned_models/cardiomegaly/resnet50".
"""

cardiomegaly_dataset_performance_prompt_4 = """
Determine whether model performance has degraded.
Training test metrics location:
new_tests/model_development/cardiomegaly/vgg16/training_output/test_metrics.json

Inference evaluation metrics location:
new_tests/model_development/cardiomegaly/vgg16/inference_output/metrics.json
Results destination: new_tests/compare_performance/cardiomegaly/vgg16/
Should significant performance degradation be detected, apply fine-tuning using data from:
splitted_data/cardiomegaly_dataset/fine_tuning_dataset/
Model file path: "new_tests/model_development/cardiomegaly/vgg16/training_output/best_model.pt".
Configuration file path:
new_tests/model_development/cardiomegaly/vgg16/training_output/model_config.json'
Store fine-tuned model at: new_tests/fine_tuned_models/cardiomegaly/vgg16
"""

**Brain tumor classification performance check and fine-tuning prompts (MRI)**

brain_tumor_new_dataset_performance_prompt_1 = """
Determine whether the model's performance has degraded by comparing the training test metrics at
"new_tests/model_development/brain_tumor_new_dataset/efficientnet/training_output/test_metrics.js
on"
with the inference evaluation metrics at
"new_tests/model_development/brain_tumor_new_dataset/efficientnet/inference_output/metrics.json"
.

Save the comparison results to:
"new_tests/compare_performance/brain_tumor_new_dataset/efficientnet/".

If a significant performance decline is detected, fine-tune the model using the dataset:
"splitted_data/brain_tumor_new_dataset/fine_tuning_dataset/".

Apply partial fine-tuning with the following settings:
- Augmentation level: basic
- Epochs: 50
- Early stopping patience: 5
- Class imbalance handling: enabled

Initialize fine-tuning from:
"new_tests/model_development/brain_tumor_new_dataset/efficientnet/training_output/best_model.pt
"
using the configuration file:
"new_tests/model_development/brain_tumor_new_dataset/efficientnet/training_output/model_config.
json".

Save the fine-tuned model to:
"new_tests/fine_tuned_models/brain_tumor_new_dataset/efficientnet".
"""

brain_tumor_new_dataset_performance_prompt_2 = """
Determine whether model performance has degraded.
Training test metrics location:
new_tests/model_development/brain_tumor_new_dataset/inceptionv3/training_output/test_metrics.jso
n
Inference evaluation metrics location:
new_tests/model_development/brain_tumor_new_dataset/inceptionv3/inference_output/metrics.json
Results destination: new_tests/compare_performance/brain_tumor_new_dataset/inceptionv3/
Should significant performance degradation be detected, apply fine-tuning using data from:
splitted_data/brain_tumor_new_dataset/fine_tuning_dataset/

Model file path:
"new_tests/model_development/brain_tumor_new_dataset/inceptionv3/training_output/best_model.pt
". Configuration file path:
new_tests/model_development/brain_tumor_new_dataset/inceptionv3/training_output/model_config.j
son'
Store fine-tuned model at: new_tests/fine_tuned_models/brain_tumor_new_dataset/inceptionv3
"""

brain_tumor_new_dataset_performance_prompt_3 = """
Compare:
-
"new_tests/model_development/brain_tumor_new_dataset/resnet50/training_output/test_metrics.json"
- "new_tests/model_development/brain_tumor_new_dataset/resnet50/inference_output/metrics.json"

Save comparison results to:
"new_tests/compare_performance/brain_tumor_new_dataset/resnet50/".

If performance has significantly declined, fine-tune using:
"splitted_data/brain_tumor_new_dataset/fine_tuning_dataset/"
with head-only fine-tuning, basic augmentation, and batch size 16.

Initialize from:
"new_tests/model_development/brain_tumor_new_dataset/resnet50/training_output/best_model.pt"
using config:
"new_tests/model_development/brain_tumor_new_dataset/resnet50/training_output/model_config.jso
n".

Save the fine-tuned model to:
"new_tests/fine_tuned_models/brain_tumor_new_dataset/resnet50".
"""

brain_tumor_new_dataset_performance_prompt_4 = """
Evaluate whether the model's performance has degraded by comparing the training test metrics at
"new_tests/model_development/brain_tumor_new_dataset/vgg16/training_output/test_metrics.json"
with the inference evaluation metrics at
"new_tests/model_development/brain_tumor_new_dataset/vgg16/inference_output/metrics.json".

Save the comparison results to:
"new_tests/compare_performance/brain_tumor_new_dataset/vgg16/".

If a significant performance decline is detected, fine-tune the model using the dataset:
"splitted_data/brain_tumor_new_dataset/fine_tuning_dataset/".

Initialize fine-tuning from the checkpoint:
"new_tests/model_development/brain_tumor_new_dataset/vgg16/training_output/best_model.pt"
and apply the configuration file:
"new_tests/model_development/brain_tumor_new_dataset/vgg16/training_output/model_config.json"
.

Save the fine-tuned model to:
"new_tests/fine_tuned_models/brain_tumor_new_dataset/vgg16".
"""

**Covid classification performance check and fine-tuning prompts (CT)**

covid_new_dataset_performance_prompt_1 = """
Evaluate whether the model's performance has degraded by comparing the training test metrics at
"new_tests/model_development/covid_new_dataset/efficientnet/training_output/test_metrics.json"
with the inference evaluation metrics at
"new_tests/model_development/covid_new_dataset/efficientnet/inference_output/metrics.json".

Save the comparison results to:
"new_tests/compare_performance/covid_new_dataset/efficientnet/".

If a significant performance decline is detected, fine-tune the model using the dataset:
"splitted_data/covid_new_dataset/fine_tuning_dataset/".

Initialize fine-tuning from the checkpoint:
"new_tests/model_development/covid_new_dataset/efficientnet/training_output/best_model.pt"
and apply the configuration file:
"new_tests/model_development/covid_new_dataset/efficientnet/training_output/model_config.json".

Save the fine-tuned model to:
"new_tests/fine_tuned_models/covid_new_dataset/efficientnet".

"""
covid_new_dataset_performance_prompt_2 = """
Assess whether the model's performance has degraded by comparing the training test metrics at
"new_tests/model_development/covid_new_dataset/inceptionv3/training_output/test_metrics.json"
with the inference evaluation metrics at
"new_tests/model_development/covid_new_dataset/inceptionv3/inference_output/metrics.json".

Save the comparison results to:
"new_tests/compare_performance/covid_new_dataset/inceptionv3/".

If a significant performance decline is detected, fine-tune the model using the dataset:
"splitted_data/covid_new_dataset/fine_tuning_dataset/".

Apply partial fine-tuning.
Initialize from the checkpoint:
"new_tests/model_development/covid_new_dataset/inceptionv3/training_output/best_model.pt"
using the configuration file:
"new_tests/model_development/covid_new_dataset/inceptionv3/training_output/model_config.json".

Save the fine-tuned model to:
"new_tests/fine_tuned_models/covid_new_dataset/inceptionv3".

"""
covid_new_dataset_performance_prompt_3 = """
Determine whether model performance has degraded.
Training test metrics location:
new_tests/model_development/covid_new_dataset/resnet50/training_output/test_metrics.json
Inference evaluation metrics location:
new_tests/model_development/covid_new_dataset/resnet50/inference_output/metrics.json
Results destination: new_tests/compare_performance/covid_new_dataset/resnet50/
Should significant performance degradation be detected, apply fine-tuning using data from:
splitted_data/covid_new_dataset/fine_tuning_dataset/
Fine-tuning configuration: train entire model. Data augmentation: moderate level. Training duration:
30 epochs. Early stopping patience: 5.

Model file path:
"new_tests/model_development/covid_new_dataset/resnet50/training_output/best_model.pt".
Configuration file path:
new_tests/model_development/covid_new_dataset/resnet50/training_output/model_config.json'
Store fine-tuned model at: new_tests/fine_tuned_models/covid_new_dataset/resnet50
"""

covid_new_dataset_performance_prompt_4 = """
Determine whether model performance has degraded.
Training test metrics location:
new_tests/model_development/covid_new_dataset/vgg16/training_output/test_metrics.json
Inference evaluation metrics location:
new_tests/model_development/covid_new_dataset/vgg16/inference_output/metrics.json
Results destination: new_tests/compare_performance/covid_new_dataset/vgg16/
Should significant performance degradation be detected, apply fine-tuning using data from:
splitted_data/covid_new_dataset/fine_tuning_dataset/
Model file path:
"new_tests/model_development/covid_new_dataset/vgg16/training_output/best_model.pt".
Configuration file path:
new_tests/model_development/covid_new_dataset/vgg16/training_output/model_config.json'
Store fine-tuned model at: new_tests/fine_tuned_models/covid_new_dataset/vgg16
"""

**PneumoniaMNIST classification performance check and fine-tuning prompts (X-ray)**

pneumoniamnist_28_performance_prompt_1 = """
Compare:
-"new_tests/model_development/pneumoniamnist_28/efficientnet/training_output/test_metrics.json"
-"new_tests/model_development/pneumoniamnist_28/efficientnet/inference_output/metrics.json"

Save comparison results to:
"new_tests/compare_performance/pneumoniamnist_28/efficientnet/".

If performance has significantly declined, fine-tune using:
"splitted_data/pneumoniamnist_28/fine_tuning_dataset/".

Initialize from:
"new_tests/model_development/pneumoniamnist_28/efficientnet/training_output/best_model.pt"
using config:
"new_tests/model_development/pneumoniamnist_28/efficientnet/training_output/model_config.json".

Save the fine-tuned model to:
"new_tests/fine_tuned_models/pneumoniamnist_28/efficientnet".
"""
pneumoniamnist_28_performance_prompt_2 = """
Determine whether model performance has degraded.
Training test metrics location:
new_tests/model_development/pneumoniamnist_28/inceptionv3/training_output/test_metrics.json
Inference evaluation metrics location:
new_tests/model_development/pneumoniamnist_28/inceptionv3/inference_output/metrics.json
Store comparison results at: new_tests/compare_performance/pneumoniamnist_28/inceptionv3/
Should significant performance degradation be detected, apply fine-tuning using data from:
splitted_data/pneumoniamnist_28/fine_tuning_dataset/
Model file path:
"new_tests/model_development/pneumoniamnist_28/inceptionv3/training_output/best_model.pt".

Configuration file path:
new_tests/model_development/pneumoniamnist_28/inceptionv3/training_output/model_config.json'
Store fine-tuned model at: new_tests/fine_tuned_models/pneumoniamnist_28/inceptionv3
"""
pneumoniamnist_28_performance_prompt_3 = """
Assess whether the model’s performance has degraded by comparing the training test metrics at
"new_tests/model_development/pneumoniamnist_28/resnet50/training_output/test_metrics.json"
with the inference evaluation metrics at
"new_tests/model_development/pneumoniamnist_28/resnet50/inference_output/metrics.json".

Save the comparison results to:
"new_tests/compare_performance/pneumoniamnist_28/resnet50/".

If a significant performance decline is detected, fine-tune the model using the dataset:
"splitted_data/pneumoniamnist_28/fine_tuning_dataset/".

Perform full-model fine-tuning with the following configuration:
- Data augmentation: heavy
- Training epochs: 20

Initialize from the checkpoint:
"new_tests/model_development/pneumoniamnist_28/resnet50/training_output/best_model.pt"
and apply the configuration file:
"new_tests/model_development/pneumoniamnist_28/resnet50/training_output/model_config.json".

Save the fine-tuned model to:
"new_tests/fine_tuned_models/pneumoniamnist_28/resnet50".
"""

pneumoniamnist_28_performance_prompt_4 = """
Evaluate whether the model’s performance has degraded by comparing the training test metrics at
"new_tests/model_development/pneumoniamnist_28/vgg16/training_output/test_metrics.json"
with the inference evaluation metrics at
"new_tests/model_development/pneumoniamnist_28/vgg16/inference_output/metrics.json".

Save the comparison results to:
"new_tests/compare_performance/pneumoniamnist_28/vgg16/".

If a significant performance decline is detected, fine-tune the model using the dataset:
"splitted_data/pneumoniamnist_28/fine_tuning_dataset/".

Configure fine-tuning with:
- Epochs: 50
- Early stopping patience: 10
- Class imbalance handling: disabled
- Monitoring metric: accuracy

Initialize fine-tuning from the checkpoint:
"new_tests/model_development/pneumoniamnist_28/vgg16/training_output/best_model.pt"
using the configuration file:
"new_tests/model_development/pneumoniamnist_28/vgg16/training_output/model_config.json".

Save the fine-tuned model to:
"new_tests/fine_tuned_models/pneumoniamnist_28/vgg16".
"""

**Traces – 3D Nodule Training, Inference, Fine-tuning**

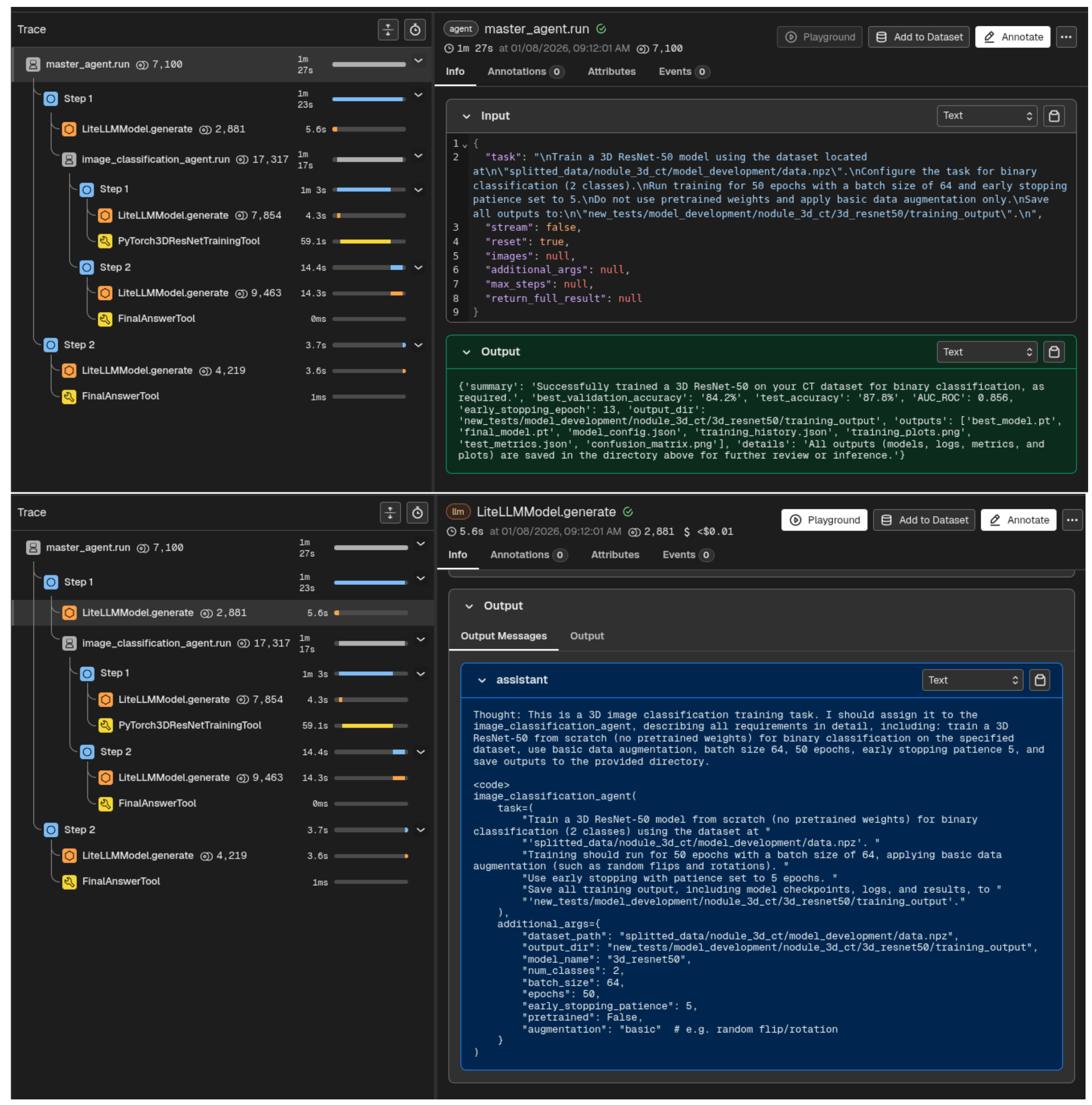

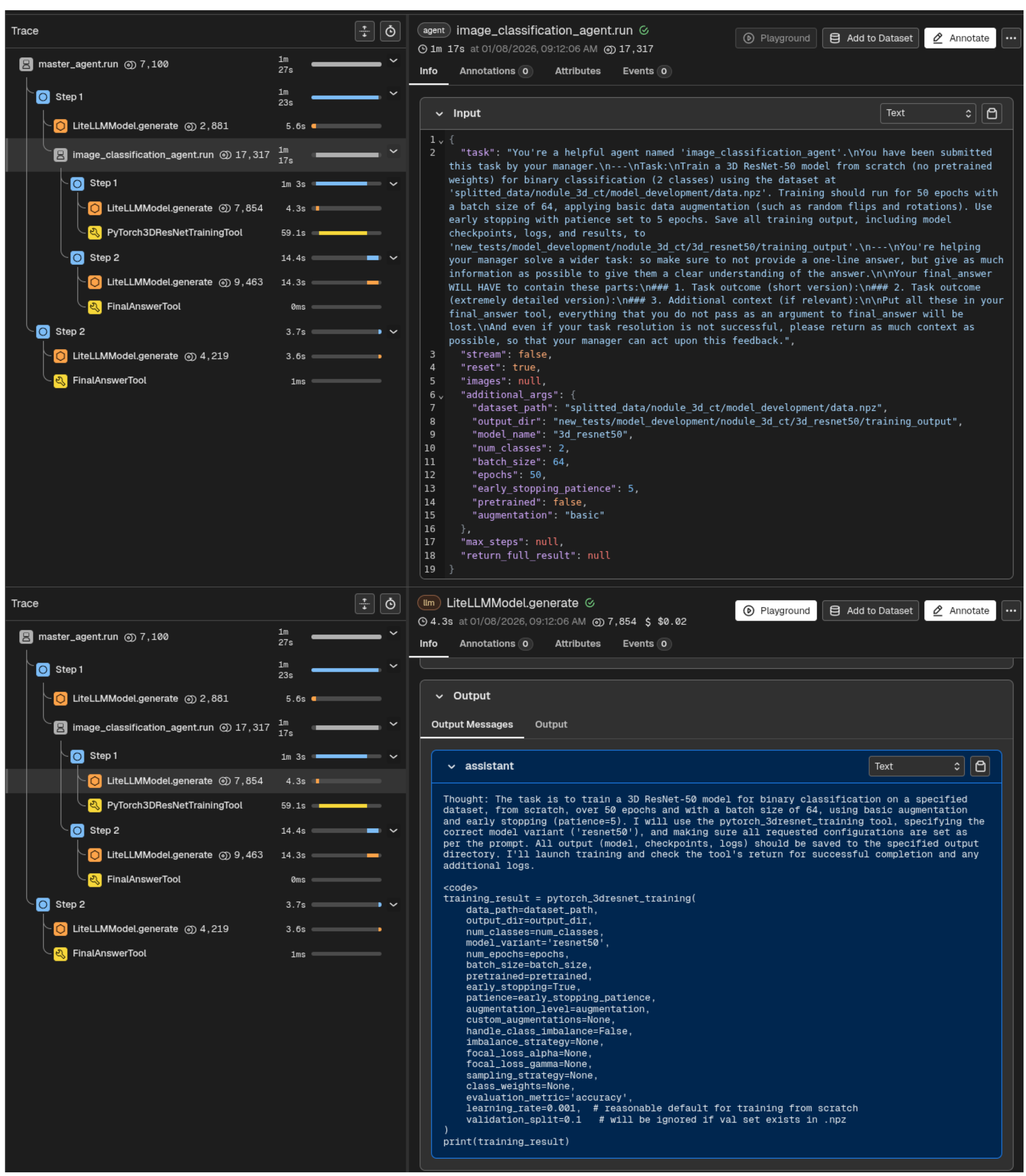

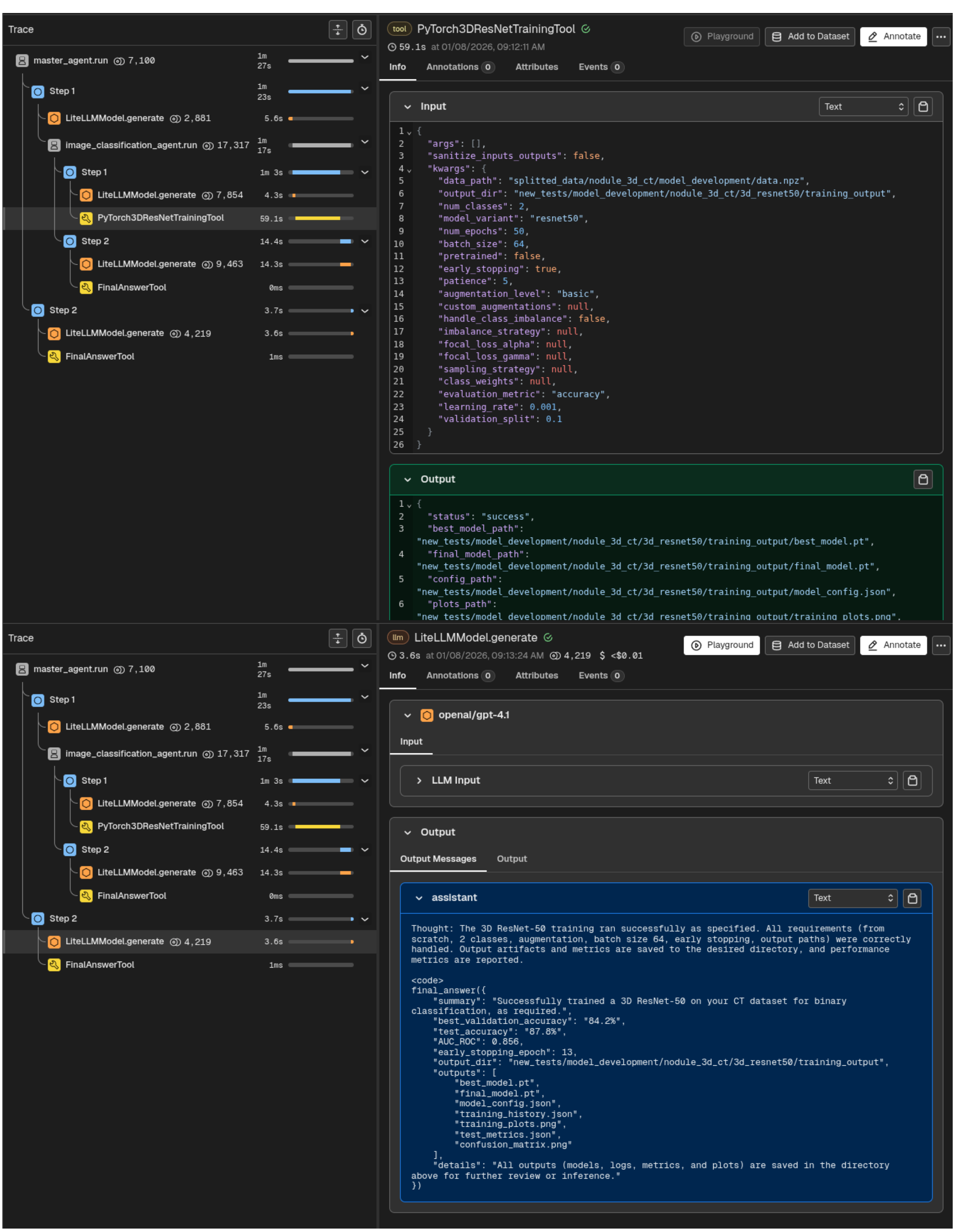

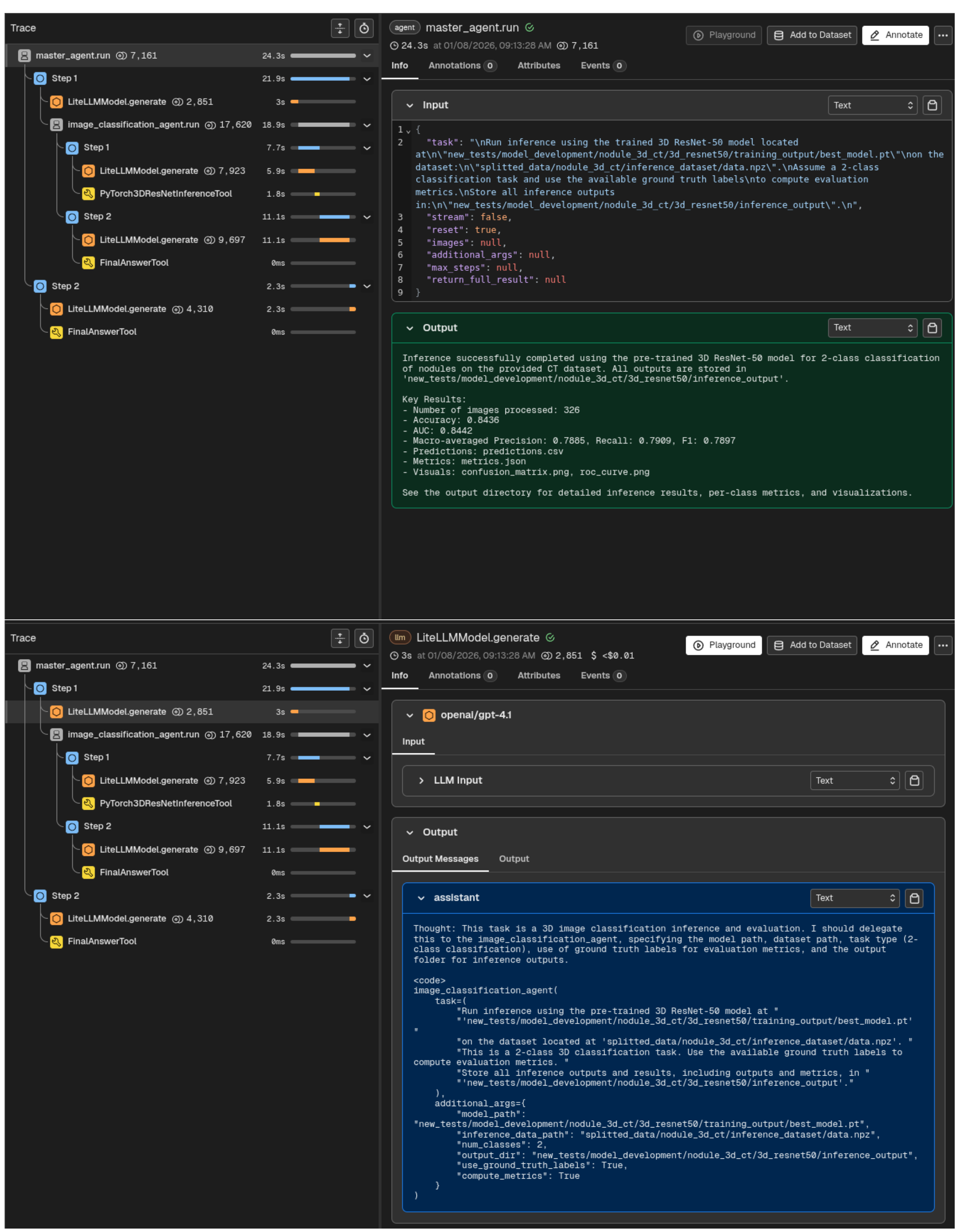

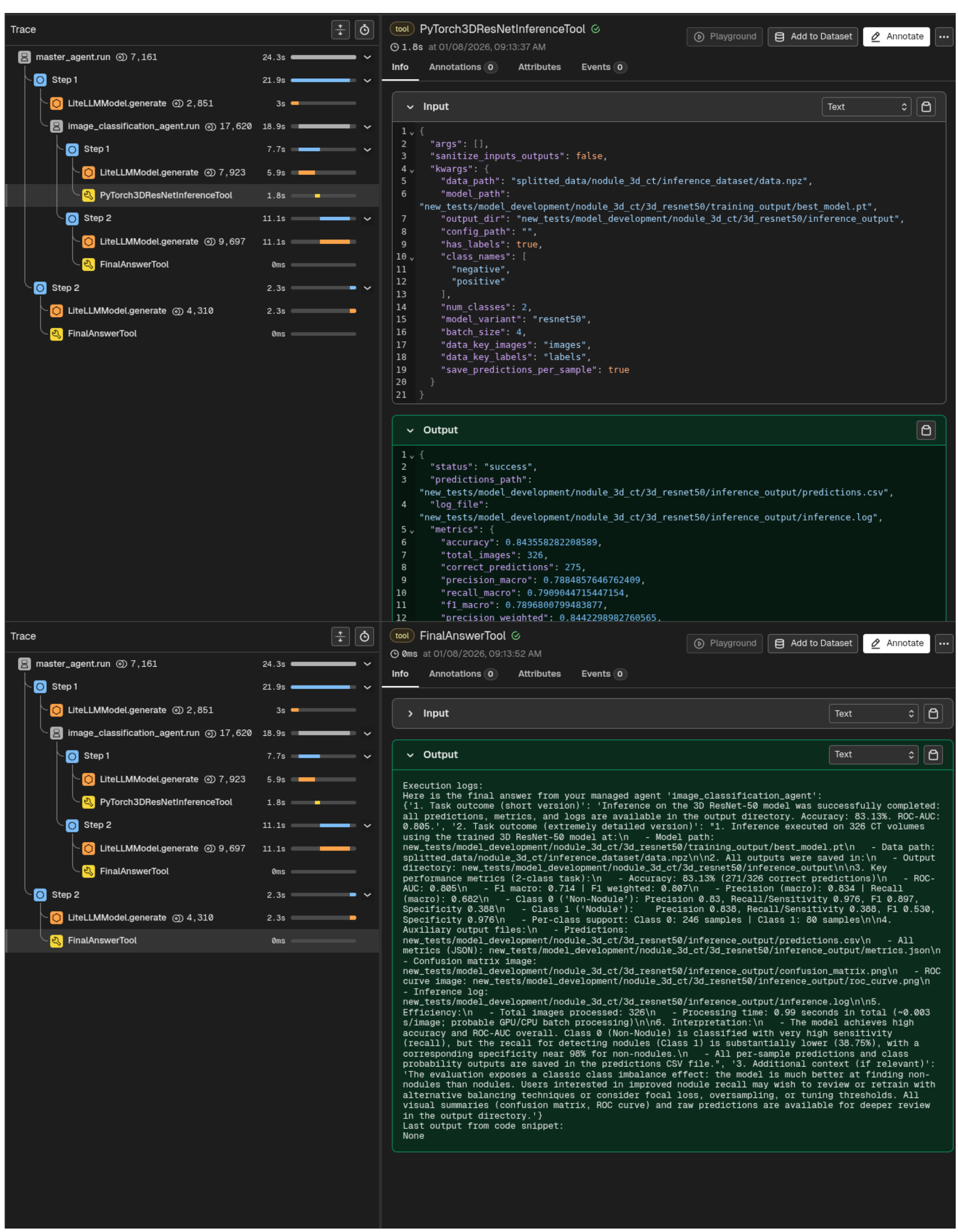

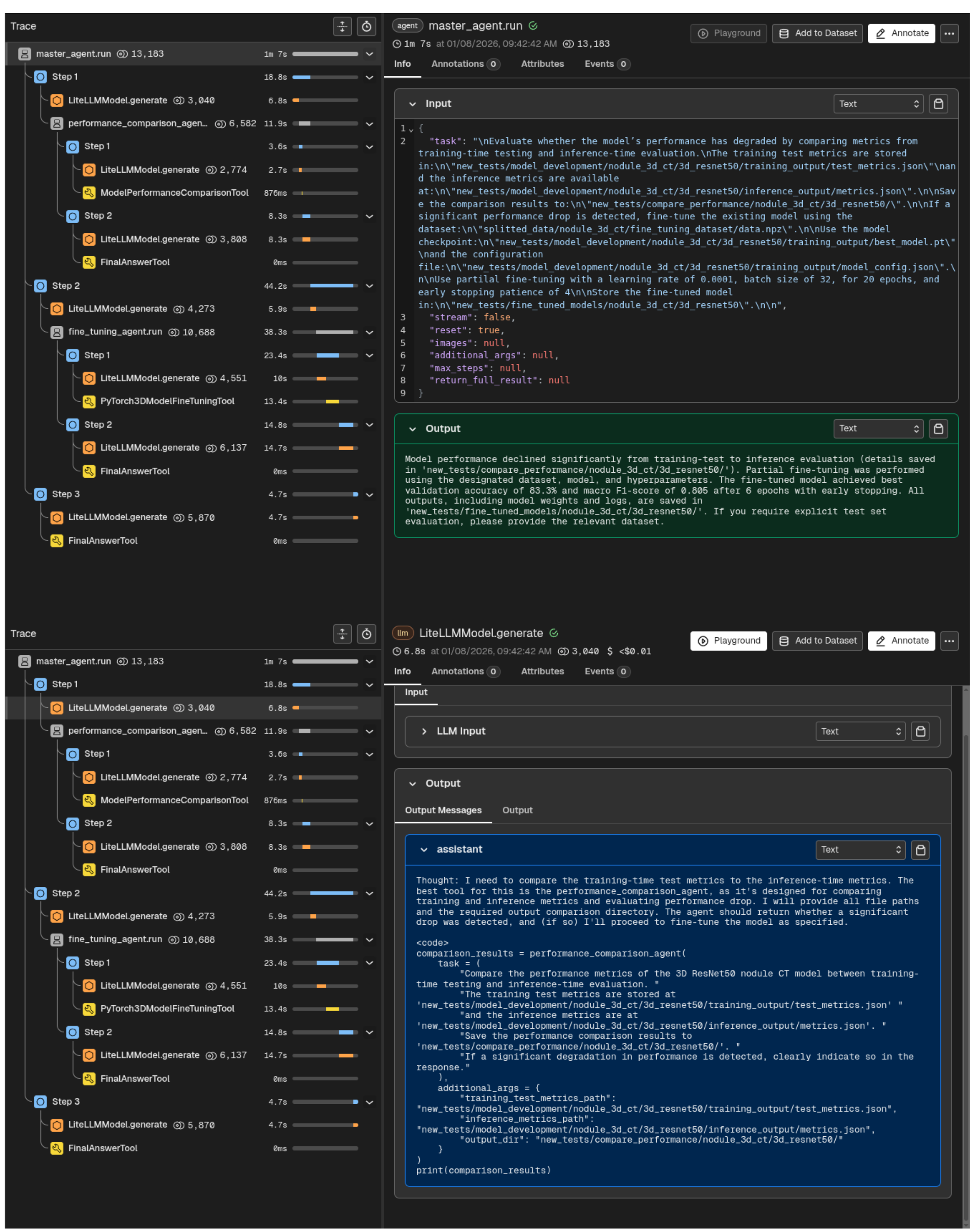

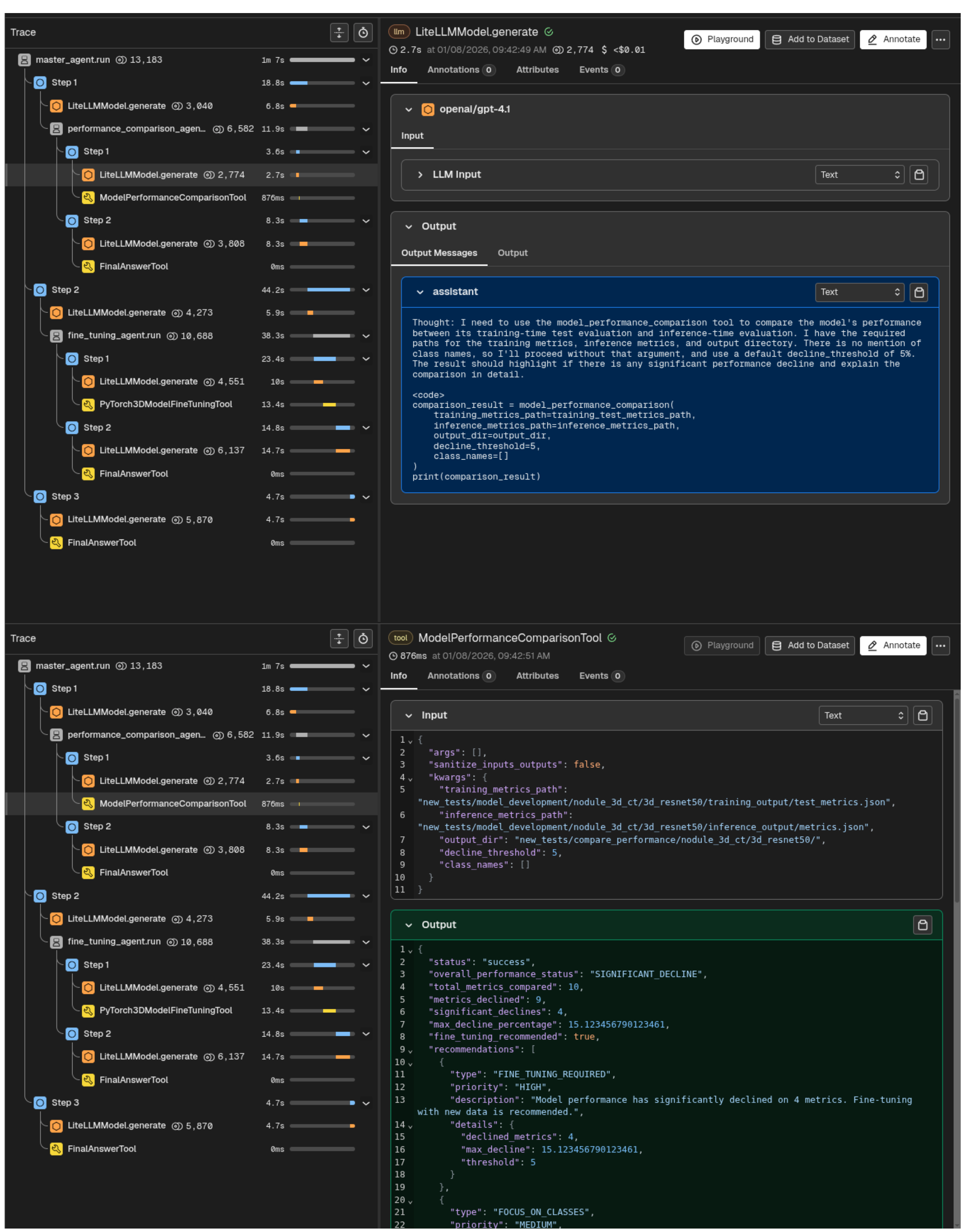

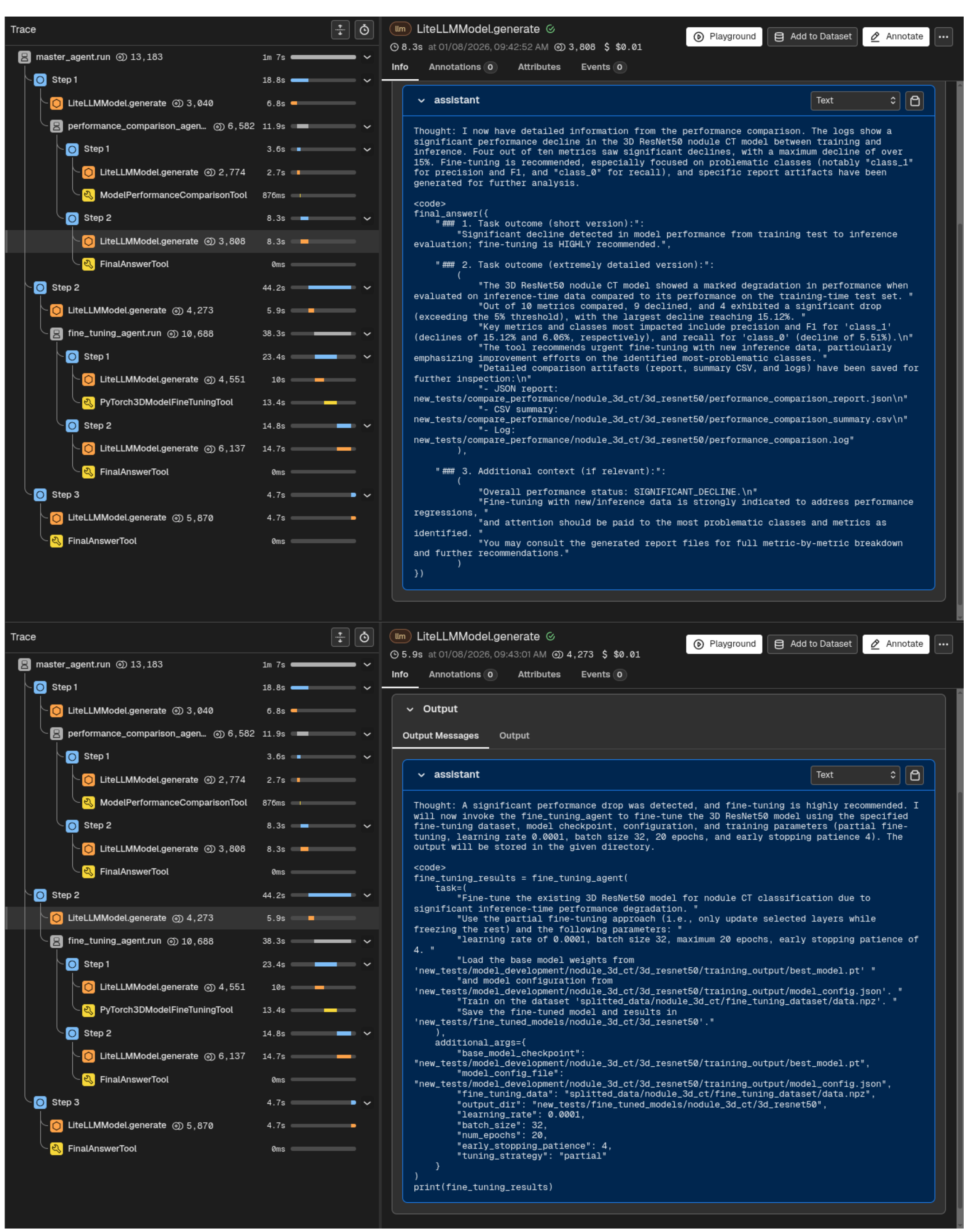

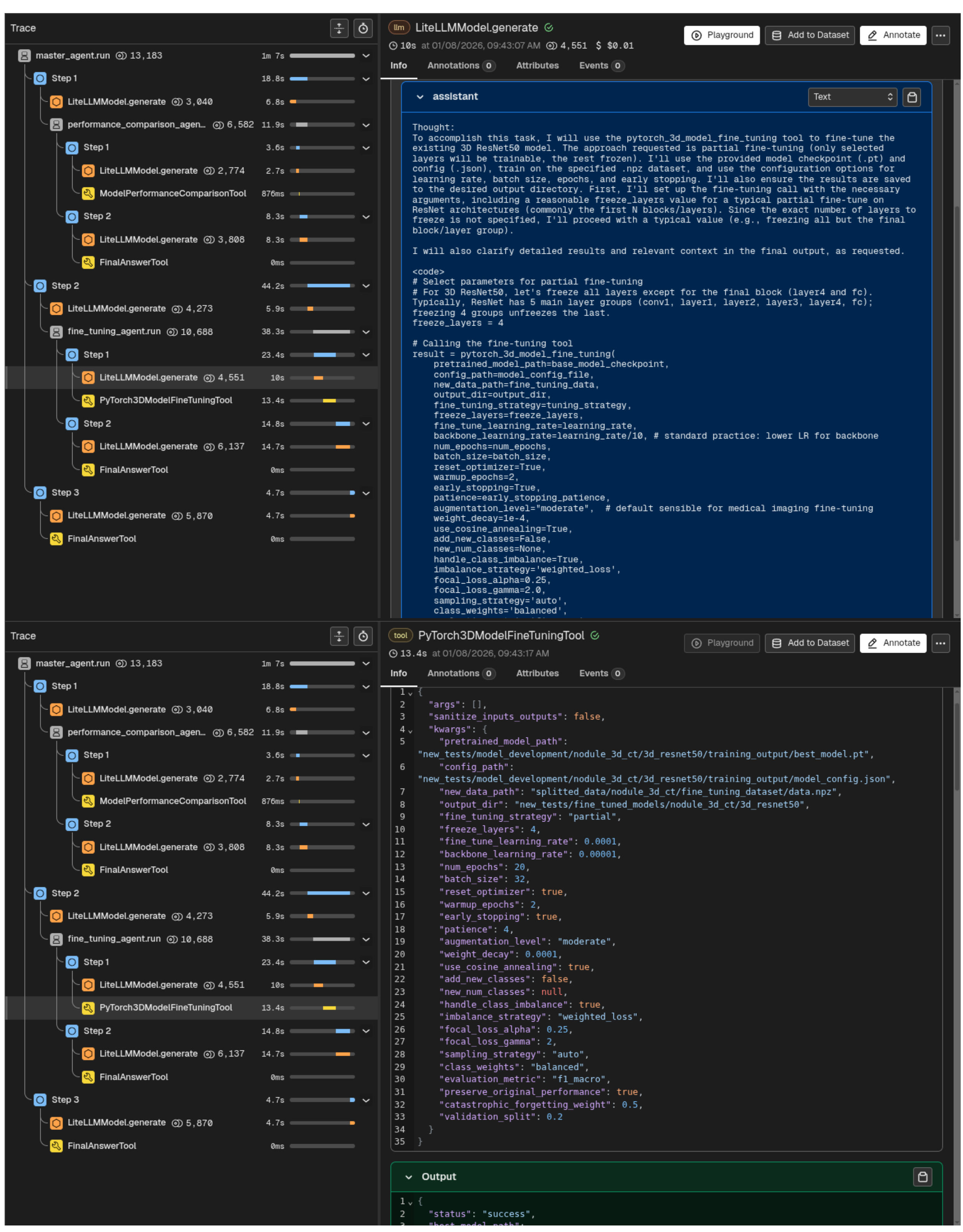

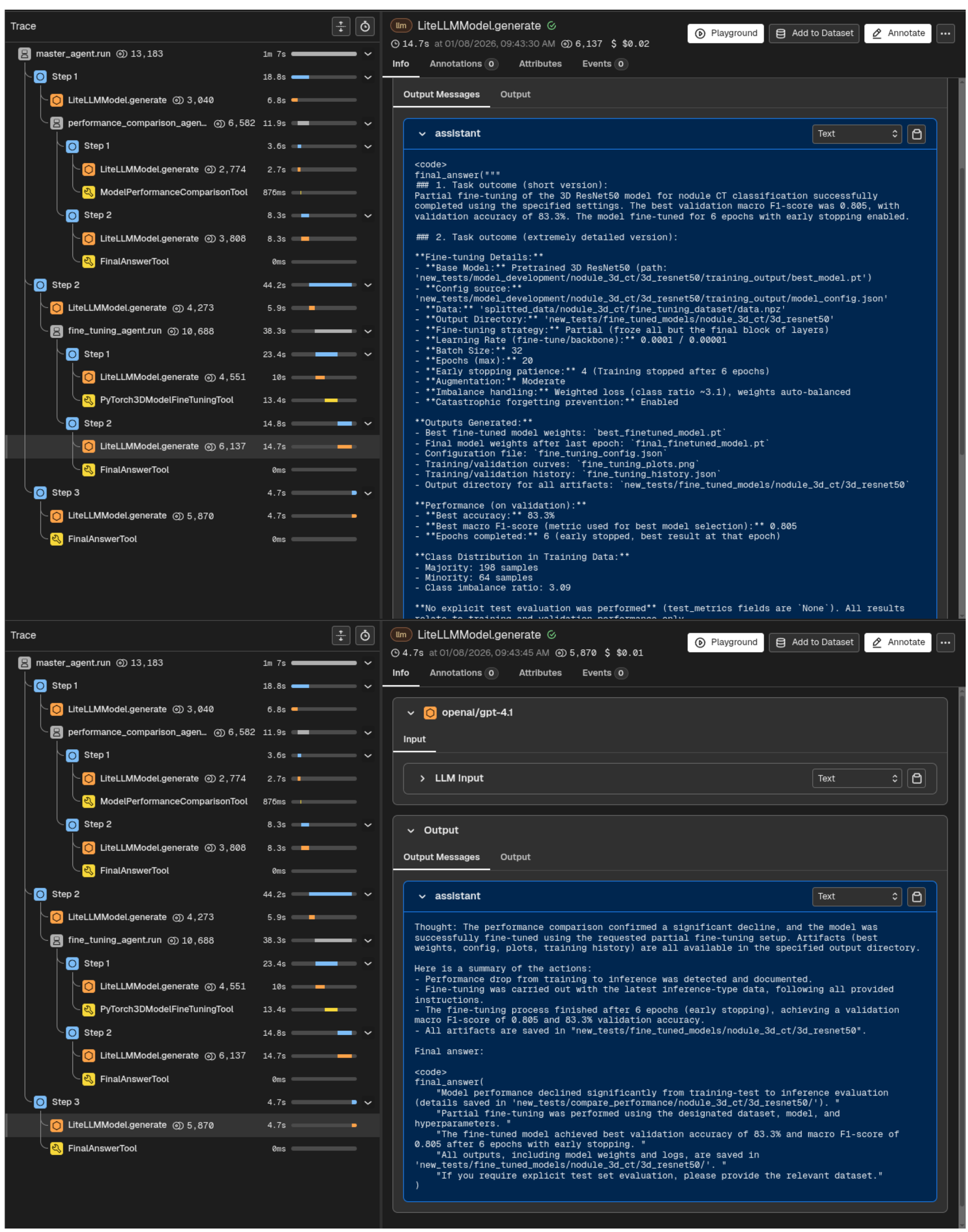

**Traces Brain tumor InceptionV3 classification – Training, Inference**

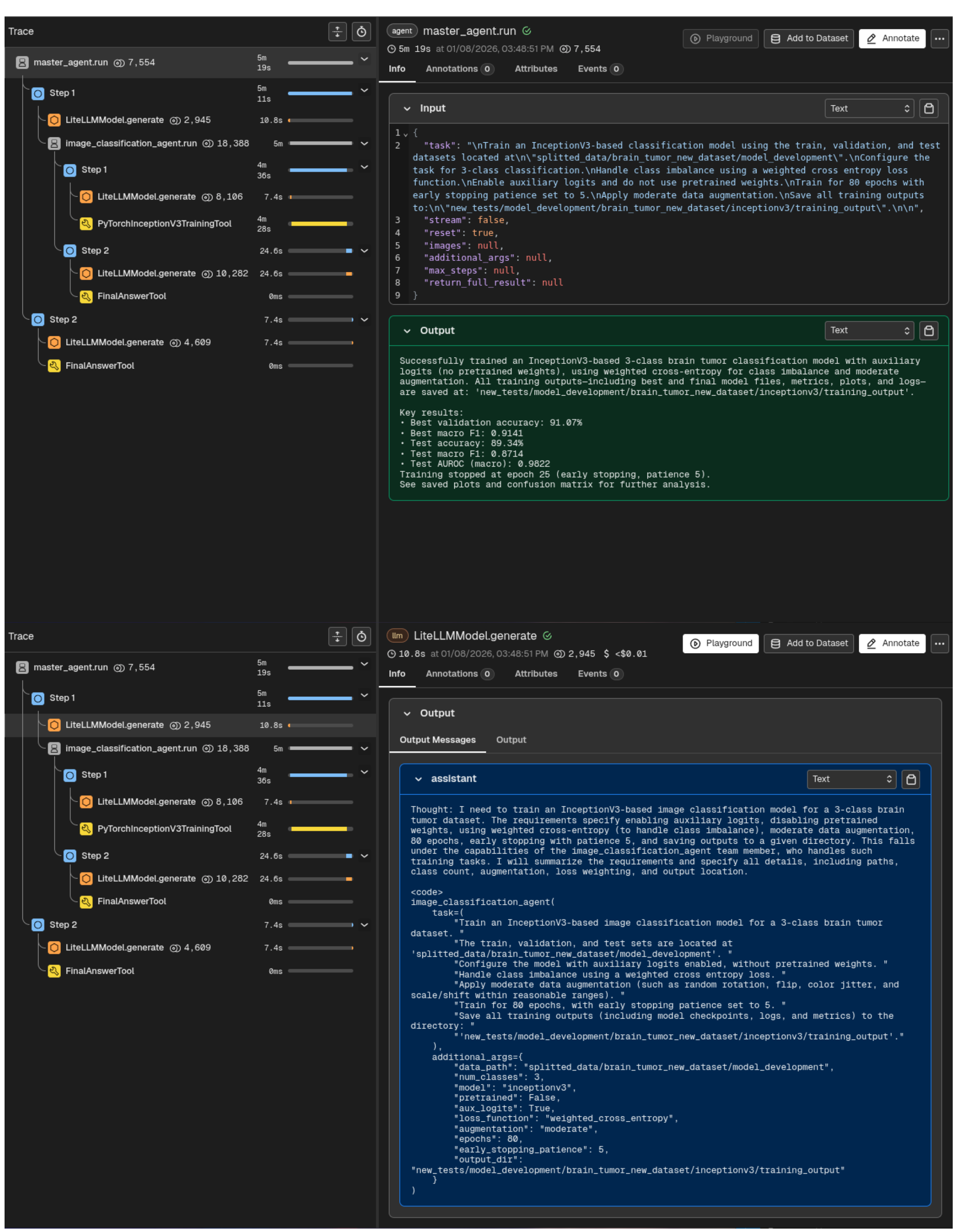

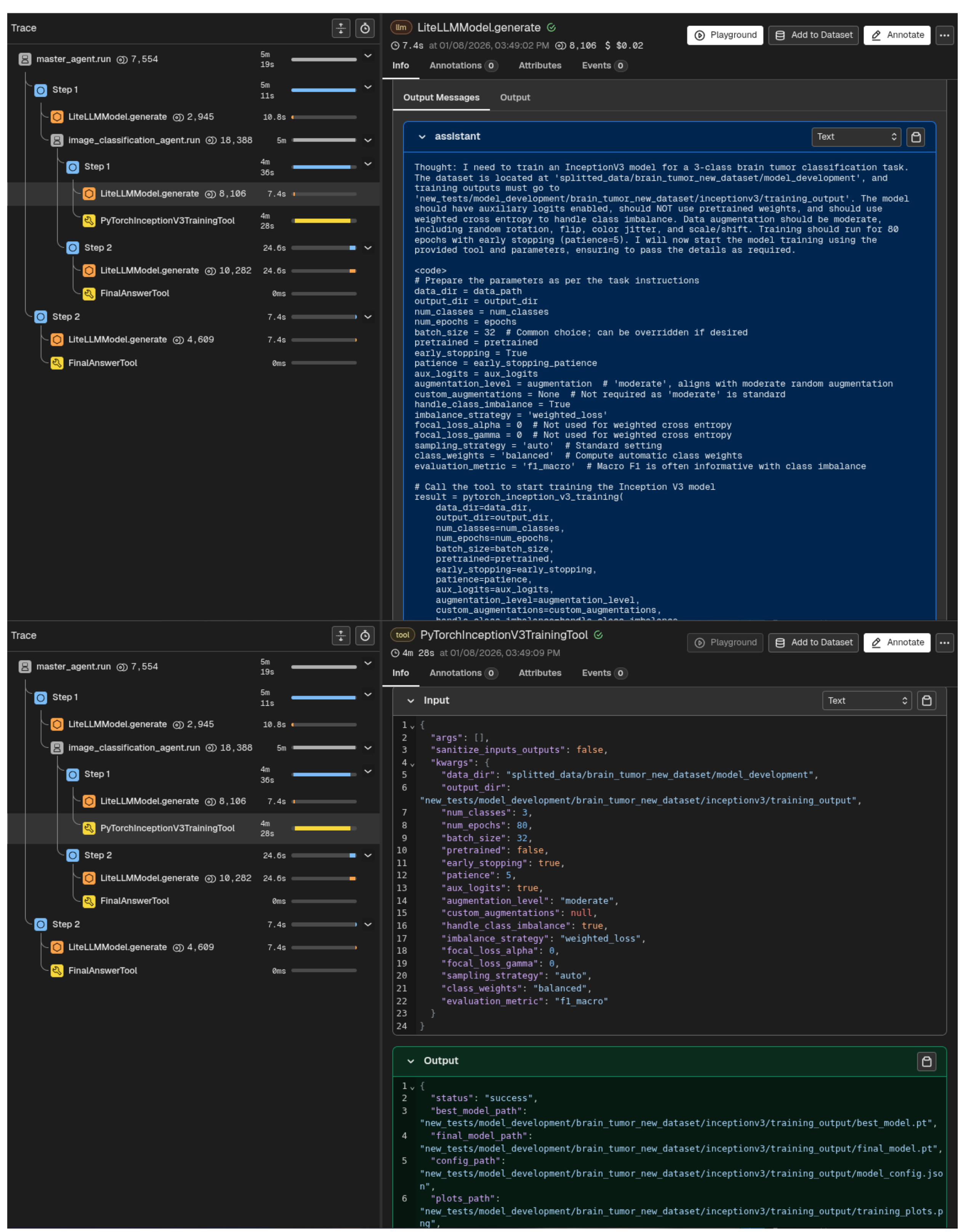

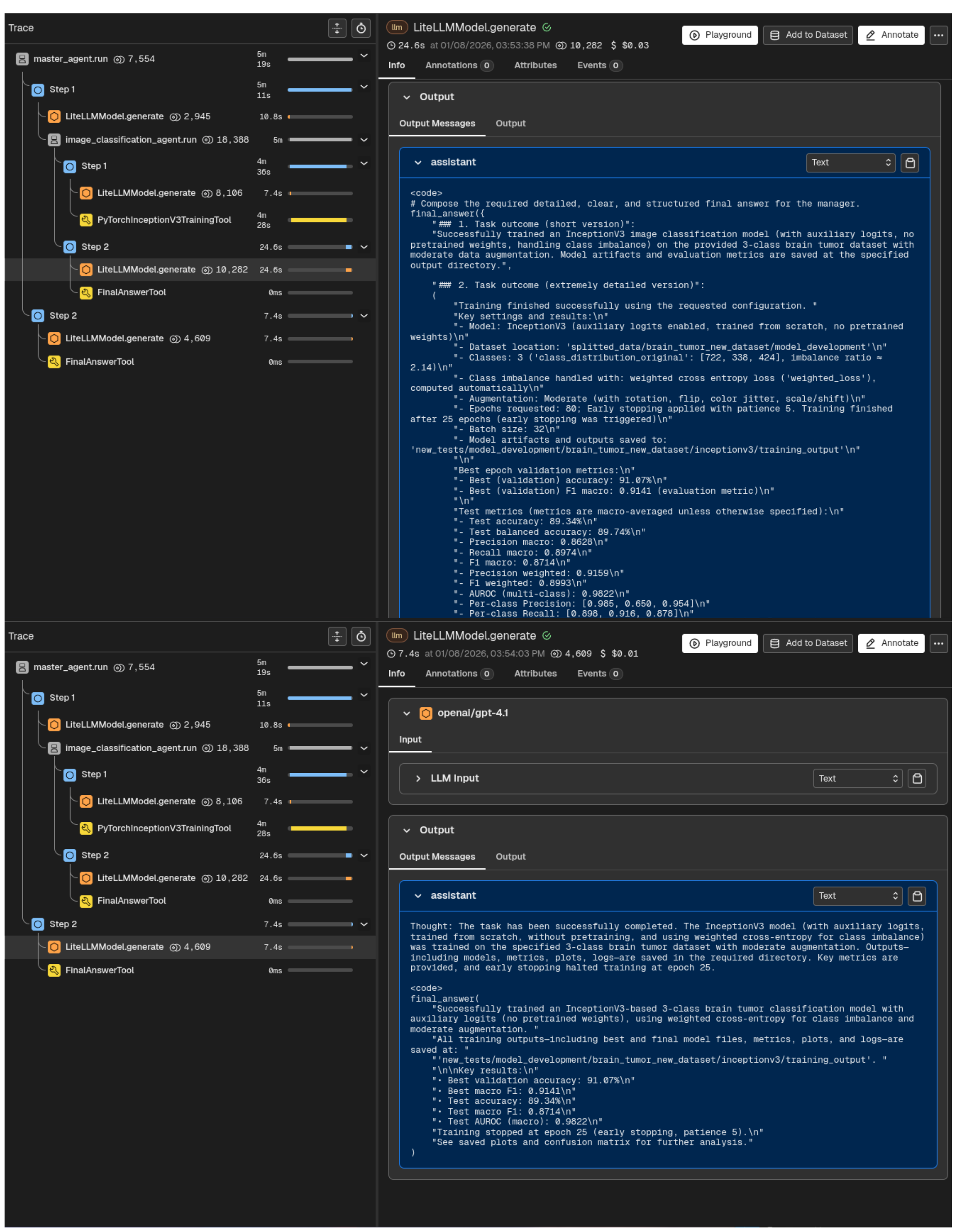

**Traces Cardiomegaly classification EfficientNet – Training**

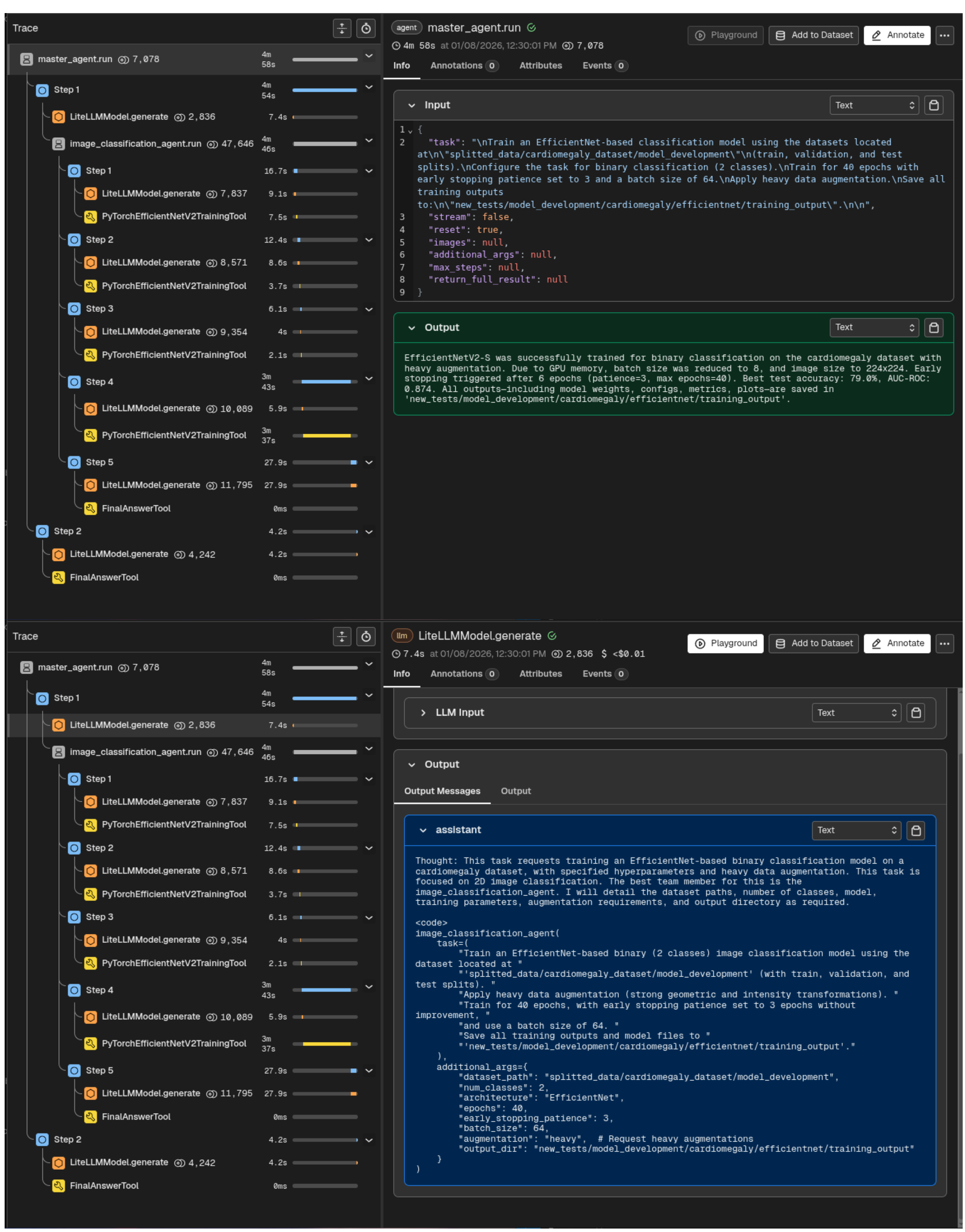

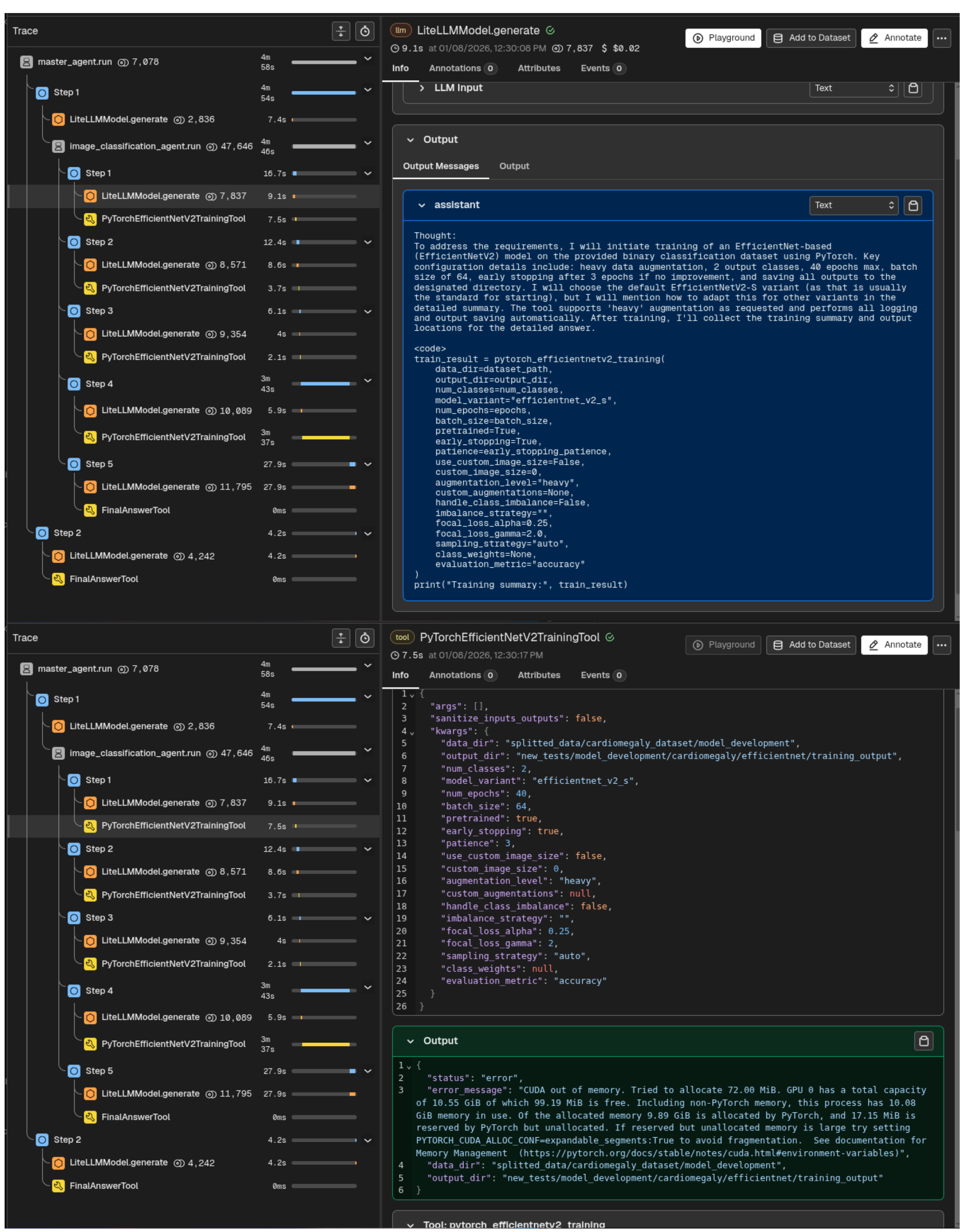

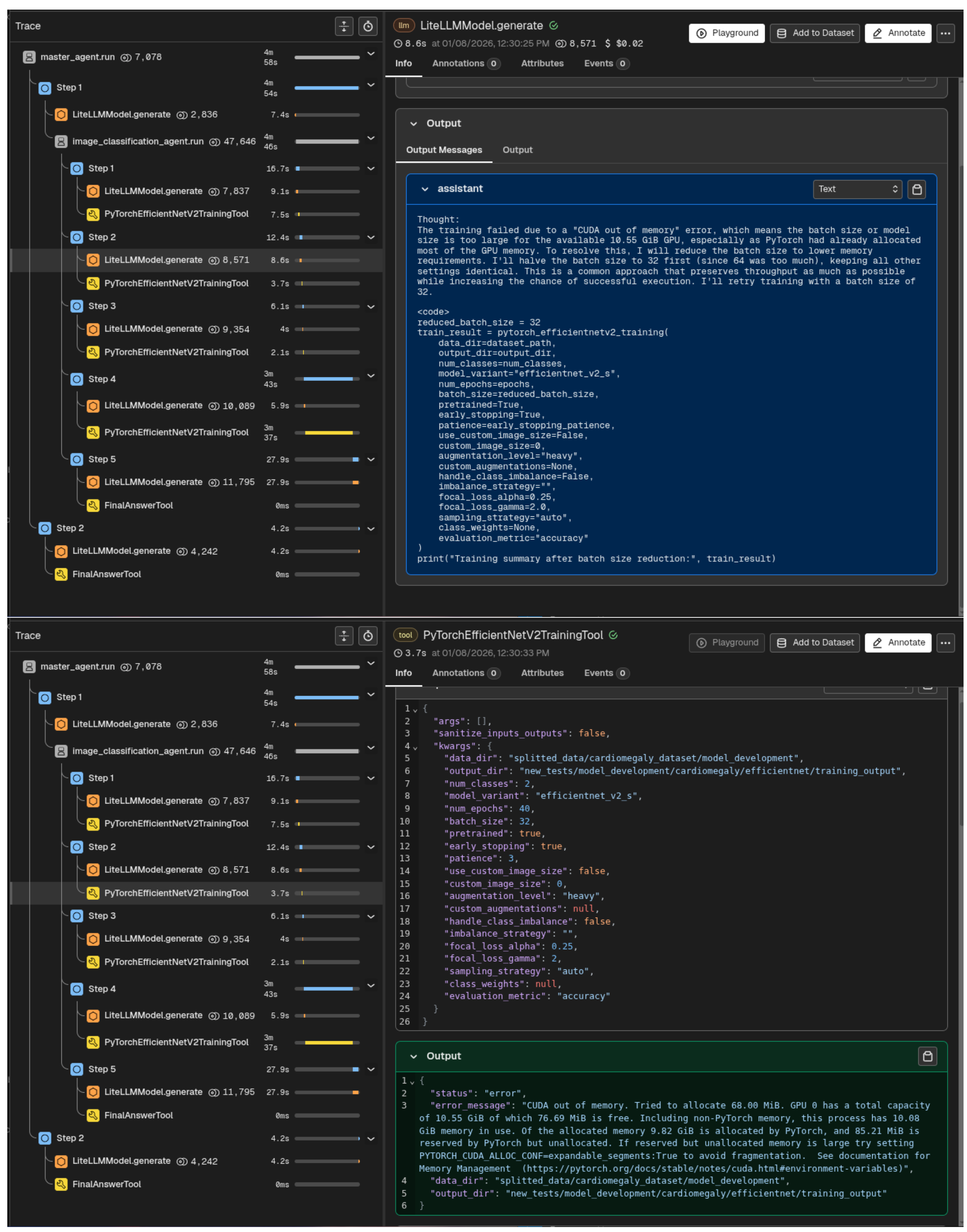

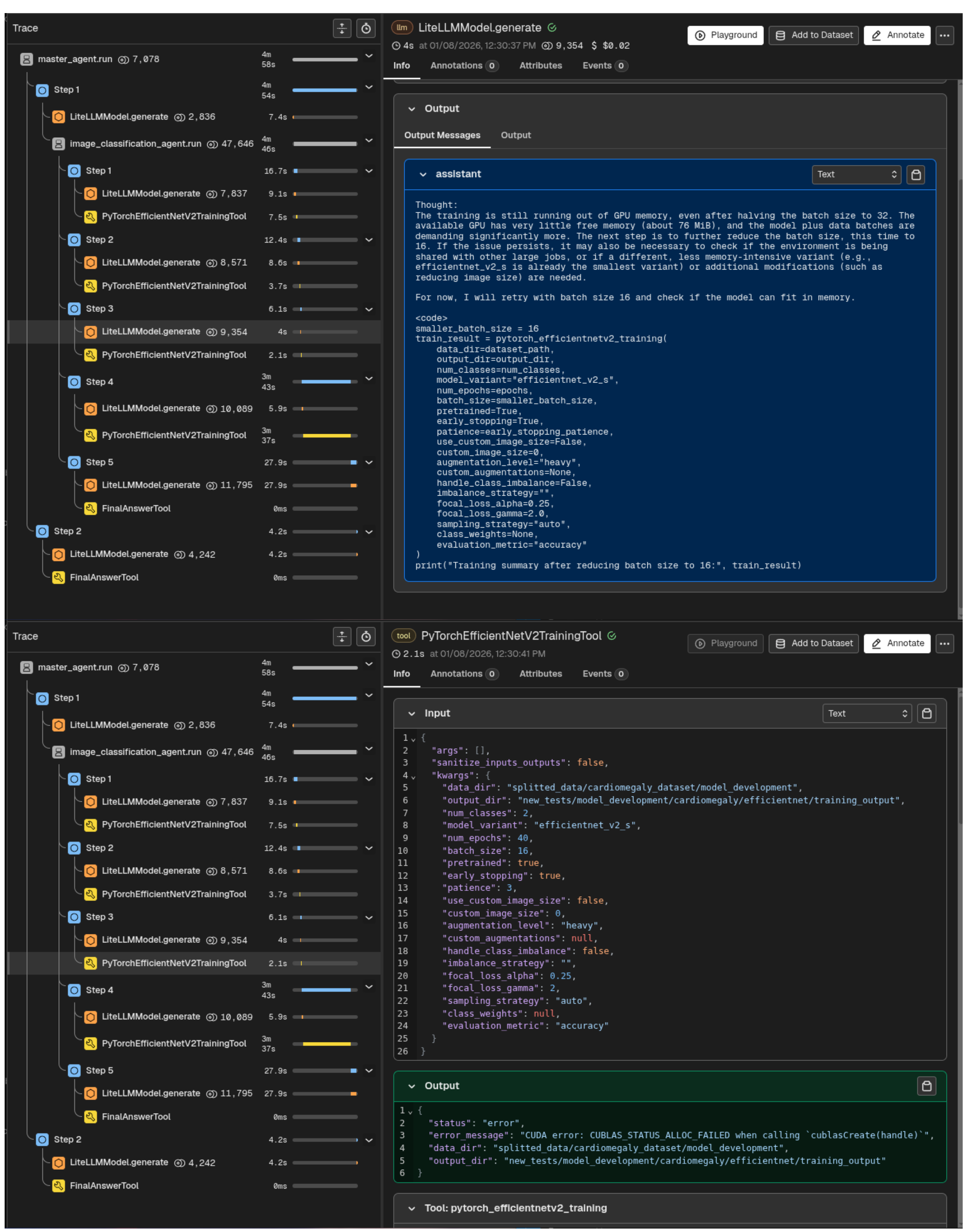

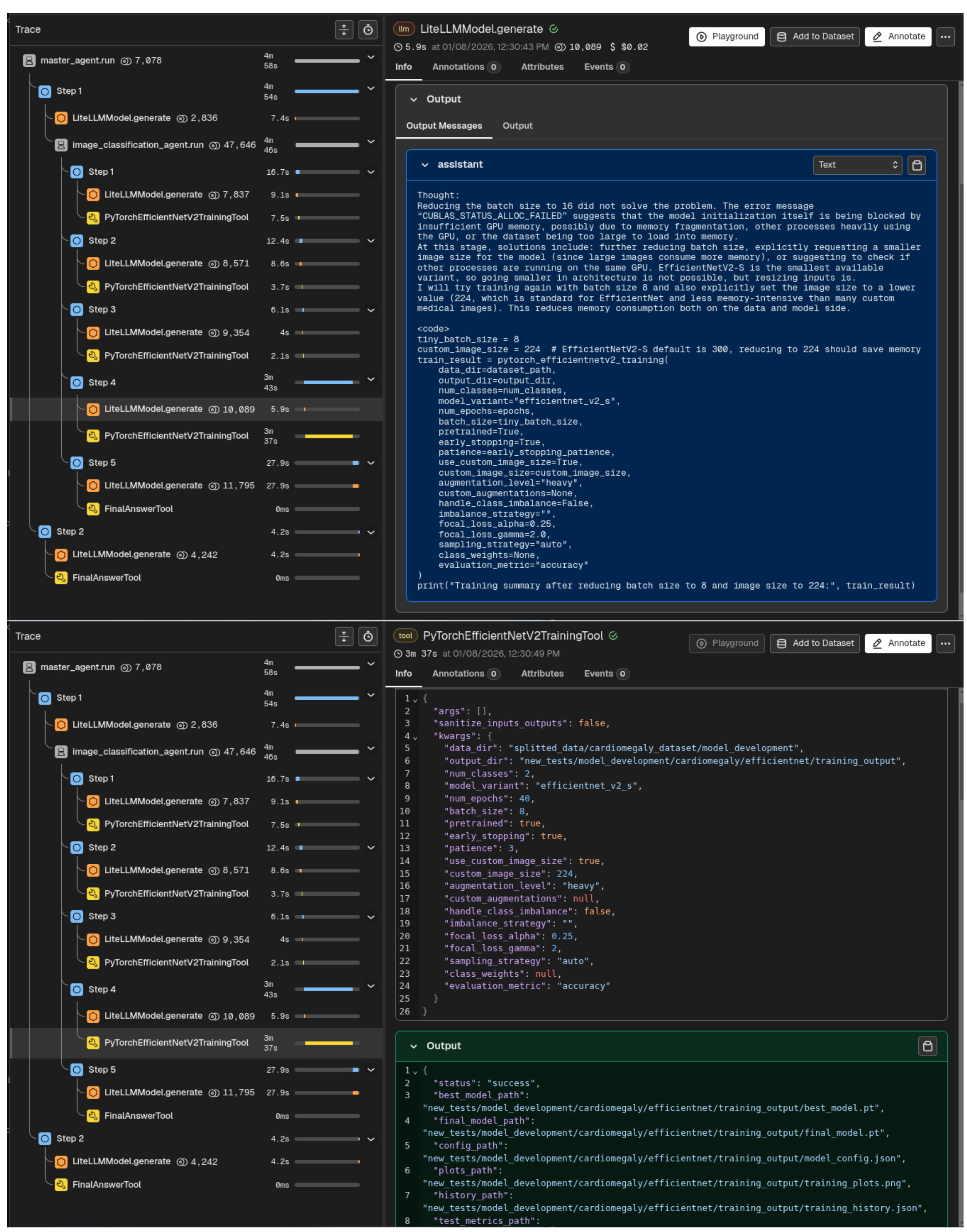

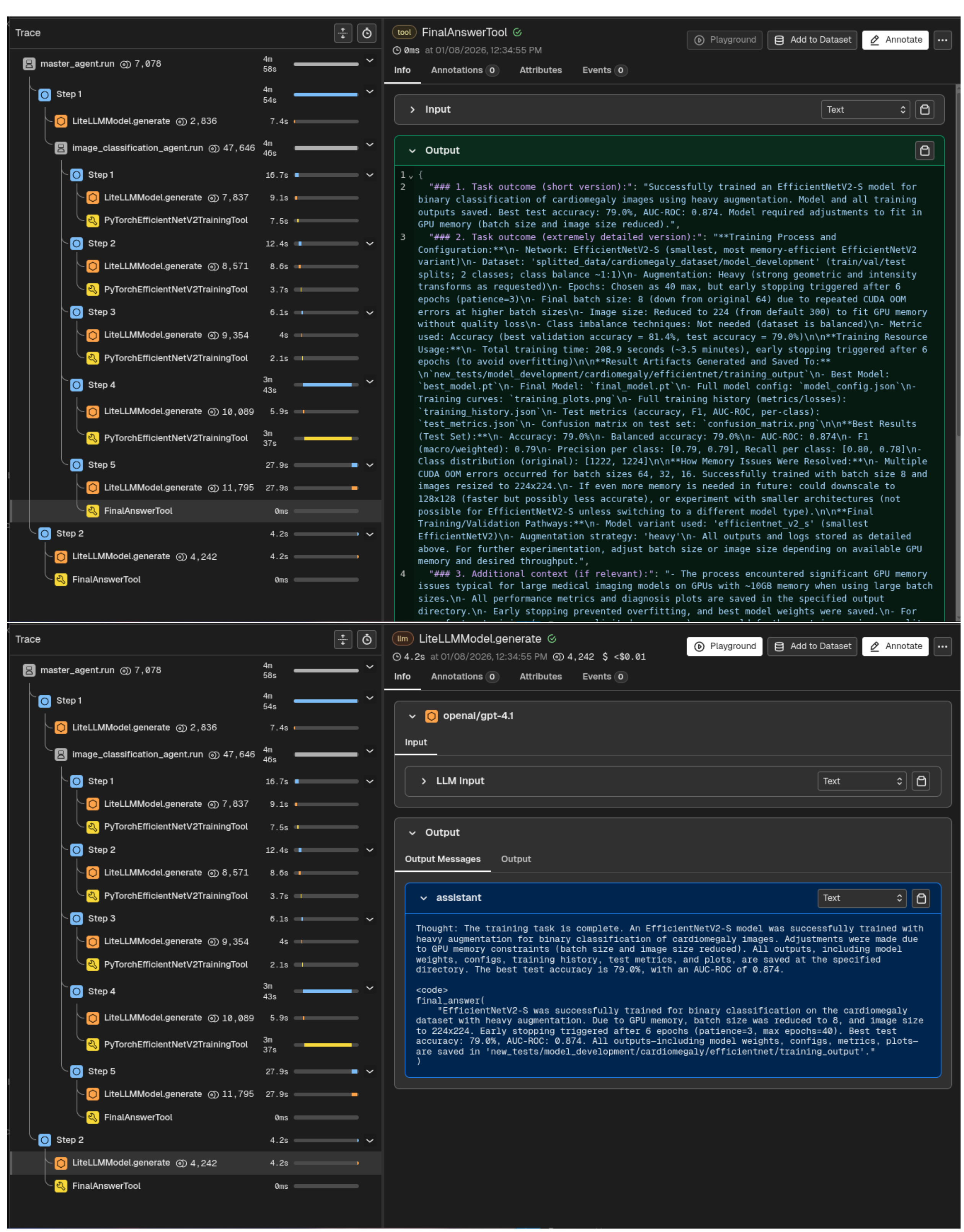

**Supplementary Table S1**. Confidence Intervals (95% CIs) for Table 1 metrics

**Brain Tumor MRI dataset**

| Model | Evaluation Dataset | Accuracy | Precision | | | | Recall | | | | F1 | | | |
|---|---|---|---|---|---|---|---|---|---|---|---|---|---|---|
| | | | Macro | Class 0 | Class 1 | Class 2 | Macro | Class 0 | Class 1 | Class 2 | Macro | Class 0 | Class 1 | Class 2 |
| **EfficientNet** | Test set | [0.949–0.982] | [0.923–0.971] | [0.984–1.000] | [0.792–0.931] | [0.965–1.000] | [0.950–0.983] | [0.912–0.974] | [0.938–1.000] | [0.963–1.000] | [0.935–0.976] | [0.951–0.985] | [0.870–0.955] | [0.970–0.995] |
| | Inference set | [0.941–0.980] | [0.941–0.980] | [0.897–0.978] | [0.920–0.991] | [0.957–1.000] | [0.937–0.979] | [0.948–1.000] | [0.871–0.972] | [0.945–1.000] | [0.939–0.980] | [0.932–0.981] | [0.908–0.972] | [0.959–0.996] |
| **Fine-tuned EfficientNet** | Inference set | [0.941–0.980] | [0.943–0.982] | [0.915–0.985] | [0.920–0.991] | [0.945–1.000] | [0.940–0.981] | [0.938–0.993] | [0.885–0.978] | [0.958–1.000] | [0.941–0.981] | [0.936–0.983] | [0.913–0.976] | [0.960–0.996] |
| **InceptionV3** | Test set | [0.951–0.982] | [0.926–0.972] | [0.976–1.000] | [0.804–0.935] | [0.973–1.000] | [0.956–0.985] | [0.936–0.986] | [0.960–1.000] | [0.934–0.988] | [0.939–0.977] | [0.961–0.990] | [0.883–0.963] | [0.959–0.990] |
| | Inference set | [0.897–0.950] | [0.898–0.950] | [1.000–1.000] | [0.812–0.929] | [0.849–0.954] | [0.901–0.952] | [0.811–0.925] | [0.937–1.000] | [0.898–0.981] | [0.895–0.950] | [0.896–0.961] | [0.882–0.954] | [0.887–0.955] |
| **Fine-tuned InceptionV3** | Inference set | [0.936–0.978] | [0.935–0.977] | [0.924–0.992] | [0.854–0.962] | [1.000–1.000] | [0.935–0.977] | [0.948–1.000] | [0.908–0.990] | [0.899–0.983] | [0.934–0.977] | [0.945–0.989] | [0.894–0.963] | [0.947–0.991] |
| **ResNet50** | Test set | [0.932–0.971] | [0.903–0.958] | [0.976–1.000] | [0.756–0.912] | [0.938–0.989] | [0.923–0.969] | [0.960–0.996] | [0.864–0.977] | [0.892–0.966] | [0.912–0.961] | [0.973–0.996] | [0.823–0.926] | [0.924–0.971] |
| | Inference set | [0.908–0.961] | [0.907–0.960] | [0.897–0.977] | [0.878–0.974] | [0.887–0.976] | [0.905–0.960] | [0.912–0.985] | [0.838–0.951] | [0.908–0.984] | [0.906–0.960] | [0.915–0.972] | [0.873–0.952] | [0.908–0.969] |
| **Fine-tuned ResNet50** | Inference set | [0.944–0.980] | [0.945–0.982] | [0.946–1.000] | [0.949–1.000] | [0.895–0.976] | [0.943–0.981] | [0.928–0.992] | [0.894–0.982] | [0.958–1.000] | [0.944–0.981] | [0.946–0.989] | [0.932–0.986] | [0.933–0.983] |
| **VGG16** | Test set | [0.922–0.963] | [0.889–0.942] | [1.000–1.000] | [0.667–0.825] | [1.000–1.000] | [0.939–0.970] | [0.872–0.950] | [1.000–1.000] | [0.921–0.980] | [0.902–0.952] | [0.932–0.974] | [0.800–0.904] | [0.959–0.990] |
| | Inference set | [0.936–0.978] | [0.935–0.977] | [0.936–0.993] | [0.843–0.955] | [1.000–1.000] | [0.935–0.978] | [0.937–0.993] | [0.922–0.991] | [0.895–0.981] | [0.935–0.977] | [0.946–0.989] | [0.894–0.964] | [0.945–0.991] |

**Covid-19 CT dataset**

| Model | Evaluation Dataset | Accuracy | Precision | | | Recall | | | F1 | | |
|---|---|---|---|---|---|---|---|---|---|---|---|
| | | | Macro | Class 0 | Class 1 | Macro | Class 0 | Class 1 | Macro | Class 0 | Class 1 |
| **EfficientNet** | Test set | [0.952–0.987] | [0.953–0.987] | [0.953–0.995] | [0.937–0.989] | [0.952–0.987] | [0.931–0.988] | [0.957–0.995] | [0.952–0.987] | [0.950–0.986] | [0.954–0.987] |
| | Inference set | [0.962–0.983] | [0.962–0.984] | [0.961–0.990] | [0.954–0.985] | [0.961–0.983] | [0.949–0.984] | [0.964–0.991] | [0.962–0.983] | [0.960–0.983] | [0.963–0.984] |
| **InceptionV3** | Test set | [0.885–0.944] | [0.885–0.944] | [0.857–0.943] | [0.888–0.962] | [0.886–0.944] | [0.879–0.959] | [0.866–0.946] | [0.885–0.944] | [0.878–0.941] | [0.886–0.945] |
| | Inference set | [0.567–0.633] | [0.732–0.790] | [0.908–1.000] | [0.530–0.600] | [0.565–0.602] | [0.138–0.212] | [0.985–1.000] | [0.473–0.542] | [0.241–0.347] | [0.691–0.749] |
| **Fine-tuned InceptionV3** | Inference set | [0.898–0.935] | [0.898–0.936] | [0.893–0.947] | [0.886–0.939] | [0.898–0.935] | [0.876–0.933] | [0.902–0.952] | [0.898–0.935] | [0.893–0.933] | [0.901–0.939] |
| **ResNet50** | Test set | [0.931–0.971] | [0.930–0.972] | [0.889–0.961] | [0.957–0.995] | [0.931–0.973] | [0.955–0.995] | [0.892–0.962] | [0.930–0.971] | [0.927–0.972] | [0.930–0.973] |
| | Inference set | [0.914–0.948] | [0.914–0.949] | [0.884–0.938] | [0.930–0.972] | [0.914–0.949] | [0.927–0.971] | [0.888–0.940] | [0.914–0.948] | [0.912–0.948] | [0.914–0.950] |
| **VGG16** | Test set | [0.952–0.987] | [0.952–0.987] | [0.938–0.989] | [0.949–0.995] | [0.952–0.987] | [0.946–0.994] | [0.942–0.990] | [0.952–0.987] | [0.950–0.986] | [0.953–0.987] |
| | Inference set | [0.952–0.977] | [0.952–0.977] | [0.945–0.982] | [0.947–0.981] | [0.952–0.977] | [0.943–0.980] | [0.949–0.984] | [0.952–0.977] | [0.949–0.977] | [0.953–0.978] |

**NoduleMNIST3D CT dataset**

| Model | Evaluation Dataset | Accuracy | Precision | | | Recall | | | F1 | | |
|---|---|---|---|---|---|---|---|---|---|---|---|
| | | | Macro | Class 0 | Class 1 | Macro | Class 0 | Class 1 | Macro | Class 0 | Class 1 |
| **3D ResNet50** | Test set | [0.784–0.899] | [0.706–0.870] | [0.832–0.947] | [0.526–0.840] | [0.702–0.866] | [0.841–0.955] | [0.500–0.824] | [0.705–0.861] | [0.853–0.937] | [0.541–0.793] |
| | Inference set | [0.807–0.883] | [0.737–0.842] | [0.874–0.946] | [0.570–0.771] | [0.756–0.861] | [0.840–0.920] | [0.642–0.833] | [0.745–0.848] | [0.867–0.923] | [0.617–0.775] |
| **DenseNet121** | Test set | [0.797–0.912] | [0.730–0.892] | [0.831–0.942] | [0.583–0.889] | [0.697–0.867] | [0.878–0.972] | [0.469–0.796] | [0.714–0.869] | [0.865–0.944] | [0.540–0.800] |
| | Inference set | [0.794–0.874] | [0.721–0.828] | [0.858–0.932] | [0.550–0.753] | [0.729–0.839] | [0.840–0.920] | [0.582–0.785] | [0.725–0.828] | [0.859–0.917] | [0.584–0.746] |
| **Fine-tuned DenseNet121** | Inference set | [0.819–0.893] | [0.756–0.864] | [0.851–0.927] | [0.627–0.833] | [0.729–0.839] | [0.889–0.954] | [0.538–0.750] | [0.742–0.846] | [0.879–0.931] | [0.596–0.767] |

**Supplementary Table S2**. Confidence Intervals (95% CIs) for Table 2 metrics

| **Cardiomegaly** | | | | | | | | | | | |
|---|---|---|---|---|---|---|---|---|---|---|---|
| **Model** | **Evaluation Dataset** | **Accuracy** | **Precision** | | | **Recall** | | | **F1** | | |
| | | | Macro | Class 0 | Class 1 | Macro | Class 0 | Class 1 | Macro | Class 0 | Class 1 |
| **EfficientNet** | Test set | [0.836–0.893] | [0.836–0.893] | [0.822–0.902] | [0.825–0.906] | [0.836–0.893] | [0.825–0.906] | [0.821–0.903] | [0.836–0.893] | [0.833–0.894] | [0.832–0.896] |
| | Inference set | [0.824–0.858] | [0.824–0.858] | [0.801–0.850] | [0.834–0.879] | [0.825–0.858] | [0.830–0.877] | [0.805–0.852] | [0.824–0.858] | [0.821–0.858] | [0.824–0.860] |
| **InceptionV3** | Test set | [0.800–0.863] | [0.801–0.865] | [0.764–0.851] | [0.812–0.901] | [0.800–0.862] | [0.826–0.910] | [0.744–0.838] | [0.800–0.862] | [0.804–0.871] | [0.789–0.858] |
| | Inference set | [0.721–0.762] | [0.777–0.812] | [0.891–0.940] | [0.650–0.700] | [0.718–0.754] | [0.483–0.551] | [0.941–0.968] | [0.704–0.748] | [0.631–0.690] | [0.772–0.809] |
| **Fine-tuned InceptionV3** | Inference set | [0.811–0.845] | [0.815–0.848] | [0.766–0.814] | [0.850–0.896] | [0.813–0.846] | [0.861–0.903] | [0.750–0.803] | [0.811–0.845] | [0.816–0.851] | [0.803–0.842] |
| **ResNet50** | Test set | [0.811–0.872] | [0.810–0.872] | [0.806–0.893] | [0.787–0.875] | [0.810–0.872] | [0.780–0.873] | [0.811–0.898] | [0.810–0.872] | [0.803–0.872] | [0.811–0.876] |
| | Inference set | [0.811–0.847] | [0.812–0.846] | [0.803–0.854] | [0.807–0.853] | [0.811–0.846] | [0.795–0.845] | [0.813–0.862] | [0.811–0.846] | [0.805–0.843] | [0.815–0.852] |
| **VGG16** | Test set | [0.828–0.887] | [0.828–0.887] | [0.810–0.894] | [0.822–0.905] | [0.829–0.887] | [0.822–0.906] | [0.808–0.893] | [0.828–0.887] | [0.827–0.890] | [0.824–0.887] |
| | Inference set | [0.805–0.841] | [0.805–0.841] | [0.792–0.843] | [0.802–0.852] | [0.805–0.841] | [0.794–0.844] | [0.802–0.851] | [0.805–0.841] | [0.798–0.838] | [0.807–0.846] |
| **Fine-tuned VGG16** | Inference set | [0.812–0.847] | [0.812–0.846] | [0.793–0.844] | [0.815–0.861] | [0.812–0.846] | [0.809–0.858] | [0.801–0.849] | [0.812–0.846] | [0.806–0.845] | [0.813–0.850] |
| **PneumoniaMNIST** | | | | | | | | | | | |
| **Model** | **Evaluation Dataset** | **Accuracy** | **Precision** | | | **Recall** | | | **F1** | | |
| | | | Macro | Class 0 | Class 1 | Macro | Class 0 | Class 1 | Macro | Class 0 | Class 1 |
| **EfficientNet** | Test set | [0.919–0.963] | [0.909–0.962] | [0.875–0.971] | [0.922–0.970] | [0.881–0.947] | [0.789–0.915] | [0.956–0.990] | [0.897–0.952] | [0.845–0.929] | [0.945–0.975] |
| | Inference set | [0.915–0.944] | [0.905–0.943] | [0.878–0.944] | [0.920–0.951] | [0.873–0.916] | [0.777–0.861] | [0.959–0.982] | [0.889–0.927] | [0.833–0.892] | [0.943–0.963] |
| **InceptionV3** | Test set | [0.919–0.963] | [0.889–0.950] | [0.807–0.930] | [0.951–0.987] | [0.908–0.960] | [0.873–0.966] | [0.923–0.973] | [0.899–0.953] | [0.854–0.933] | [0.943–0.975] |
| | Inference set | [0.932–0.958] | [0.916–0.950] | [0.875–0.939] | [0.945–0.971] | [0.907–0.945] | [0.849–0.921] | [0.955–0.978] | [0.912–0.946] | [0.871–0.922] | [0.953–0.972] |
| **ResNet50** | Test set | [0.910–0.956] | [0.880–0.943] | [0.805–0.924] | [0.935–0.979] | [0.888–0.948] | [0.829–0.944] | [0.923–0.971] | [0.885–0.944] | [0.832–0.920] | [0.936–0.969] |
| | Inference set | [0.910–0.939] | [0.886–0.927] | [0.829–0.906] | [0.930–0.960] | [0.881–0.923] | [0.812–0.891] | [0.937–0.966] | [0.885–0.923] | [0.830–0.889] | [0.938–0.959] |
| **VGG16** | Test set | [0.928–0.968] | [0.906–0.962] | [0.850–0.956] | [0.944–0.984] | [0.906–0.962] | [0.853–0.954] | [0.942–0.984] | [0.907–0.960] | [0.865–0.943] | [0.949–0.978] |
| | Inference set | [0.921–0.950] | [0.913–0.947] | [0.885–0.951] | [0.926–0.957] | [0.883–0.926] | [0.794–0.877] | [0.961–0.984] | [0.898–0.934] | [0.847–0.903] | [0.947–0.967] |
| **Fine-tuned VGG16** | Inference set | [0.927–0.955] | [0.915–0.949] | [0.883–0.946] | [0.935–0.963] | [0.895–0.935] | [0.821–0.898] | [0.959–0.981] | [0.906–0.941] | [0.860–0.913] | [0.951–0.969] |